\documentclass[aps,
prd,
english,
nofootinbib,
reprint]{revtex4-2}

\usepackage{amsfonts,amsmath,amssymb}
\usepackage{graphicx}
\usepackage[utf8]{inputenc}
\usepackage{hyperref}
\usepackage{babel}
\usepackage[shortlabels]{enumitem}
\usepackage[table]{xcolor}
\usepackage{booktabs} 

\usepackage{lipsum}
\newcommand\e{{\rm e}}
\newcommand{\dd}{\mathrm{d}}
\newcommand{\ii}{\mathrm{i}}

\newcommand{\Qh}{\widehat Q}
\newcommand{\ch}{\chi}
\newcommand{\vac}{\mathrm{vac}}
\newcommand{\NS}{\mathrm{NS}}
\newcommand{\R}{\mathrm{R}}
\newcommand\be{\begin{equation}}
\newcommand\ee{\end{equation}}
\newcommand\bea{\begin{eqnarray}}
\newcommand\eea{\end{eqnarray}}

\begin{document}

\begin{flushright}
\phantom{
{\tt arXiv:2024.$\_\_\_\_$}
}
\end{flushright}

{\flushleft\vskip-1.4cm\vbox{\includegraphics[width=1.15in]{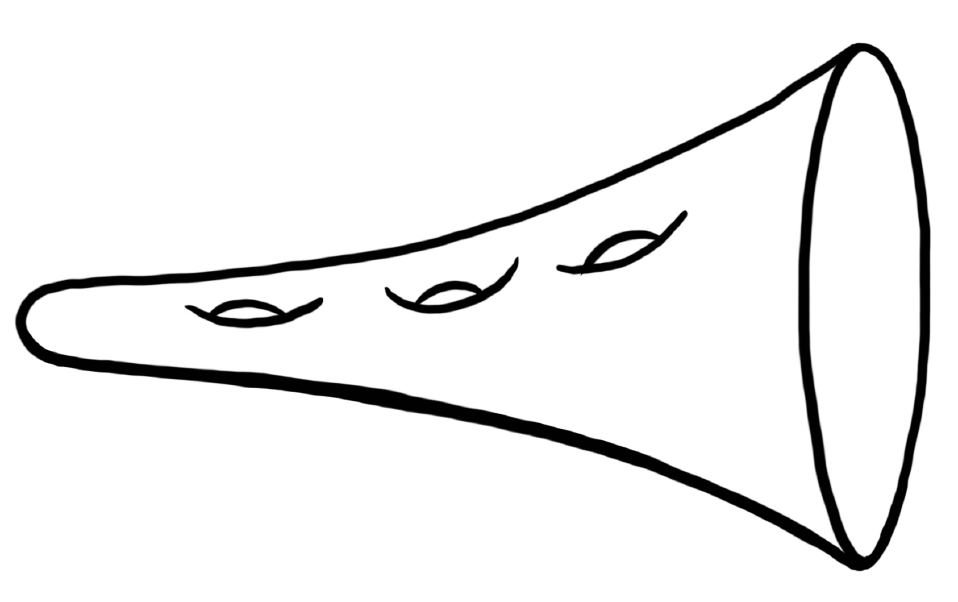}}}

\title{
Supersymmetric Virasoro Minimal Strings:\\ Unified Perturbative and Non-Perturbative Treatment}
\author{Clifford V. Johnson}
\email{cliffordjohnson@ucsb.edu}

\affiliation{Department of Physics, Broida Hall,   University of California, 
Santa Barbara, CA 93106, U.S.A.}


\begin{abstract}
There are four known ${\cal N}{=}1$ supersymmetric  Virasoro minimal string theories, naturally coming in two pairs and labeled as 0A$^\pm$ and 0B$^\pm$. We observe here that all four are  naturally  described  in terms of solutions of an ordinary differential equation that describes certain random matrix models with positive spectra. Two new additional 0A$^-$ theories, which are unorientable, are naturally described by the framework. The formulation allows for simple computation of all perturbative amplitudes, and provides complete non-perturbative information as well. Our approach makes it clear why type~0A$^-$  can be deformed to include a gas of Ramond vertex insertions, which we fully describe,  while type~0A$^+$  cannot.   While it is known that the $c{\to}\infty$ limit reduces the 0A$^-$ and 0B$^-$ models to the familiar 0A and 0B JT supergravities of Stanford and Witten, the same limit for the plus pair reduces to  non-perturbative completions of ordinary JT gravity, one of which is known.  Large families of  ${\cal N}{=}2$ and ${\cal N}{=}4$ supersymmetric analogues of the Virasoro minimal strings are also proposed and explored in detail, using the same approach.
\end{abstract}


\maketitle

\section{Introduction}
\label{sec:introduction}

The  four known ${\cal N}{=}1$ supersymmetric  Virasoro minimal string theories   naturally come as two pairs. The original  pair was  constructed by the Author in refs.~\cite{Johnson:2024fkm,Johnson:2025vyz} using random matrix model techniques. We will denote them here as 0A$^-$ and 0B$^-$, since they are associated  to  the trace with $(-1)^F$ inserted, as will be explained later. The  0A$^+$/0B$^+$ pair was recently identified by Eberhardt~\cite{Eberhardt:2026hfh} in the context of chiral 3D supergravity. That  work also noted the existence of the minus pair\footnote{Due to a different set of conventions, the use of the ``A" and ``B" ends up being reversed when translating between the two bodies of work. Our conventions here and in refs.~\cite{Johnson:2024fkm,Johnson:2025vyz} are such that upon the classical ($c\to\infty$) limit of the models, they align with the ``A" and ``B" choices of the respective JT supergravities in ref.~\cite{Stanford:2019vob}, as well as those of ref.~\cite{Klebanov:2003wg} for minimal strings.}, confirming the observations made in refs.~\cite{Johnson:2024fkm,Johnson:2025vyz} that their partition function and trumpet amplitudes   precisely describe features of  2D superconformal field theory (SCFT)  expected to be holographically dual to a 3D supergravity system, while also confirming the vanishing of all perturbative $n$--point amplitudes (for $n\geq3$) that appears quite naturally in the random matrix model critical ``two-cut'' formulation of the 0B$^+$ theory.

 It is worth taking a moment to reflect upon the striking picture that has emerged in the construction of the (bosonic) Virasoro minimal string (VMS) and the supersymmetric (SVMS) theories. It is well known that random matrix models are powerful tools for computing the topological sum (and sometimes beyond) defining various 2D gravity theories. From one perspective they are regulated versions of the Polyakov~\cite{Polyakov:1981rd,Polyakov:1981re} world-sheet path integral, where the double scaling limit~\cite{Douglas:1990ve,Brezin:1990rb,Gross:1990vs,Gross:1990aw} is a prescription for taking a continuum limit and extracting universal physics. From a holographic perspective, the study of 2D gravity should involve some kind of 1D holographic dual and the matrix model plays that role, appearing as an ensemble of 1D Hamiltonians that are in a sense ``dual'' to the 2D gravity.   The random matrix models that are used to formulate the VMS and SVMS theories are  ensembles of Hamiltonians whose leading spectral density is the Cardy density of states $\rho_0(P)$
 that arises from writing the $S$-transform of a vacuum character  in a Liouville CFT as a sum over  characters with momentum~$P$.
 There is one such object for the VMS, and hence one such string, while for the supersymmetric case there are two such characters (arising from the fact there are both  the NS$^+$ and the NS$^-$ vacua) and hence (once the separate 0A and 0B projections are made) two pairs of SVMS, broadly speaking. 

The random matrix model then  provides,  on the one hand, a self-consistent definition of a string theory by doing what it does best: Effectively performing the integration of Liouville-dressed matter\footnote{Of course the total central charge of Liouville plus matter is $26$ so as to cancel the $-26$ from the ghost sector (with the usual adjustments to $15$ and $-15$ for the supersymmetric case.) } correlators  (times the appropriate ghost factor) over the moduli space ${\cal M}_{g,n}$ of the genus $g$ Riemann surface with $n$ punctures, $\Sigma_{g,n}$. 
The special feature for these VMS constructions is that ``matter" here happens to be the timelike Liouville CFT (instead of, say, a $c<1$ minimal model,  used when building the more traditional minimal string theory).

On the other hand, this turns out to be a computation in a precise quantization of chiral three-dimensional gravity, sometimes called Virasoro TQFT~\cite{Collier:2023fwi,Teschner:2003em}.  In that setting, a  Riemann surface $\Sigma_{g,n}$ naturally arises as the boundary of a three-manifold. The Hilbert space associated with it is spanned by Virasoro conformal blocks of Liouville CFT:  In a pair-of-pants decomposition of $\Sigma_{g,n}$, the states are labelled by the Liouville momenta~$P_i$ propagating through its internal plumbing, and, the natural inner product on the Hilbert space involves integrating over each of the $P_i$ with a measure containing that {\it same} Cardy density of states $\rho_0(P)$.  This is why the matrix model, defined with $\rho_0(P)$ as its density of states, and its inherent structures that   perform integrations over the moduli space ${\cal M}_{g,n}$ with such data, is the most natural tool for implementing that inner product.  (Indeed, it is hard to think of any tool more natural!)

Arguably the most basic observable of the random matrix model is the leading macroscopic loop amplitude. Its Laplace transform is the spectral density. From the string theory perspective this is the D-brane  amplitude that is useful for probing the structure of the closed string theory. For the VMS, the remarkable result is that in the conformal field theory (spacelike and timelike Liouville plus ghosts), the appropriately defined boundary state  integrated over moduli space, gives precisely the Cardy density!  It is not the Cardy density referring to the worldsheet CFT, but instead it is natural to associate it to the  holographic boundary dual of the 3D gravity. Much the same structure goes through for the ${\cal N}{=}1$ SVMS, as observed in ref.~\cite{Johnson:2025vyz}, and more recently in ref.~\cite{Eberhardt:2026hfh}. Much more is to be uncovered  in this paper.

Given all that, it would be willfully neglectful to {\it not}  suggest that the Cardy density for  superconformal field theories with extended supersymmetry ({\it i.e.,} ${\cal N}\ge2$)
likely also define string theories and chiral 3D gravities, by constructing  the random matrix models that have them as their spectral densities. Of course, we will indeed suggest and explore this.

With all that in mind, we can state this paper's five main overarching objectives:
\begin{enumerate}
    \item To show that all four ${\cal N}{=}1$ SVMS theories listed above fit naturally together as solutions to the natural ``string equations'' used to formulate the appropriate {\it positive} random matrix models that are natural for this supersymmetric setting.  It can all be done in terms of solutions of a single special equation. This is done in {\bf Section~\ref{sec:CFT-RMM}}.
    
    \item  To show that, due to recent work~\cite{Johnson:2026twg,Johnson:2026jls,Johnson:2026jgs},  this framework allows for precise understanding of and swift computation  of the amplitudes (often called ``quantum volumes" in this context~\cite{Collier:2023cyw}\footnote{This terminology is natural since in a classical limit they become the Weil--Petersson volumes of the  moduli space of bordered Riemann surfaces.}). In fact,  closed form expressions for them will  be written down by simply tailoring general expressions derived in ref.~\cite{Johnson:2026twg}. This is presented in {\bf Section~\ref{sec:Amplitudes-Neveu-Schwarz-insertions}}.
    
    \item Moreover, crucially, the  turning on of certain Ramond sector insertions is very natural in this framework, allowing for easy  computation of the associated quantum volumes. It is explained how that works here, and once again,  various closed form formulae written in ref.~\cite{Johnson:2026jls} can be readily adapted for use here! These results are  in {\bf Section~\ref{sec:Amplitudes-with-Ramond-insertions}}.

\end{enumerate}

Finally, in view of the  paragraph above about making new string theories based on Cardy formulae for extended supersymmetry:

\begin{enumerate}
\setcounter{enumi}{3}
    \item {\bf Section~\ref{sec:an-N=2-generalization}} defines and explores  such new  supersymmetric strings  by constructing families of  ${\cal N}{=}2$ strings, exploring several example models, which show a rich assortment of features. Notably,   ${\cal N}{=}1$ SVMS appear as special $U(1)_R$ charge sectors. One of the models, at $\hat c{=}2$, has an enhanced symmetry that allows an ${\cal N}{=}4$ string (at level $1$) to be defined.  

    \item {\bf Section~\ref{sec:an-N=4-generalization}} goes even further and constructs a family of (small) ${\cal N}{=}4$ SVMS models, for arbitrary level $\hat k$. It is confirmed that the ${\hat k}{=}1$ model and the enhanced ${\cal N}{=}2$ model of Section~\ref{sec:an-N=2-generalization} coincide. 
    
    \bigskip It should be mentioned that the modular bootstrap works of Eguchi, Sugawara, and Taormina~\cite{Eguchi:2003ik,Eguchi:2008ct,Eguchi:2006tu} were  key to these latter two sections.
\end{enumerate}

The paper concludes with some brief discussion in {\bf Section~\ref{sec:closing}}, including remarks about future directions.

\section{Conformal Field Theory Meets Multicriticality}
\label{sec:CFT-RMM}
Let us quickly see why the known members of the quartet of ${\cal N}{=}1$ supersymmetric Virasoro minimal string theories, 
\{0A$^-$, 0B$^-$, 0A$^+$, 0B$^+$\}, all fit nicely into  the same intersecting underlying matrix model formulation.  (During the proceedings, it will become clear why the 0A$^-$ case naturally contains three distinct theories, so we're really dealing with a sextet.)

\subsection{Superconformal field theory reminder}

\noindent  First let us recall some conformal field theory basics, which will be useful and also help set our conventions.~\footnote{For useful reviews, from an older perspective as well as a more modern  conformal bootstrap perspective, see {\it e.g.} refs.~\cite{DiFrancesco:1997nk,Ginsparg:1988ui,Ribault:2024rvk,Kusuki:2024gtq}. For use in later sections, a review on  ${\cal N}{=}2$ superconformal methods is ref.~\cite{Lin:2016gcl}.}
The Virasoro algebra generated by the usual $L_m$:
\begin{equation}
 [L_m,L_n]
 =(m-n)L_{m+n}
 +\frac{c}{12}(m^3-m)\delta_{m+n,0}\ ,
 \label{eq:Virasoro}
\end{equation}
where $c$ is the central charge and $L_0$ measures the conformal weight $h$ of a state: $L_0|h\rangle{=}h|h\rangle$, 
is enlarged by the supergenerators $G_r$ to complete the superconformal algebra:
\begin{align}
 [L_m,G_r]
 &=\left(\frac m2-r\right)G_{m+r}\ ,
 \nonumber
 \\
 \{G_r,G_s\}
 &=2L_{r+s}
 +\frac c3\left(r^2-\frac14\right)\delta_{r+s,0}\ ,
 \label{eq:super-extension}
\end{align}
with  two standard choices of moding of $G_r$,  Neveu-Schwarz (NS), with~$r{\in}\mathbb{Z}{+}\frac12$, and Ramond (R) with~$r{\in}\mathbb{Z}$.

The CFT central charge in Liouville theory, where the energies $h$ form a continuum, is given in terms of the  standard parameter $Q$  for the supersymmetric case as: 
\begin{equation}
    c=\frac32+3Q^2\ ,\quad Q\equiv b^{-1}+b\ ,\quad \widehat Q\equiv b^{-1}-b\ ,
    \label{eq:c-Q-Qh}
\end{equation}
and we have also defined the quantity $\Qh$, which  plays a natural role.
The conformal weights of states in the SCFT are, in the NS and R Hilbert spaces: \begin{align}
  h_P=\frac{c-\frac32}{24}+\frac{P^2}{2}+\frac{\delta}{16}\ ,
  \quad
  \delta=
  \left\{\begin{matrix}
     0&(\NS) \\ 1&(\R)
  \end{matrix}
  \right. \ .\label{eq:conformal-weights}
\end{align}
In constructing characters, superscripts $+$ and $-$  distinguish the ordinary trace
from the trace with $(-1)^F$ inserted (``supertrace''). With $q\equiv\e^{2\pi\ii\tau}$ we have:
\begin{align}
&\chi_P^{\NS+}(\tau)
\equiv
{\rm Tr}^{{\NS}}
_P\left[ 
q^{L_0-\frac{c}{24}}\right]=
q^{\frac{P^2}{2}}\,\vartheta_3(\tau)^{\frac12}\eta(\tau)^{-\frac32}\ ,\nonumber\\
&\chi_P^{\NS-}(\tau)
\equiv
{\rm Tr}^{\NS}_P
\left[(-1)^Fq^{L_0-\frac{c}{24}}\right]=
q^{\frac{P^2}{2}}\,\vartheta_4(\tau)^{\frac12}\eta(\tau)^{-\frac32}\ ,\nonumber\\
&\chi_P^{\R+}(\tau)
\equiv
{\rm Tr}^{{\R}}_P\left[ 
q^{L_0-\frac{c}{24}}\right]
=2^{\frac12}q^{\frac{P^2}{2}}\,\vartheta_2(\tau)^{\frac12}\eta(\tau)^{-\frac32}\ ,
\end{align}
where we use the standard Jacobi $\vartheta$-functions and Dedekind's $\eta$-function.
(Note that a generic $\R+$ character has two states at lowest energy. Since the Ramond zero mode $G_0$ anticommutes with $(-1)^F$ and obeys $G_0^2{=}P^2/2\neq0$, it pairs bosonic and fermionic states at every level. Consequently, the corresponding $\R-$ character vanishes.)

In constructing the vacuum characters, $Q$ and $\widehat Q$ are seen to play a special role  since at imaginary Liouville momenta $P{=}\ii Q/2$ and $P{=}\ii {\widehat Q}/2$ we have $h^\NS_{\ii Q/2}{=}0$ and $h^\NS_{\ii {\widehat Q}/2}{=}\frac12$.
 The degenerate vacuum characters are, correspondingly:
\begin{align}
  \ch_{\vac}^{\NS+}(\tau)
  &=\ch_{\frac{\ii}{2}(b+b^{-1})}^{\NS+}(\tau)
   -\ch_{\frac{\ii}{2}(b-b^{-1})}^{\NS+}(\tau)\ ,
  \label{eq:NSplus-vac}\\
  \ch_{\vac}^{\NS-}(\tau)
  &=\ch_{\frac{\ii}{2}(b+b^{-1})}^{\NS-}(\tau)
   +\ch_{\frac{\ii}{2}(b-b^{-1})}^{\NS-}(\tau)\ ,
  \label{eq:NSminus-vac}\\
  \ch_{\vac}^{\R+}(\tau)&=\frac12\ch_0^{\R+}(\tau)\ ,
  \label{eq:Rplus-vac}
\end{align}
with the $\frac12$ in the last line giving one state in the $R+$ vacuum.
Under an $S$ transformation, the characters transform according to:
\begin{align}
 \chi_{P_0}^{\NS+}\!\left(-\frac1\tau\right)
 &=
 \int_{-\infty}^{\infty}\dd P\,
 \cos(2\pi P_0P)\,
 \chi_P^{\NS+}(\tau)\ ,
 \label{eq:S-transform-NSplus}\\
 \chi_{P_0}^{\NS-}\!\left(-\frac1\tau\right)
 &=
 \frac1{\sqrt2}
 \int_{-\infty}^{\infty}\dd P\,
 \cos(2\pi P_0P)\,
 \chi_P^{\R+}(\tau)\ ,
 \label{eq:S-transform-NSminus}
 \end{align}
 and:
 \begin{equation}
 \chi_{P_0}^{\R+}\!\left(-\frac1\tau\right)
 =
 \sqrt2
 \int_{-\infty}^{\infty}\dd P\,
 \cos(2\pi P_0P)\,
 \chi_P^{\NS-}(\tau)\ .
 \label{eq:S-transform-Rplus}
\end{equation}
Focusing on the vacuum characters, 
after inserting imaginary momenta into the transformation kernel, we  get one or other of the combinations:
\begin{equation}
  \cosh(\pi QP)+\xi\cosh(\pi\widehat QP)=\left\{\begin{matrix}
      2\sinh(\pi bP)\sinh(\pi b^{-1}P)\\2\cosh(\pi bP)\cosh(\pi b^{-1}P)
  \end{matrix} \right.\ ,
  \label{eq:pm-choices}
  \end{equation}
  where the choices are $\xi=\mp1$. The result is that:
\begin{widetext}
\begin{eqnarray}
 \hskip-0.6cm 
 \ch_{\vac}^{\NS+}\!\left(-\frac1\tau\right)
  =\int_0^\infty \!\!\dd P\,
  4\sinh(\pi bP)\sinh(\pi b^{-1}P)\,\ch_P^{\NS+}(\tau)\, 
\label{eq:NS-spectrum}
\end{eqnarray}
 and
\begin{eqnarray}
 \hskip-0.3cm 
\ch_{\vac}^{\NS-}\!\left(-\frac1\tau\right)
  =\int_0^\infty\!\! \dd P\,
  2\sqrt2\cosh(\pi bP)\cosh(\pi b^{-1}P)\,\ch_P^{\R+}(\tau)\, 
  \label{eq:R-spectrum}
\end{eqnarray}
    
\end{widetext}

The modular integral kernels  displayed above define measures in the
SCFT momentum $P$. To relate this to the matrix-model
momentum/energy variables,  define:
\begin{equation}
 p=\frac{P}{2}\ ,\quad
 E=p^2=\frac{P^2}{4}\ ,
 \quad
 P=2\sqrt{E}\ ,\quad
 \frac{\dd P}{\dd E}=\frac{1}{\sqrt{E}}\ ,
 \label{eq:CFT-to-matrix-momentum}
\end{equation}
and with an  extra factor of $1/\sqrt{2}$ to achieve a symmetric presentation, we define the spectral densities as:
\begin{eqnarray}
\label{eq:super-NSplus-density}
\rho^{+}_0(E)=\e^{S_0}2\sqrt{2}\frac{\sinh(2\pi b\sqrt{E})\sinh(2\pi b^{-1}\sqrt{E})}{\sqrt{E}} \ ,
\end{eqnarray}
 and
\begin{eqnarray}
 \label{eq:super-NSminus-density}
\rho^{-}_0(E)=\e^{S_0}2\sqrt{2}\frac{\cosh(2\pi b\sqrt{E})\cosh(2\pi b^{-1}\sqrt{E})}{\sqrt{E}}\ . 
\end{eqnarray}

 So there are two possible choices for leading spectral density of a supersymmetric Virasoro minimal string. One (\ref{eq:super-NSplus-density})  resembles the familiar bosonic case~\cite{Collier:2023cyw} while the other (\ref{eq:super-NSminus-density}) is the one upon which the supersymmetric strings of refs.~\cite{Johnson:2024fkm,Johnson:2025vyz} were built. 
 
   What will become clear in a moment is that both values of $\xi$ in equation~(\ref{eq:pm-choices}) are possible as natural choices in  the {\it same} random matrix model framework. Moreover, the  form~(\ref{eq:pm-choices}) strongly suggests that we interpret~$\xi$ as a variable interpolating parameter. While that lives outside what is normally done in ${\cal N}{=}1$ SCFT, it will be quite natural from the matrix model and 2D gravity perspectives, as well as ${\cal N}{=}2$ SCFT, where it is connected to spectral flow and $U(1)_R$-charge. We  will discuss this more later in Section~\ref{eq:chat-equals-3-strings}.

\subsection{Multicritical random matrix model connection}

Briefly recalling the background,   models of random matrices of size $N$ yield various gravity theories when~$N$ is taken large while  parameters in the polynomial potential ({\it i.e.} the function specifying the probability of a draw) are tuned to certain universal ``multicritical'' values.  A given gravity theory can be described as an admixture of  different amounts of the basic multicritical potentials, set by the values of the parameters called~$t_k$. The  spectral densities~(\ref{eq:super-NSplus-density}) and~(\ref{eq:super-NSminus-density}) 
correspond to  random matrix models with two specific  sets of $t_k$:
\begin{equation}
    t_k^{\pm}=2\sqrt{2}\pi\frac{\pi^{2k}}{(k!)^2}\left(Q^{2k}\mp{\widehat Q}^{2k}\right)\ ,
    \label{eq:tk-formulae}
\end{equation}
The $t_k^+$ family (which has  the  difference) first appeared in the study of the ordinary (bosonic) VMS~\cite{Johnson:2024bue,Castro:2024kpj} while the $t_k^-$ family (which has the sum) was originally identified in ref.~\cite{Johnson:2024fkm} as relevant to the SVMS.

The fact that the matrix model formulae for the bosonic and supersymmetric $t_k$ differed only by the swop of a sign seemed remarkable at first, but the reason behind it is now rather clear: The two cases are choices between the NS$^+$ and NS$^-$ vacua (and their resulting Cardy densities) of the previous section.  
  
To find a natural setting for all this we turn to the random matrix models in a bit more depth.  So very much of this was explored in detail in this very context in ref.~\cite{Johnson:2025vyz}, and so we will simply borrow and adapt results as needed, without burdening the Reader with too much repetition of lengthy exposition. 

With holographic lessons in mind, these random matrix models are to be thought of as exploring random Hamiltonians. Hence the relevant matrix models for this supersymmetric setting should naturally be of positive matrices.  There are two  natural formulations. One arises  from a Wishart-type construction, generalized to include multicritical behaviour arising at the endpoint of a double-scaled density of eigenvalues. This gives rise to type A models. The other way is to study multicriticality arising from the endpoints of two densities of eigenvalues as they collide and merge. This gives type B models. The ``A'' and ``B'' terms match the conventions concerning whether one sums over worldsheets with $(-1)^\zeta$ inserted or not, where $\zeta\in\mathbb Z_2$ is the mod-two index of the Dirac operator: $\zeta{=}0$ for an even spin structure and $\zeta{=}1$ for an odd one~\cite{Stanford:2019vob}.

Let us focus on the type~A case first. An efficient way of formulating everything  is to use a set of orthogonal polynomials $P_n$ (where $n$ is an index) that satisfy a recursion relation. All physical quantities can be  expressed in terms of the $P_n$ and so computing everything boils down to determining them. The matrix model's specific potential is encoded in a difference equation for the recursion coefficients, and in the double-scaling limit~\cite{Brezin:1990rb,Douglas:1990ve,Gross:1990vs} at universality, the ratio $n/N$ becomes a continuous parameter $X\in(0,1)$. Criticality is at $X=1$, and universal physics is found by scaling into  the infinitesimal neighbourhood of this value. Introducing a book-keeping parameter $\delta$ that goes to zero as $N\to\infty$, the scaling  part of $X$ away from unity is parameterized by the coordinate~$x$: $X=1+(x-\mu)\delta^{2k}$ where $\mu$ will be discussed below and $x\in\mathbb{R}$. In a similar fashion,  a  combination of the recursion  coefficients, which have index $n$, become a continuous function of $X$ at large $N$, ultimately yielding a scaling function $u(x)$ in the neighbourhood of the critical point. The fate of the difference equation of the recursion coefficients is a non-linear ODE for $u(x)$ called the string equation.

A core point is that while solutions of the ODE live on the whole line $-\infty\leq x\leq+\infty$, it is only a semi-infinite range of $x$ running from $-\infty$ to the endpoint value $\mu$ that defines the matrix model.\footnote{This corresponds to the fact that while the orthogonal polynomials come in infinite families, we only use $N$ of them to build the matrix model. (We do take $N\to\infty$, but there's still a notion of a ``top'' one; it is indexed by $\mu$.)} The leading spectral density is given by:
\begin{equation}
    \label{eq:spectral-density-leading}
\rho^{(b)}_{0}(E) \!= \frac{1}{2\pi\hbar}\int_{-\infty}^\mu\!\frac{\Theta(E{-}u_0(x)) dx}{\sqrt{E-u_0(x)}}\ ,
\end{equation}
where, for $\Gamma=0$, $u_0(x)$ is the leading piece in an $\hbar$ (genus) expansion of the function $u(x)$: $u(x)=u_0(x)+\sum_{g=1}^\infty u_{2g}(x)\hbar^{2g}$, where $\hbar$ is the scaling part of $\frac{1}{N}$.

So having the function $u_0(x)$ is equivalent to knowing the leading spectral density. As will be reviewed shortly, $u_0(x)$ is determined by the parameters $t_k$ (describing the (critical) potential of the model). Actually, the parameter $\mu$ (that highest value of $x$ in the integral~(\ref{eq:spectral-density-leading})) is actually $t_0$ (see below), and this will be important when interpreting the formulae~(\ref{eq:tk-formulae}) for $t_k$ determined by the Cardy density.

The full function $u(x)$ determines the free energy according to:
\begin{equation}
    \label{eq:0A-free}
    2\hbar^2\frac{\partial^2 F^{\rm 0A}}{\partial x^2}=u(x)\ ,
\end{equation}
where it is understood that  after the integrals are performed the result is evaluated at $x=\mu$. Indeed once $u_0(x)$ is determined, the string equation (to be written shortly) determines its genus   corrections $u_{2g}(x)$ in terms of $u_0$'s derivatives, and hence through~(\ref{eq:0A-free}) so are the perturbative corrections $F^{\rm 0A}_g$ to the free energy. 

Consequently, all perturbative $n$-point correlators of the random matrix model in this limit depend only on $u_0(x)$ and its derivatives evaluated at $x=\mu$. The value of the function itself there, $u_0(\mu)$, is the threshold energy~$E_0$ of the density function. Note that ref.~\cite{Johnson:2026twg}  recently has shown how to quickly write precise formulae for the $n$-point correlators at any genus $g$. We shall use them a lot presently.

In the case of the relevant random matrix models, the ODE for the function $u(x)$ is~\cite{Morris:1990bw,Dalley:1991qg,Dalley:1992br}:
\begin{equation}
\label{eq:big-string-equation}
u{\cal R}^2-\frac{\hbar^2}2{\cal R}{\cal R}^{\prime\prime}+\frac{\hbar^2}4({\cal R}^\prime)^2=\hbar^2\Gamma^2\ .
\end{equation}
 Here:
 \begin{equation}
      {\cal R}{\equiv}\sum_{k=1}^\infty t_k R_k[u]{+}x\ ,
      \label{eq:R-def}
 \end{equation}
where 
$R_k[u]{=}u^k+\cdots+\# \hbar^{2k-2}u^{(2k-2)}$ is the $k$th Gel'fand-Dikii polynomial~\cite{Gelfand:1975rn} in~$u(x)$ and its $x$-derivatives, normalized here so that the purely polynomial part is unity. (Here, $u^{(m)}$ means the $m$th $x$-derivative, and intermediate terms involve products of lower derivative orders.) The first four are:
 \begin{eqnarray}
 \label{eq:GD-polynomials}
 &&R_0[u]{=}1\ ,\quad R_1[u]{=}u\ ,\quad  R_2[u]{=}u^2{-}\frac{\hbar^2}{3}u^{\prime\prime}\ , \quad \text{and}\nonumber\\
 &&R_3[u]{=}u^3{-}\frac{\hbar^2}{2}(u^\prime)^2{-} {\tiny \hbar^2}uu^{\prime\prime}{+}\frac{\hbar^4}{10}u^{\prime\prime\prime\prime}\ .
 \end{eqnarray}   
Higher $R_k[u]$ can be obtained using a recursion relation, but it is  not needed here. It will be  enough to know that at leading order in a small $\hbar$ expansion, all terms except the  $u^k$ in $R_k[u]$ can be discarded.  

It should be clear now that  (\ref{eq:GD-polynomials}) and~(\ref{eq:R-def}) are consistent with the fact that $t_0$ sets a value of $x$ and that is in fact the $\mu$ seen in (\ref{eq:spectral-density-leading}).\footnote{This is also consistent with the underlying KdV flows: $\frac{\partial u}{\partial t_k}{=}\frac{\partial}{\partial x}R_{k+1}[u]$, which won't be used explicitly in this paper.} This will be important in a moment. 
Taking $\Gamma{=}0$ for now (we will turn it on later) we see that at leading order we must solve:
\begin{equation}
     u_0{\cal R}_0^2=0\ ,\quad \text{with}\quad {\cal R}_0\equiv\sum_{k=1}^\infty t_k u_0^k+x\ ,
     \label{eq:classical-one}
 \end{equation}
 and the non-trivial solutions of interest have ${\cal R}_0=0$ for $x<0$, and $u_0(x){=}0$ for $x\geq0$.
 The $u_0$ that solves this (given a set $t_k$) determines the leading spectral density $\rho_0(E)$ through~(\ref{eq:spectral-density-leading}). 
 
 For many applications, the logic can be run  in the other direction: Given a leading spectral density $\rho_0(E)$, perhaps determined by a separate gravity computation,  what is the matrix model function $u_0(x)$ (and hence the~$t_k$) that corresponds to it? (It is this procedure that yielded the $t_k$ given in equation (\ref{eq:tk-formulae}).)

 Finally we highlight a crucial feature that is very relevant to the physics studies in this paper. There are two classic kinds of low energy behaviour at the endpoints of spectral densities, the Wigner-type square-root ``soft-edge'' fall off, or the Wishart-type inverse square-root ``hard-edge'' divergence.   In fact, the  string equation admits solutions that yield both classes of behaviour, and moreover shows how to migrate from one class of behaviour to the other.

 The classic soft-edge random matrix models (such as those that usually appear in bosonic systems) have $\mu{\leq}0$, with $u_0(x)$ and its derivatives having  values there that determine  the correlators. The lowest energy behaviour is controlled  by $t_1$, and so (\ref{eq:classical-one}) gives $u_0{\simeq} {-}x/t_1{+}\cdots$. Putting this into (\ref{eq:spectral-density-leading}) yields the classic square root  Wigner behaviour: $\rho_0(E){\simeq}\sqrt{E-E_0}+\cdots$ where $E_0{=}{-}\mu/t_1$.  
 
 Hard edge systems by contrast  have $\mu{>}0$. This is naturally captured by having the leading solution for the function be  $u_0(x)=0$, and  from~(\ref{eq:spectral-density-leading})  the resulting spectral density near the endpoint has
 the classic $\rho_0{\simeq}1/\sqrt{E}$ behaviour as a result.

 For completeness, note that for $\Gamma=0$, the string equation~(\ref{eq:big-string-equation}) has the solution ${\cal R}=0$ for all $x$. This is the string equation of the original ordinary Hermitian matrix models. The leading order solution is then ${\cal R}_0=0$  for all $x$, in contrast to solution class described just below~(\ref{eq:classical-one}) where that is only true for $x\leq0$.  If ${\cal R}_0=0$  for all $x$, then there is no finite $x$ region where, the   $u_0(x)$ vanishes, and so only soft edge behaviour is available for such models. Turning back to the full equation and considering perturbation theory, if developing it for $\mu\leq0$,  the perturbative solutions with leading form ${\cal R}_0{=}0$ for either scheme are identical at every order, and hence the soft edge perturbative physics is indistinguishable from the bosonic case if they share the same~$t_k$. How solutions transition to the $x>0$  regime ({\it i.e.,} solving either ${\cal R}_0{=}0$ or $u_0{=}0$ there) is entirely a non-perturbative matter. This is at the heart of how one can supply well-behaved non-perturbative {\it matrix model motivated} completions to perturbation theory that originated in poorly behaved bosonic contexts.\footnote{\label{fn:non-pert}This goes all the way back to the work of refs.~\cite{Dalley:1991qg,Dalley:1991vr}, and in the current context is precisely what was used to provide a matrix model method for a non-perturbative completion of JT gravity~\cite{Johnson:2019eik,Johnson:2020exp,Johnson:2022wsr}, allowing for reliable computations of key features that depend on non-perturbative physics such as the complete spectral form factor, free energy {\it etc}.}

 With all that in mind, turning back to the  formulae~(\ref{eq:tk-formulae}) for the $t_k$ of the SVMS we see some immediate implications. For the plus case, $\mu=t_0$ must be zero, and therefore we must expand the physics around $\mu=0$, resulting in soft-edge-like results. So (in view of footnote~\ref{fn:non-pert}) perturbatively the results will be {\it identical} to solutions that come from just solving and perturbing around the bosonic string equation ${\cal R}_0=0$, but will have different non-perturbative physics: The full solution for all $x$ asymptotes to $0$ in the $x>0$ regime. 
 
 Meanwhile, for the minus case in (\ref{eq:tk-formulae}), $\mu=t_0=4\sqrt{2}\pi$. This pushes perturbation theory into the $x>0$ regime where the leading solution is the hard-edge $u_0(x)=0$ solution, and the resulting correlators are of a very different sort from the plus case.

 In short, we see that the string equation ``knows'', from the matching to the conformal field theory results, exactly how to self-consistently pick the two different classes of solution.
The difference between the two classes of solution is made even more stark when moving away from just NS insertions, as we shall see in Section~\ref{sec:Amplitudes-with-Ramond-insertions}.

For the 0A$^-$ case where $\mu>0$, we need  the  $x>0$ expansion of the  $u(x)$ solution of the string equation~(\ref{eq:big-string-equation}), and some example orders are (now with $\Gamma$ turned on):
\begin{widetext}
\begin{align}
u(x)
={}&0+
\left(\Gamma^2-\frac14\right)
\frac{\hbar^2}{x^2}
\Bigg\{
1
-2t_1\left(\Gamma^2-\frac94\right)
\frac{\hbar^2}{x^3}
+\left(\Gamma^2-\frac94\right)
\frac{\hbar^4}{x^6}
\left[
7t_1^2\left(\Gamma^2-\frac{21}{4}\right)
-2t_2x\left(\Gamma^2-\frac{25}{4}\right)
\right]
+\cdots 
\Bigg\}\ ,
\label{eq:positive-x-u-expansion}
\end{align}
\end{widetext}
showing the leading $u_0(x)=0$ classical solution already discussed. But there is an additional feature: For $\Gamma=\frac12$, every order in the $\hbar^2$ genus expansion vanishes. This has the same consequences already noted in this context~\cite{Johnson:2020heh} for JT supergravity: In addition to the $\Gamma=0$ theory, the matrix model naturally knows about  two 0A$^-$ SVMS theories. These $\Gamma=\pm\frac12$ cases correspond to the two choices~\cite{Stanford:2019vob} of pin projection that can be introduced in summing over {\it unorientable} worldsheets. We have therefore enlarged the roster of SVMS theories from four to six. By virtue of the fact that $u(x)=0$ to all orders for these cases, the two additional theories have {\it all perturbative amplitudes vanishing}.

Before proceeding further, let's see how we can already compute amplitudes for NS insertions  by adapting  the formulae that ref.~\cite{Johnson:2026twg} wrote for $W_{g,n}$ random matrix correlators.  We must specialize by using the appropriate~$t_k$, (equivalently, knowledge of $u_0(x)$ at $x{=}\mu$) and then convert them to quantum volumes using the following conventions for the Laplace transform relations for the minus sector:
\begin{equation}
\label{eq:quantum-volume-conversion}
    W_{g,n}(\{ z_i\}) = \prod_{i=1}^n\Biggl[\int_0^\infty 4\pi P_i \dd P_i \e^{-4\pi P_i z_i}\Biggr]V^{(b)}_{g,n}(\{P_i\})\ .
\end{equation}
For the plus sector we  instead use the convention with an extra factor of $\sqrt2$ per leg:
\begin{equation}
\label{eq:quantum-volume-conversion-plus}
    W_{g,n}(\{ z_i\}) = \prod_{i=1}^n\Biggl[\int_0^\infty 4\sqrt2\pi P_i \dd P_i \e^{-4\pi P_i z_i}\Biggr]V^{(b)}_{g,n}(\{P_i\})\ ,
\end{equation}
where the extra per-leg factor is consistent with  the scaling choice made below equation~(\ref{eq:CFT-to-matrix-momentum}) in order to present the spectral densities $\rho_0^\pm(E)$ with matching factors (along with a tacit rescaling of $\hbar$ between the plus and minus theories). It will result in the standard unit normalization of  the three point function:
$V^{(b)}_{0,3}$ in the plus sector (see below).

\phantom{.}

\section{Amplitudes: Neveu-Schwarz Insertions}
\label{sec:Amplitudes-Neveu-Schwarz-insertions}

\subsection{Some 0A$^+$ Amplitudes}
For the 0A$^{+}$ case, with the soft edge, we use the $t_k^{+}$ choice from~(\ref{eq:tk-formulae}) (with the relative minus sign), and  work out the first few derivatives of $u_0(x)$ at $x=\mu=0$ to be:
\begin{align}
 u_0'(0)&=-\frac{\sqrt2}{16\pi^3}\ ,\quad
 u_0''(0)=-\frac{2c-15}{768\pi^4}\ ,\nonumber\\
 u_0^{(3)}(0)&=-\frac{\sqrt2(20c^2-300c+1077)}{147456\pi^5}\ ,\nonumber\\
 u_0^{(4)}(0)&=-\frac{(2c-15)(244c^2-3660c+12429)}{21233664\pi^6}\ .
 \label{eq:bosonic-u0-and-derivatives}
\end{align}
From this, using the general formulae from ref.~(\cite{Johnson:2026twg}) we get the following examples of $W^{(b)}_{g,n}$.
At genus zero:

\begin{align}
 &W^{(b)}_{0,3}(z_1,z_2,z_3)
 =\frac{\sqrt2}{32\pi^3}\prod_{i=1}^3\frac1{z_i^2}\ ,
 \\&W^{(b)}_{0,4}(z_1,z_2,z_3,z_4)
 =\prod_{i=1}^4\frac1{z_i^2}
 \left[
 \frac{2c-15}{1536\pi^4}
 +\frac{3}{512\pi^6}\sum_{i=1}^4\frac1{z_i^2}
 \right]\ ,
\end{align}
and at genus one:
\begin{widetext}
\begin{align}
\nonumber\\
 &\hskip-1.3cm W^{(b)}_{1,1}(z)
 =\frac{\sqrt2(2c-15)}{2304\pi}\frac1{z^2}
 +\frac{\sqrt2}{256\pi^3}\frac1{z^4}\ ,\nonumber\\
 &\hskip-1.3cm W^{(b)}_{1,2}(z_1,z_2)
={}\frac1{z_1^2z_2^2}
\left[
\frac{(2c-19)(2c-11)}{73728\pi^2}
+\frac{2c-15}{6144\pi^4}\sum_{i=1}^2\frac1{z_i^2}
+\frac{5}{4096\pi^6}\sum_{i=1}^2\frac1{z_i^4}
+\frac{3}{4096\pi^6}\frac1{z_1^2z_2^2}
\right]\ .
\end{align}
 Using the plus-sector Laplace transform conventions in~(\ref{eq:quantum-volume-conversion-plus}) we get the amplitudes (``quantum volumes"):
\begin{align}
& V^{(b)}_{0,3}=1\ ,\nonumber
\\& V^{(b)}_{0,4}(P_1,P_2,P_3,P_4)=\frac{2c-15}{24}+\sum_{i=1}^4P_i^2\ ,\nonumber\\
 &V^{(b)}_{1,1}(P)
 =\frac{2c-15}{576}+\frac{P^2}{24}\ ,\nonumber\\
 &V^{(b)}_{1,2}(P_1,P_2)
 =\frac{(2c-19)(2c-11)}{9216}
 +\frac{2c-15}{288}(P_1^2+P_2^2)
 +\frac{(P_1^2+P_2^2)^2}{48}\ .
 \label{eq:N1-quantum-volumes-0Aplus}
\end{align}
It is easy to compute many more $V^{(b)}_{g,n}$ using ref.~\cite{Johnson:2026twg}, but we will stop here.

\end{widetext}

\subsection{Some 0A$^-$ Amplitudes}
For the 0A$^{-}$ case, with the hard edge, for the $\Gamma{=}\pm\frac12$ theories we already saw that all amplitudes vanish for all $(g,n)$ (except the universal $(0,2)$ cylinder case), because of the vanishing of $u(x)$ (see equation~(\ref{eq:positive-x-u-expansion})) to all orders. For $\Gamma{=}0$ where the $u(x)$ perturbative  series is non-vanishing we can simply use the closed form formulae derived  in Section III of ref.~\cite{Johnson:2026twg} for $W_{g,n}(\{t_k\})$ and substitute our $t_k^-$ (with $\Gamma{=}0$). We will need $t^-_k$  $(k=1,2,3)$ for up to genus 4. Converting those~$t_k$ to expressions involving~$c$, we get:

\begin{align}
\label{eq:tk-minus-in-terms-of-c}
 &t_1^-=\frac{2\sqrt2\pi^3}{3}(2c-15)\ ,
 \\
 &t_2^-=\frac{\sqrt2\pi^5}{36}(4c^2-60c+369)\ ,
 \\
 &t_3^-=\frac{\sqrt2\pi^7}{1944}(2c-15)(4c^2-60c+657)\ .
\end{align} 
\begin{widetext}
The  explicit $W^{(b)}_{g,n}$ for $g=0,1,2,3$ are given in ref.~\cite{Johnson:2026twg}, and don't need repetition. We merely need to inverse Laplace transform them with the new convention
(\ref{eq:quantum-volume-conversion}) for writing them in terms of momenta $P_i$, with the resulting  quantum volumes:

\begin{align}
 & V^{(b)}_{0,n}=0\ ,\, (n\geq3)\ ,\quad
 V^{(b)}_{1,n}
 =(-1)^n\frac{(n-1)!}{8}(4\pi)^n\ ,\quad 
 \nonumber\\&V^{(b)}_{2,n}
 =(-1)^n\frac{3(n+1)!}{512}(4\pi)^n
 \left[4t_1^-(n+2)+(4\pi)^2\sum_{i=1}^nP_i^2\right]\ ,
 \nonumber\\
\text{and} \nonumber \\
 &V^{(b)}_{3,n}
=(-1)^n\frac{(n+3)!}{10240}(4\pi)^n
\Bigg[
(n+4)\{21(n+5)(t_1^-)^2-50t_2^-\}
+\frac{63}{6}(n+4)t_1^-(4\pi)^2\sum_{i=1}^nP_i^2
+\frac{375}{4}\frac{(4\pi)^4}{120}\sum_{i=1}^nP_i^4
\nonumber\\
&\hspace{12.5cm}
+\frac{189}{2}\frac{(4\pi)^4}{36}\sum_{1\leq i<j\leq n}P_i^2P_j^2
\Bigg]\ .
\label{eq:amplitudes-hard-edge}
\end{align}
(Note that a few of these for select $g$ and $n$ were obtained in ref.~\cite{Johnson:2025vyz}, but in a convention where there is an extra  factor of~$2\pi$ per leg.)
There is a closed form formula for $V^{(b)}_{4,n}$ that was computed in ref.~\cite{Johnson:2026twg}. It is on the longer side, so we will not write it here. The current convention introduces a factor of $(4\pi)^n$,  and  we must set $\Gamma$ to zero and of course substitute $t_k$ there to the $t^-_k$ here and $b_i{\to}P_i$.


\end{widetext}
\subsection{Some 0B$^{-}$ Amplitudes} The relevant string equation for type 0B models comes from working with double-scaled two-cut models. This was successfully worked out in detail in ref.~\cite{Johnson:2021owr} for the 0B supergravity model of Stanford and Witten~\cite{Stanford:2019vob}, and further developed for the 0B SVMS more recently in ref.~\cite{Johnson:2025vyz}. We focus on the symmetric perturbations of the two-cut system, for which the  string equations are the Painlev\'{e}~II heirarchy~\cite{Periwal:1990gf,Periwal:1990qb,Crnkovic:1990mr}, for a function $r(x)$ (the analogue of $u(x)$ for the 0A system), from which the free energy is derived~{\it via}:
\begin{equation}
    \label{eq:0B-free}
    \hbar^2\frac{\partial^2 F^{0B}}{\partial x^2}=r^2\ .
\end{equation}
We need not show the Painlev\'e~II system of ODEs for $r(x)$ here because happily, the solutions for $r(x)$ of that system can be made by combining solutions $u(x)$ of the string equation~(\ref{eq:big-string-equation}). In fact in ref.~\cite{Johnson:2025vyz}  it was shown that:
\begin{equation}
    \label{eq:u0B-defintion}
    r^2=\frac12\left[u_{\Gamma=+\frac12}+u_{\Gamma=-\frac12}\right]\ .
\end{equation}
The payoff for writing things in this way is that correlators $W_{g,n}$ for the models will come from   using each of the $u$ functions in the sum in the manner made clear in ref.~\cite{Johnson:2026twg}, {\it i.e.,} expanding the function in $\hbar$ powers and seeing that they can all be written in terms of the leading solution and its derivatives evaluated at $x=\mu$.

Now, the key observation is that  for $x>0$, the $u_{\Gamma=\pm\frac12}$ functions actually {\it vanish} to all orders in perturbation theory, as we saw,  and hence so does $r^2$.  Since the 0B$^-$ model has all of its correlators defined in terms of this function and its derivatives at $x=\mu>0$, we immediately see why the correlators of 0B$^-$ all vanish for any $n$ and~$g$ aside from the universal $W_{0,2}$ (the cylinder). This was a key prediction of ref.~\cite{Johnson:2025vyz}, confirmed recently in ref.~\cite{Eberhardt:2026hfh} using alternative methods.

\subsection{Some 0B$^{+}$ Amplitudes}

This case is new territory for this approach, since now we must study the expansion of $r^2$ for $x<0$. This can be developed  systematically by noting that $x<0$  solutions of (\ref{eq:big-string-equation}) are perturbatively equivalent to solving 
\begin{equation}
    {\cal R}-2\hbar\Gamma{\widehat R}=0\ ,
\end{equation}
where ${\widehat R}$ solves the Gelfand-Dikii equation:
\begin{equation}
    \label{eq:gelfand-dikii}
    4[u(x)-\sigma] {\widehat R}^2-2\hbar^2{\widehat R}{\widehat R}^{\prime\prime}+\hbar^2({\widehat R}^\prime)^2=1\ ,
\end{equation}
with energy $\sigma=0$. After a bit of work, one can develop the expansions for $u_\Gamma$ in this regime, expressing everything in terms of the leading solution $u_0(x)$ that solves ${\cal R}_0=0$, where~${\cal R}_0$ is defined in~(\ref{eq:classical-one}). The result is:
\begin{widetext}
    \begin{align}
u_{\Gamma}(x)
={}&u_0
-\hbar\Gamma\,\frac{u_0'}{\sqrt{u_0}}
+\hbar^2\frac{\partial^2}{\partial x^2}
\left[
-\frac{1}{12}\log\!\left(u_0'\right)
+ \frac{\Gamma^2}{2}\log u_0\right]
+\hbar^3\frac{\partial^2}{\partial x^2}
\left[
\frac{\Gamma}{12}
\frac{u_0''}{\sqrt{u_0}\,u_0'}
-\frac{\Gamma\left(1+4\Gamma^2\right)}{24}
\frac{u_0'}{u_0^{3/2}}
\right]
\nonumber\\[2mm]
&\hskip1cm+\hbar^4\frac{\partial^2}{\partial x^2}
\Bigg\{
\frac{u_0^{(4)}}{288\left(u_0'\right)^2}
-\frac{7u_0''u_0'''}{480\left(u_0'\right)^3}
+\frac{\left(u_0''\right)^3}{90\left(u_0'\right)^4}
+\Gamma^2
\left(
-\frac{u_0'''}{24u_0u_0'}
+\frac{\left(u_0''\right)^2}
       {24u_0\left(u_0'\right)^2}
+\frac{u_0''}{12u_0^2}
-\frac{7\left(u_0'\right)^2}{96u_0^3}
\right)
\nonumber\\
&\hskip10cm
+\Gamma^4
\left(
\frac{u_0''}{24u_0^2}
-\frac{\left(u_0'\right)^2}{12u_0^3}
\right)
\Bigg\}
+O\!\left(\hbar^5\right)\ .
\label{eq:uGamma-expansion-hbar4}
\end{align}
Adding the cases with opposite sign of~$\Gamma$ gives:
\begin{align}
\frac12\left[u_{+\Gamma}+u_{-\Gamma}\right]
={}&
u_0
+\hbar^2\frac{\partial^2}{\partial x^2}
\left[-\frac{1}{12}\log u_0'
+\frac{\Gamma^2}{2}\log u_0
\right]
+
\hbar^4\frac{\partial^2}{\partial x^2}
\Bigg\{
\frac{(u_0'')^3}{90(u_0')^4}
-
\frac{7u_0'''u_0''}{480(u_0')^3}
+
\frac{u_0^{(4)}}{288(u_0')^2}
\nonumber\\
&
+
\Gamma^2
\left[
-\frac{7(u_0')^2}{96u_0^3}
+
\frac{u_0''}{12u_0^2}
+
\frac{(u_0'')^2}{24(u_0')^2u_0}
-
\frac{u_0'''}{24u_0'u_0}
\right]
+
\Gamma^4
\left[
-\frac{(u_0')^2}{12u_0^3}
+
\frac{u_0''}{24u_0^2}
\right]
\Bigg\}
+O(\hbar^6)\ ,
\label{eq:minus-x-expansion-r-squared}
\end{align}
where we left $\Gamma$ explicit for a moment to show the structure of the solution more clearly. In particular, all odd powers of $\Gamma$ cancel in the sum.
\vfill\eject
\end{widetext}
Note that for $k=1$ we have $u_0=-x$ and this becomes (putting $\Gamma^2=\frac14$):
\begin{align}
r^2
={}&
-x
-\frac{\hbar^2}{8x^2}
+\frac{9\hbar^4}{32x^5}
+O(\hbar^6)\ ,
\end{align}
which is in fact the asymptotic expansion of the Hastings-Mcleod~\cite{Hastings-McLeod} solution of Painlev\'e~II.

The key  points at this juncture are that  (1) the $\Gamma$-independent terms are precisely the same as the soft edge expansion used for the bosonic theories, and (2) the terms involving  $\Gamma$  are in fact divergent when  $u_0\to0$, and since~$u_0$ vanishes at $x=\mu=0$, where genus perturbation theory will be defined for the 0B$^+$  case, such terms
lie outside the realm of what can be captured by the genus expansion. The divergence of those terms does not signal a sickness of the theory, but just the process of perturbatively expanding and using the solution all the way up to $x=0$. The full non-perturbative theory is well defined, and in fact $r^2$ and its derivatives are finite and non-zero at $x=0$.  A key point here is that these terms are not artifacts, but key components of the physics distinguishing it from the bosonic theory it resembles perturbatively. Further aspects of these  terms are discussed in footnote~\ref{fn:the-extra-terms}, including how to perturbatively regulate them with a simple physical deformation.\footnote{\label{fn:the-extra-terms}In fact, it is straightforward to regulate the divergent terms so that they can be part of the genus expansion. If instead one uses non-zero $\sigma$ in equation~(\ref{eq:gelfand-dikii}), all the pure powers of $u_0$ in the $\Gamma$-dependent terms in the expansion~(\ref{eq:minus-x-expansion-r-squared}) become  powers of $u_0-\sigma$ instead. So at $x=0$ where we evaluate correlators, these terms will now constitute a series of {\it finite} closed-string-like terms at every genus. Remembering that $|\Gamma|=\frac12$ here, something that in the type~0A context is connected to an unorientable theory, these terms bring to mind the purely closed string sector of an unoriented theory.}

At this point, looking at the $\Gamma$-independent
sector that gives a nice genus perturbation theory again, using the~$t_k^+$ again we have correlators/amplitudes built from the identical $u_0(x)$ as we had for the 0A$^+$ case, and hence again we see that {\it at least as far as perturbation theory goes} the 0B$^+$ theory again resembles the ordinary VMS perturbation theory. 

There is an important normalization factor however. Compare the free energies for the 0A and 0B sectors given in equations~(\ref{eq:0A-free}) and~(\ref{eq:0B-free}), and recall that $r^2$ has been constructed here from solutions  $u$ of the string equation. The relative factor of two between the free energies amounts to a rescaling of the string coupling between the 0A$^+$ and 0B$^+$ theories: 
\begin{equation}
    \hbar_{\rm 0B}=\sqrt{2}\hbar_{\rm 0A}\ .
\end{equation}
Since all the universal formulae in ref.~\cite{Johnson:2026twg} for correlators and hence amplitudes $V^{(b)}_{g,n}$ are based on $u_0$ and its derivatives, this translation results in a factor relating the amplitudes and free energies, {\it i.e.,} genus by genus:
\begin{align}
&F_g^{0B^+}=2^g F_g^{0A^+},
\\
&V_{g,n}^{(b),0B^+}(P_1,\ldots,P_n)
=
2^g\,
V_{g,n}^{(b),0A^+}(P_1,\ldots,P_n)\ ,
\label{eq:A-to-B-scaling}
\end{align}
and as we saw, the $V_{g,n}^{(b),0A^+}$ are identical to the bosonic VMS amplitudes (after translating central charges appropriately).
Note that the resulting relative rescaling~(\ref{eq:A-to-B-scaling}) appeared in the approach of ref.~\cite{Eberhardt:2026hfh} as a result of the fact that the A and B theories differ by whether or not the sum over spin structure is weighted by $(-1)^\zeta$, which introduces the relative factor $2^g$. See also ref.~\cite{Stanford:2019vob}.

That perturbatively 0B$^{+}$ and 0A$^{+}$  have soft-edge (bosonic-looking) correlators was observed in ref.~\cite{Eberhardt:2026hfh} using techniques in intersection theory. We have recovered that here, although it is clear here that this does not mean that the theories are identical to the bosonic VMS as matrix integrals. Rather, within the positive matrix models that are appropriate for supersymmetry  there is a natural choice of parameters that admits soft edge expansions that yield amplitudes that are equivalent to the bosonic case.  Beyond perturbation theory the models are quite different. This isn't a contradiction with the computational results of ref.~\cite{Eberhardt:2026hfh} so much as it is a reminder that the key techniques used there for exploring correlators can really only see physics that can be arranged geometrically into genus perturbation theory. 

Indeed, we've seen how these solutions naturally sit alongside the hard edge models yielding the physics of the 0A$^-$/0B$^-$ sectors. Moreover, all of these models are non-perturbatively well-defined, as already made clear in refs.~\cite{Johnson:2024fkm,Johnson:2025vyz}, in stark contrast to the ordinary VMS models (except the $b{=}1$ VMS case, which has good non-perturbative behaviour~\cite{Collier:2023cyw,Johnson:2024bue}).

\section{Amplitudes: Adding Ramond Insertions}
\label{sec:Amplitudes-with-Ramond-insertions}

Equation~(\ref{eq:big-string-equation}), which  we've been using to  organize all the physics, for 0A or 0B, has parameter $\Gamma$ on the right hand side. So far we have had $\Gamma$ fixed: For defining the 0A theories we had $\Gamma=0$ (although it is clear that the non-oriented $\Gamma{=}\pm\frac12$ SVMS theories also exist in their own right!), while 0B theories were built from combining the $\Gamma=+\frac12$ and $\Gamma={-}\frac12$ $u(x)$-solutions together.

Recent work~\cite{Johnson:2026twg,Johnson:2026jls,Johnson:2026jgs} has shown how turning on large~$\Gamma$ yields a deformation of the classical theory that corresponds to turning on Ramond backgrounds. In fact, this framework turns out to be very powerful for studying that sector, especially as compared to 
other approaches. For computing the volumes of the moduli space of super-Riemann surfaces with both NS boundaries and R punctures, it enables many
closed-form expressions  to be readily derived. Moreover, a deformation of the standard spectral curve was derived this way, allowing the topological recursion techniques to be applied if so desired. 

All those results can fruitfully  be applied here. First we note that varying $\Gamma$ is simply not available for the~0B$^\pm$ cases, since the map to the appropriate $r^2$ solutions of Painlev\'e~II/mKdV (symmetric Zakharov-Shabat system~\cite{Periwal:1990gf,Periwal:1990qb,Hollowood:1992xq,Crnkovic:1990mr}) requires the combined $\Gamma{=}\pm\frac12$ pair.\footnote{Combinig solutions with opposite signs of {\it general} $\Gamma$ certainly does give interesting solutions, but they do not seem to readily pertain to two-cut solutions any more. This seems clear from the fact that the $x>0$ branch no longer has $r^2$ vanishing to all orders in perturbation theory. This only happens for $|\Gamma|=\frac12$.} This fits precisely with the constructions in ref.~\cite{Eberhardt:2026hfh}, where the 0B projection (beware, recall again that it is called 0A there) removes the Ramond sector. 
So we turn to the 0A$^\pm$ cases, which we take in turn.

\begin{widetext}
\subsection{0A$^-$ SVMS Amplitudes with Ramond insertions}
\subsubsection{Some examples} Using the formalism described above, ref.~\cite{Johnson:2026jls} already computed many closed form formulae for the quantities $W_{g,n}^{(2m)}$, corresponding to genus $g$, $n$ NS boundaries, and $2m$ Ramond punctures. The formulae were written in a quite general way so as to accommodate different theories, so we can use them by just substituting the $t_k^-$ we already computed. Some of those are given in terms of the central charge $c$ in equation~(\ref{eq:tk-minus-in-terms-of-c}).
We then convert to amplitudes (quantum volumes) $V^{(b)(2m)}_{g,n}$ using equation~(\ref{eq:quantum-volume-conversion}), and some examples are:

\begin{align}
V_{0,n}^{(b)(2)}(\{P_i\})
={}&(-1)^{n+1}\frac{(n-1)!}{2}(4\pi)^n\ ,
\\
V_{0,n}^{(b)(4)}(\{P_i\})
={}&(-1)^n\frac{(n+1)!}{6\cdot4^2}(4\pi)^n
\left[
4t_1^-(n+2)
+(4\pi)^2\sum_{i=1}^nP_i^2
\right]\ ,
\\
V_{1,n}^{(b)(2)}(\{P_i\})
={}&(-1)^{n+1}\frac{5(n+1)!}{192}(4\pi)^n
\left[
4t_1^-(n+2)
+(4\pi)^2\sum_{i=1}^nP_i^2
\right]\ ,
\\
V_{1,n}^{(b)(4)}(\{P_i\})
={}&(-1)^{n+1}\frac{(n+3)!}{5760}(4\pi)^n
\Bigg[
-31(n+4)(n+5)(t_1^-)^2
+70(n+4)t_2^-
\nonumber\\
&\hspace{1.4cm}
-\frac{31}{2}(n+4)t_1^-(4\pi)^2\sum_{i=1}^nP_i^2
-\frac{35}{32}(4\pi)^4\sum_{i=1}^nP_i^4
-\frac{31}{8}(4\pi)^4
\sum_{1\leq i<j\leq n}P_i^2P_j^2
\Bigg]\ ,
\\
V_{2,n}^{(b)(2)}(\{P_i\})
={}&(-1)^{n+1}(n+3)!(4\pi)^n
\Bigg[
\frac{73}{7680}(n+4)(n+5)(t_1^-)^2
-\frac{259}{11520}(n+4)t_2^-
\nonumber\\
&\hspace{1.4cm}
+\frac{73}{15360}(n+4)t_1^-(4\pi)^2\sum_{i=1}^nP_i^2
+\frac{259}{737280}(4\pi)^4\sum_{i=1}^nP_i^4
+\frac{73}{61440}(4\pi)^4
\sum_{1\leq i<j\leq n}P_i^2P_j^2
\Bigg]\ .
\end{align}
\end{widetext}

There are many more explicit closed-form formulae in \cite{Johnson:2026jls} that can  be readily converted, and we only show a few results here. More can be converted  as desired.

\subsubsection{The spectral curve}
It is straightforward to define the spectral curve from which, if desired, individual examples of the volumes computed above can be computed using topological recursion~\cite{Chekhov:2006vd,Eynard:2007kz} techniques (it will not give the closed form expressions however!).

The method follows closely what was done already~\cite{Johnson:2026jls} to deform the topological recursion data consisting of the standard (NS) curve:
\begin{equation}
    X=\hat z^2\ ,\quad  Y = -\frac{\cos(2\pi \hat z)}{2\hat z}\ , 
    \end{equation} 
    and standard (Cauchy form) Bergman kernel:
    \begin{equation}
    \label{eq:bergman-cauchy} B(\hat z_1,\hat z_2)=\frac{d\hat z_1 d\hat z_2}{(\hat z_1-\hat z_2)^2}\ ,
    \end{equation} to include R-punctures. A deformed curve {\it and} Bergman kernel resulted.  Now we want to deform the curve:
\begin{equation}
X=\hat z^2\ ,\quad  Y =- 2\sqrt{2}\pi\frac{\cos(2\pi b \hat z)\cos(2\pi b^{-1} \hat z)}{\hat z}\ ,
\end{equation} 
Here the normalization follows from the fact that
$Y(\hat z){=}-\pi\hbar i\,\rho_0^-(-\hat z^2)$. The analytic continuation converts the hyperbolic cosines in the density into ordinary cosines.
As we will see, a deformed  Bergman kernel will  arise again.

The key is to solve the full classical limit of the string equation~(\ref{eq:big-string-equation}) for~$u_0$, where now $\widetilde\Gamma=\hbar\Gamma$ is held fixed in the $\hbar\to0$ classical limit: 
\begin{equation}
  \label{eq:compact-string-equation}
u_0{\cal R}_0^2={\widetilde\Gamma}^2\ ,
\end{equation}
where
\begin{eqnarray}
{\cal R}_0&\equiv&\sum_{k=1}^\infty t_k^-u_0^k+x\\ 
&=&
2\sqrt{2}\,\pi
\left[
I_0(2\pi Q\sqrt{u_0})
+
I_0(2\pi\widehat Q\sqrt{u_0})
-2
\right]+x\ ,\nonumber
\end{eqnarray}    
Recall that, given a solution $u_0(x)$, the spectral density $\rho_0(E)$ has the following integral representation:
\begin{equation}
\label{eq:leading-representation-2}
    \rho_0(E)=\frac{1}{2\pi\hbar}\int_{-\infty}^\mu \frac{\Theta(E-u_0(x))}{\sqrt{E-u_0(x)}} dx\ ,
\end{equation}
where here $\mu=t_0^-=4\sqrt{2}\pi$.
It is useful to work with $u_0$ as a coordinate, and so, after computing the Jacobian:
\begin{equation}
\label{eq:nice-form-u0}
    \rho_0(E) = \frac{1}{2\pi \hbar} \int_{E_0}^E \frac{f_0(u_0) du_0}{\sqrt{E-u_0}}  + \frac{\widetilde\Gamma}{4\pi \hbar} \int_{E_0}^E \frac{du_0}{u_0^{3/2} \sqrt{E-u_0}}\ , 
\end{equation}
where:
\begin{eqnarray}
f_0 &=& \sum_k kt_k^-u_0^{k-1}\\
&=&
\frac{2\sqrt{2}\,\pi^2}{\sqrt{u_0}}
\left[
Q\,I_1\!\left(2\pi Q\sqrt{u_0}\right)
+
\widehat Q\,I_1\!\left(2\pi\widehat Q\sqrt{u_0}\right)
\right]\ ,
\nonumber
\end{eqnarray}
and $E_0\equiv u_0(\mu)$. This threshold energy  is defined by putting $u_0=E_0$ and $x=\mu$ into~(\ref{eq:compact-string-equation}), giving:
\begin{equation}
\widetilde \Gamma=
2\sqrt{2}\,\pi \sqrt{E_0}
\left[
I_0(2\pi Q\sqrt{E_0})
+
I_0(2\pi\widehat Q\sqrt{E_0})
\right]\ .
\label{eq:endpoint-equation}
\end{equation}
Inverting this relation gives:
\begin{widetext}
\begin{align}
E_0(\widetilde\Gamma)
=
\frac{\widetilde\Gamma^2}{32\pi^2}
-\frac{Q^2+\widehat Q^2}{1024\pi^2}\widetilde\Gamma^4
+
\frac{
7\left(Q^2+\widehat Q^2\right)^2
-\left(Q^4+\widehat Q^4\right)
}{131072\pi^2}\widetilde\Gamma^6
+O(\widetilde\Gamma^8)\ .
\label{eq:e0-expansion}
\end{align}
We then can write the final form for the spectral density as:
\begin{equation}
\begin{split}
\rho_0(E)
={}&
\frac{1}{2\pi\hbar}
\Bigg[
2\sqrt{2}\,\pi^2
\int_{E_0}^{E}
\frac{
Q\,I_1\!\left(2\pi Q\sqrt{u_0}\right)
+\widehat Q\,I_1\!\left(2\pi\widehat Q\sqrt{u_0}\right)
}
{\sqrt{u_0}\sqrt{E-u_0}}\,du_0
\\
&\hspace{3.8cm}
+2\sqrt{2}\,\pi
\left[
I_0\!\left(2\pi Q\sqrt{E_0}\right)
+I_0\!\left(2\pi\widehat Q\sqrt{E_0}\right)
\right]
\frac{\sqrt{E-E_0}}{E}
\Bigg]\ ,
\qquad E>E_0\ .
\end{split}
\label{eq:exact-density-i0}
\end{equation}
where the integral form seems to be the most compact presentation of the result. 
Just as a check, sending ${\widetilde\Gamma}\to0$ sends $E_0\to0$ while sending ${\widetilde\Gamma}/\sqrt{E_0}\to4\sqrt{2}\pi$. With lower limit at $0$, inside the bracket the integral gives
$2\sqrt{2}\pi[\cosh(2\pi Q\sqrt{E})+\cosh(2\pi\widehat Q\sqrt{E})-2]/\sqrt{E}$,
while the second term gives $4\sqrt{2}\pi/\sqrt{E}$ and the undeformed density~(\ref{eq:super-NSminus-density}) is recovered.

\end{widetext}
To define the  spectral curve  directly in a hard edge coordinate we'll denote as $\hat z$, the first step is to rewrite the spectral curve by defining:
\begin{equation}
    X(\hat z)=\hat z^2\ ,\quad 
    Y(\hat z)=-\pi\hbar i\,\rho_0(-\hat{z}^2)\ ,
    \label{eq:new-Ramond-curve}
\end{equation}
where again $\rho_0(E)$ is given in equation~(\ref{eq:exact-density-i0}). The hard edge disc data are now in $\omega_{0,1}{=}Y(\hat z)dX(\hat z)$.  Finally,  the Bergman kernel to use is obtained by transforming away from the simple Cauchy case to:
\begin{eqnarray}
\hskip-0.5cm {\widehat B}({\hat z}_1,{\hat z}_2)
=
\frac{{\hat z}_1{\hat z}_2}{z_1z_2}
\frac{d{\hat z}_1d{\hat z}_2}{(z_1-z_2)^2}\ ,\,\,
\text{with}\,\,
z_i=\sqrt{{\hat z}_i^{\,2}+E_0} \ ,   
\end{eqnarray}
and we used that $z'({\hat z})={{\hat z}}/{z}$. Expanding in $E_0$ gives:
\begin{widetext}
\begin{equation}
{\widehat B}({\hat z}_1,{\hat z}_2)
=
\left[
\frac{1}{({\hat z}_1-{\hat z}_2)^2}
-\frac{E_0}{2{\hat z}_1^{\,2}{\hat z}_2^{\,2}}
+\frac{3E_0^2({\hat z}_1^{\,2}+{\hat z}_2^{\,2})}
{8{\hat z}_1^{\,4}{\hat z}_2^{\,4}}
+O(E_0^3)
\right]d{\hat z}_1d{\hat z}_2\ ,
\end{equation}
and after substituting the solution's expansion of $E_0$ given in~(\ref{eq:e0-expansion})
we obtain:
\begin{equation}
\begin{split}
{\widehat B}({\hat z}_1,{\hat z}_2)
={}&
\Bigg[
\frac{1}{({\hat z}_1-{\hat z}_2)^2}
-\frac{\widetilde\Gamma^2}
{64\pi^2{\hat z}_1^{\,2}{\hat z}_2^{\,2}}
+\widetilde\Gamma^4
\left(
\frac{Q^2+\widehat Q^2}
{2048\pi^2{\hat z}_1^{\,2}{\hat z}_2^{\,2}}
+
\frac{3({\hat z}_1^{\,2}+{\hat z}_2^{\,2})}
{8192\pi^4{\hat z}_1^{\,4}{\hat z}_2^{\,4}}
\right)
+O(\widetilde\Gamma^6)
\Bigg]
d{\hat z}_1d{\hat z}_2\ .
\end{split}
\end{equation}
\end{widetext}
Using this, $Y(\hat z)$ defined in (\ref{eq:new-Ramond-curve}), and the involution $\sigma(\hat z)=-\hat z$, the standard topological recursion kernel:
\begin{equation}
 K(\hat z_1,\hat z)=
 -\frac{\displaystyle\int_{\sigma(\hat z)}^{\hat z}{\widehat B}(\hat z_1,\zeta)}
 {2\bigl(Y(\hat z)-Y(\sigma(\hat z))\bigr)dX(\hat z)}\ ,
 \label{eq:kernel-def}
\end{equation}
can be expanded in powers of $\widetilde\Gamma^2$, and for any $(g,n)$ proceeding order by order yields the $W^{(2m)}_{g,n}$, the correlator with $2m$ Ramond punctures. Laplace transforming then gives the amplitudes $V^{(b)(2m)}_{g,n}(\{P_i\})$. It is worth noting here again that topological recursion isn't really needed here at all. For this class of theory, the approach of ref.~\cite{Johnson:2026twg} will yield these results more quickly and allow (as we've seen) more powerful general closed form expressions to be derived rather simply.

\subsection{0A$^+$ SVMS Amplitudes with Ramond insertions}

One might assume that since turning on $\Gamma$ yields the appropriate Ramond amplitudes in the previous case, it should work again here, but there is a subtlety. The genus expansion that included $\Gamma$ terms in the previous $x>0$ case is of quite a different character from the  $x<0$ expansion. While the $x>0$ case has the interpretation as closed string vertex operator insertions, the expansion for $x<0$, given in equation~(\ref{eq:uGamma-expansion-hbar4}), has two distinct features: 
\begin{enumerate}[(1)]
    \item It has odd powers of $\hbar$, consistent with an open string (D-brane) background interpretation.

\item Again, those terms have explicit inverse (and log) $u_0$ terms in them again, so precisely where we compute correlators ($\mu=0$ where $u_0=0$) the resulting $W_{g,n}$ (and  their Laplace transforms) are divergent and do not fall into genus perturbation theory.
\end{enumerate}
So, at least for this manner in which the random matrix model captures Ramond backgrounds, we see again that genus perturbation theory  resembles the usual bososonic soft edge form for 0A$^+$, which is again consistent with observations made in ref.~\cite{Eberhardt:2026hfh} for this case.

\bigskip

At this point we end our study of ${\cal N}{=}1$ supersymmetric Virasoro minimal strings. We've shown how interconnected they are, and how well they fit together in the same framework.  In the next Sections, we shall see that examples of the 0A$^\pm$ variants, so seemingly distinct in their character, appear together as even and odd charge subsectors of the new and richer ${\cal N}{=}2$ supersymmetric theories that we shall construct.

\section{New ${\cal N}{=}2$ Supersymmetric Strings}
\label{sec:an-N=2-generalization}

In this section we deliver on the promise made in the Introduction: There ought to be  VMS-type string theories with {\it extended} supersymmetry, naturally constructed by building matrix models that have the
${\cal N}\geq2$ SCFT Cardy densities for their leading spectral densities. For the case of ${\cal N}{=}2$, things are already quite rich, and so we will begin our explorations of this idea there. In fact,  a small ${\cal N}{=}4$  supersymmetric case will naturally appear as a special point of enhanced symmetry, and will be discussed in Section~\ref{sec:surprise-N=4}. This will then invite us to construct more general small ${\cal N}{=}4$  strings in Section~\ref{sec:an-N=4-generalization}.

\subsection{More superconformal field theory reminders}

\noindent 
The ${\cal N}=2$ superconformal algebra has two supergenerators~$G^\pm_r$ where the $\pm$ denotes charges under the $U(1)_R$ R-symmetry group, which itself has generators $J_n$. The resulting superconformal algebra enlarges the Virasoro~algebra (\ref{eq:Virasoro}) to:
\begin{align}
 [L_m,J_n]&=-nJ_{m+n}\ ,
\nonumber \\
 [J_m,J_n]&=\frac c3m\delta_{m+n,0} ,
\nonumber
 \\
 [L_m,G_r^\pm]
 &=\left(\frac m2-r\right)G_{m+r}^\pm\ ,
 \nonumber
 \\
 [J_m,G_r^\pm]&=\pm G_{m+r}^\pm\ ,
 \nonumber
 \\
 \{G_r^+,G_s^-\}
 &=2L_{r+s}+(r-s)J_{r+s}
 +\frac c3\left(r^2-\frac14\right)\delta_{r+s,0}\ ,
\nonumber
 \\
 \{G_r^+,G_s^+\}&=\{G_r^-,G_s^-\}=0\ .
 \label{eq:N=2-superconformal-algebra}
\end{align} In addition to their conformal weight, states are also labelled by their $U(1)_R$ charge $Q_R$: $J_0|h,Q_R\rangle=Q_R|h,Q_R\rangle$.
The usual integer moding on $r$ denotes the  Ramond~(R) sector and half-integer is Neveu-Schwarz~(NS).

Key to what is to follow is the fact that these sectors can be mapped into each other by spectral flow: There is an action on all generators parameterized by  $\eta$:
\begin{align}
 L_n&\longrightarrow
 L_n+\eta J_n+\frac c6\eta^2\delta_{n,0}
  ,
 \nonumber
 \\
 J_n&\longrightarrow
 J_n+\frac c3\eta\delta_{n,0}\ ,
 \nonumber
 \\
 G_r^\pm&\longrightarrow G_{r\pm\eta}^\pm\ ,
 \label{eq:spectral-flow}
\end{align}
where $\eta$ is real. 
As a result, a state with conformal weight~$h$ and $R$-charge $Q_R$ is mapped to one with:
\begin{equation}
 h_\eta=h+\eta Q_R+\frac c6\eta^2\ ,
 \qquad
 Q_{R,\eta}=Q_R+\frac c3\eta\ .
 \label{eq:N2-flow-weights}
\end{equation}
Integer $\eta$-flow preserves NS or R moding, while half-integer flow exchanges them. 
A generic $\eta$ produces a more generally twisted sector.

For the NS vacuum, $h{=}Q_R{=}0$.  After $\eta=\pm\frac12$ spectral flow:
\begin{equation}
 h_{\pm\frac12}=\frac c{24},
 \qquad
 Q_{R,\pm\frac12}=\pm\frac c6\ .
 \label{eq:N2-vac-to-R}
\end{equation}
These states are   Ramond (R) ground states.  Recall that  in the R sector the supercurrents have integer moding, and the algebra~(\ref{eq:N=2-superconformal-algebra}) gives:
\begin{equation}
\{G_0^+,G_0^-\}
=2L_0-\frac{c}{12}.
\label{eq:N2-R-zero-mode-algebra}
\end{equation}
Unitarity therefore implies the Ramond-sector bound
$h_{\rm R}\geq \frac{c}{24}$. The states~(\ref{eq:N2-vac-to-R}) hence saturate this bound, while carrying
$Q_{R,\pm1/2}=\pm c/6$.  They are therefore Ramond ground states.  So these two distinguished Ramond ground states, with charges $\pm c/6$, are connected directly to the NS vacuum by half-unit spectral flow.

This of course lifts to the full characters that can be written. In addition to dependence on $q=\e^{2\pi\ii\tau}$, there is also a natural dependence on $y=\e^{2\pi\ii z}$ and for a representation $\mathcal R$ in the NS or R Hilbert space, define
\begin{align}
 \chi_{\mathcal R}^{\NS^+}(\tau,z)
 &={\rm Tr}_{\mathcal H_{\mathcal R,\NS}}
 q^{L_0-c/24}y^{J_0},
 \nonumber\\
 \chi_{\mathcal R}^{\NS^-}(\tau,z)
 &={\rm Tr}_{\mathcal H_{\mathcal R,\NS}}
 (-1)^Fq^{L_0-c/24}y^{J_0},
 \nonumber\\
 \chi_{\mathcal R}^{\R^+}(\tau,z)
 &={\rm Tr}_{\mathcal H_{\mathcal R,\R}}
 q^{L_0-c/24}y^{J_0},
 \nonumber\\
 \chi_{\mathcal R}^{\R^-}(\tau,z)
 &={\rm Tr}_{\mathcal H_{\mathcal R,\R}}
 (-1)^Fq^{L_0-c/24}y^{J_0}.
 \label{eq:Rminus-def-2}
\end{align}
 Notice that $e^{i\pi J_0}\sim(-1)^F$, which follows from the superconformal algebra bracket between $J_0$ and $G_r^\pm$: {\it i.e.} $e^{i\pi J_0}G_r^\pm e^{-i\pi J_0}=-G_r^\pm$ while a similar action on all other generators 
 gives a plus sign.
 
 As a result, there are relations between these  characters under special shifts in $z$:
 \begin{align}
 \chi^{\NS^-}(\tau,z)
 &\sim \chi^{\NS^+}\!\left(\tau,z+\frac12\right),
 \nonumber\\
 \chi^{\R^+}(\tau,z)
 &\sim q^{{\hat c}/8}y^{{\hat c}/2} \chi^{\NS^+}\!\left(\tau,z+\frac{\tau}{2}\right),
 \nonumber\\
 \chi^{\R^-}(\tau,z)
 &\sim q^{{\hat c}/8}y^{{\hat c}/2}
\chi^{\NS^+}\!\left(\tau,z+\frac{\tau+1}{2}\right),
 \label{eq:special-relations}
\end{align}
where ${\hat c}=c/3$.  
The shift
$ z\rightarrow z+\frac12$
swaps the  insertion of $(-1)^F$ while 
$z\rightarrow z+\frac\tau2$ implements a half-unit spectral flow. However, these shifts are exchanged by modular $S$-transformation.
The consequence of all this is that while the $S$-transformation of the vacuum character in each of the  different sectors again mixes them according to $\NS^+{\leftrightarrow}\NS^+$, $\NS^-{\leftrightarrow}\R^+$, and $\R^-{\leftrightarrow}\R^-$, the modular kernels that appear in the transformations are all of the {\it same form} in order to have consistency with the special relations~(\ref{eq:special-relations}). There is a continuous part, and a discrete part that mixes in the massless (BPS) representations. The choice of conventions we shall make in this paper is to write the disc density in the same energy units as the ${\cal N}{=}2$ conventions that were used by Turiaci and Witten~\cite{Turiaci:2023jfa} for ${\cal N}{=}2$ JT supergravity, to which our results will collapse in the classical limit.

The relation between the central charge and the ${\cal N}{=}2$ Liouville screening charge ${\cal Q}$ is:  \begin{equation}
    \hat c \equiv \frac{c}{3}=1+{\cal Q}^2\ ,
\end{equation} where ${\cal Q}^2$ is, in general, positive and real. 
For the charge variable that appears  in the continuum kernel we write below it is
useful to use the centered convention
\begin{equation}
 q\equiv Q_R-\frac{{\mathcal Q}^2}{2},
 \qquad
 E_0=\frac{q^2}{4}.
 \label{eq:N2-centered-charge}
\end{equation}
This is the magnitude of shift naturally associated with the half-unit NS-to-R spectral flow of~(\ref{eq:N2-vac-to-R}).  

\subsubsection{The massive sector}
For a multiplet carrying a charge~$Q_R$ under the $U(1)_R$ symmetry,  the continuum part of the density is:
\begin{widetext}

\begin{equation}
\rho_{0,q}^{(\mathcal Q)}(E)
=
\frac{1}{8\pi^3\hbar}\,
\sinh\!\left(2\pi\sqrt{E-E_0}\right)
\sum_{n\in\mathbb Z}
\frac{1}{
E-E_0+
\left(
\frac{q}{2}
+n\frac{\mathcal Q^2}{2}
\right)^2
}\,
\Theta(E-E_0),\quad\text{where}\quad
\qquad
E_0=\frac{q^2}{4}\ .
\label{eq:N=2-density-form-one}
\end{equation}
This is the vacuum representative of the continuum modular coefficient ${\cal K}$ from ref.~\cite{Eguchi:2003ik}, in charge sector $q$, with a specific normalization choice to be explained shortly.
This form makes explicit the {\it extended} character structure used in ref.~\cite{Eguchi:2003ik}, where good modular behaviour is ensured by performing a sum over an infinite number of spectral images. This appears very naturally at rational values of ${\mathcal Q}^2=2K/N$, where
$K$ and $N$ are positive integers. Then, the natural extended characters are essentially $\Theta$-functions with integral level $NK$, which transform nicely.  Particular rational points can exhibit  enhanced symmetry, and ``resonant'' charge sectors satisfying $2q/{\mathcal Q}^2\in{\mathbb Z}$ can have special $E=0$ behaviour, as we shall see.
Using the  Poisson resummation formula and  a Fourier series~\cite{Abramowitz:1964}: \begin{equation}
 \sum_{n\in\mathbb Z}
 \frac{1}{
 p^2+\left(\frac q2+n\frac{{\mathcal{Q}}^2}{2}\right)^2}
 =
 \frac{2\pi}{{\mathcal{Q}}^2p}
 \left[
 1+2\sum_{m=1}^{\infty}
 e^{-4\pi m p/{\mathcal{Q}}^2}
 \cos\!\left(\frac{2\pi m q}{{\mathcal{Q}}^2}\right)
 \right]= \frac{2\pi}{{\mathcal{Q}}^2p}
 \left[
 \frac{
 \sinh\!\left(\frac{4\pi p}{{\mathcal{Q}}^2}\right)
 }{
 \cosh\!\left(\frac{4\pi p}{{\mathcal{Q}}^2}\right)
 -\cos\!\left(\frac{2\pi q}{{\mathcal{Q}}^2}\right)
 }
 \right]\ ,
 \label{eq:Poisson-resummed-image-sum}
\end{equation} where $p=\sqrt{E-E_0}$,  gives the form:
\begin{equation}
\rho_{0,q}^{(\mathcal Q)}(E)
=
\frac{1}{8\pi^{3}\hbar}\,
\frac{
\displaystyle \frac{2\pi}{\mathcal Q^{2}}\,
\sinh\!\left(2\pi\sqrt{E-E_0}\right)
\sinh\!\left(
\frac{4\pi}{\mathcal Q^{2}}\sqrt{E-E_0}
\right)
}{
\displaystyle
\sqrt{E-E_0}\,
\left[
\cosh\!\left(
\frac{4\pi}{\mathcal Q^{2}}\sqrt{E-E_0}
\right)
-
\cos\!\left(\frac{2\pi q}{\mathcal Q^{2}}\right)
\right]
}\,
\Theta(E-E_0)\ ,
\qquad
E_0=\frac{q^2}{4}.
\label{eq:N=2-density-form-two}
\end{equation}
     \end{widetext}
It is straightforward using form~(\ref{eq:N=2-density-form-one}) to see that in the classical limit ($\mathcal{Q}\to\infty$) we recover the ${\cal N}{=}2$ JT supergravity form given in ~\cite{Turiaci:2023jfa,Mertens:2017mtv,Stanford:2017thb} since the $\mathcal{Q}^4$ in the denominator in the sum over almost all images collapses the density to zero. All except the $n=0$ case, for which the $E_0$ cancels against $\frac{q^2}{4}$ in the denominator, and we recover:
\begin{equation}
    \rho_0^{\text{${\cal N}{=}2$ JT}} = 
    \frac{1}{8\pi^3\hbar}\frac{\sinh\!\left(2\pi\sqrt{E-E_0}\right)}{E}\ .
\end{equation}
A similar result can be obtained from form~(\ref{eq:N=2-density-form-two}) by simply expanding in large ${\mathcal Q}$. 
(Our normalization choice above in (\ref{eq:N=2-density-form-one}) was made to ensure that we recover ref.~\cite{Turiaci:2023jfa}'s  normalization.)

An interesting detail of the large ${\cal Q}^2$ limit is worth considering here. The charges $q$ that ref.~\cite{Turiaci:2023jfa} have, in  ${\cal N}{=}2$ JT supergravity, form a discrete lattice, with spacing set by $1/{\hat q}$, where ${\hat q}$ is  the $U(1)_R$ charge of the supercharge. The charges are discrete because the $U(1)_R$  is compact in  the underlying BF formulation. 

A natural way to arrive at that here is to start with rational $Q^2=2K/N$, where  as already stated $K$ and $N$ are positive  integers. 
The resulting lattice of charges $q$ turns out to be in units of $1/N$. The large ${\cal Q}^2$ limit can then be achieved by taking $K$ large, holding $N={\hat q}$ fixed, thereby recovering the lattice of ref.~\cite{Turiaci:2023jfa}.

It will be useful later to list ref.~\cite{Eguchi:2003ik}'s extended character labelling conventions here. The sums in~(\ref{eq:N=2-density-form-one}) and~(\ref{eq:Poisson-resummed-image-sum}) are over  integer~$n$ labelling the extended characters, where:
\begin{equation}
 n\in r+N\mathbb Z,
 \qquad r\in\mathbb Z_N\ ,
 \label{eq:n-and-r}
\end{equation}
with $r$ labelling $N$ spectral images.
A massive representation and its charge  starts out labelled with $j_0$ where:
\begin{equation}
 0\leq j_0\leq 2K-1\ ,
 \qquad Q_R^{(0)}=\frac{j_0}{N}\ ,
 \label{eq:jzero-range}
\end{equation}
and this gets extended by the $N$ spectral flow copies into the label:
\begin{equation}
 j=j_0+2Kr\pmod{2NK}\ ,\qquad Q_R=\frac{j}{N}\ .
 \label{eq:j-combination}
\end{equation}
The centered charge of  (\ref{eq:N2-centered-charge})  then is:
\begin{equation}
 q_j=Q_R-\frac{\mathcal Q^2}{2}
 =\frac{j-K}{N},
 \qquad E_j=\frac{q_j^2}{4}.
 \label{eq:centered-q}
\end{equation}
Using (\ref{eq:j-combination}) gives the charge  $q_{j_0,r}$ of the $r$th family:
\begin{equation}
 q_{j_0,r}
 =\frac{j_0-K}{N}+r\,\mathcal Q^2,
 \qquad r=0,\ldots,N-1.
 \label{eq:q-family}
\end{equation}
So there are $N$ spectral flow images, separated by charge unit ${\cal Q}^2=2K/N$. Often therefore we will just talk about the properties of the $2K$ ``seed''  families at $r=0$, each represented by $q_{j_0,0}$. 

Keeping track of the available charges is important since the spectral density~(\ref{eq:N=2-density-form-two}) will adjust its shape significantly as one runs over the available $q_j$ for a given value of ${\cal Q}$. This will give much richer behaviour than seen in the JT supergravity limit, as will be clear in examples to come.

\subsubsection{The massless sector}

The vacuum modular transform 
also contains massless (BPS) states with labels:
\begin{equation}
 r\in\mathbb Z_N,
 \qquad 1\leq s\leq N+2K-1.
 \label{eq:full-massless-range}
\end{equation}

The vacuum $S$-transform involves only a subset of this family, labelled by: 
\begin{equation}
 r'\in\mathbb Z_N,
 \qquad
 s'=K+1,\ldots,K+N-1,
 \label{eq:visible-range}
\end{equation}
and the associated centered charges are:
\begin{equation}
q_{s'}=\frac{s'-K}{N}
\in\left\{\frac1N,\frac2N,\ldots,\frac{N-1}{N}\right\}.
\label{eq:N2-short-centered-charge}
\end{equation}
Much of this labelling will become clearer in examples to appear shortly.

The part of the density pertaining to those contributions is contained in the modular coefficient denoted ${\cal D}_0$ in ref.~\cite{Eguchi:2003ik}, whose form we will recall shortly. It is instructive however to take a different route to it, using the random matrix model methods developed for this in refs.~\cite{Johnson:2023ofr,Johnson:2024tgg,Johnson:2025oty,Turiaci:2023jfa}.

In the positive random matrix model of the type discussed earlier, with string equation given by~(\ref{eq:big-string-equation}), it is the parameter  $\Gamma$ that counts degenerate states at $E=0$. Here, for the sector with charge $q$ we will relabel it as $\Gamma_q$ and we have the BPS contribution:
\begin{equation}
 \rho_{{\rm BPS},q}(E)=\Gamma_q\,\delta(E)\ ,
 \label{eq:N2-BPS-density}
\end{equation}
and as before we will define $\widetilde\Gamma_q\equiv\hbar\Gamma_q$ as the parameter held fixed in the classical limit $\hbar\to0$. Ref.~\cite{Johnson:2025oty} then relates this to the classical continuum part of the density {\it via}:
\begin{equation}
 |\widetilde\Gamma_q|
 =2\pi\hbar
 \left|\operatorname*{Res}_{E=0}\rho_{0,q}^{({\mathcal Q})}(E)\right|\ ,
 \label{eq:N2-residue-Gamma}
\end{equation}
where we are anticipating that we will use density (\ref{eq:N=2-density-form-two}) as our random matrix model's leading spectral density. (We  ignore the $\Theta$-function in this part, since we are extracting information by  analytically continuing to $E{=}0$.)  So let us do that. Simplifying first by writing:
\begin{equation}
p=\sqrt{E-\frac{q^2}{4}}\ ,
\qquad
a=\frac{4\pi}{{\mathcal Q}^2},
\qquad
\phi=\frac{2\pi q}{{\mathcal Q}^2}\ ,
\end{equation}
 as $E\to0$, $p\to iq/2$, and so $\cosh(ap)\to\cos\phi$, and the last factor  in the denominator of (\ref{eq:N=2-density-form-two}) is going to zero.  For
$\sin\phi\neq0$ the zero is simple, but we'll retain the next order for later use:
\begin{eqnarray}
&&\hskip-0.3cm \cosh(ap)-\cos\phi
=
\frac{a}{q}\sin\phi\,E
+\left[
\frac{a}{q^3}\sin\phi
-\frac{a^2}{2q^2}\cos\phi
\right]E^2
\nonumber\\
&&\hskip4.05cm +\,O(E^3)\ .
\label{eq:E-expansion-one}
\end{eqnarray}
Meanwhile we have
$\sinh(2\pi p)\to i\sin(\pi q)$, and  also: 
\begin{equation}
    \sinh(ap)= i\sin\phi + \frac{a\cos\phi}{iq}E+O(E^2)\ ,
    \label{eq:E-expansion-two}
\end{equation} and if $\sin\phi{\neq}0$
the residue  becomes:
\begin{equation}
 \operatorname*{Res}_{E=0}\rho_{0,q}^{({\mathcal Q})}(E)
 =\frac{i}{8\pi^3\hbar}\sin(\pi q)\ ,
 \qquad
 \sin\!\left(\frac{2\pi q}{{\mathcal Q}^2}\right)\neq0\ .
 \label{eq:N2-generic-residue}
\end{equation}
This is the case for generic charge. 
When $2q/{\mathcal Q}^2\in{\mathbb Z}$ with
$q\notin{\mathbb Z}$ the $\sin\!\left(\frac{2\pi q}{{\mathcal Q}^2}\right)\equiv\sin\phi$ vanishes and so we must use the next order in the $E$ expansions~(\ref{eq:E-expansion-one}) and~(\ref{eq:E-expansion-two}),
with the result that the coefficient of the pole is doubled,
\begin{equation}
 \operatorname*{Res}_{E=0}\rho_{0,q}^{({\mathcal Q})}(E)
 =\frac{i}{4\pi^3\hbar}\sin(\pi q)\ ,
 \qquad
 \sin\!\left(\frac{2\pi q}{{\mathcal Q}^2}\right)=0\ .
 \label{eq:N2-resonant-residue}
\end{equation}
This is the special ``resonant'' case, an example of which we will encounter later.
So finally we have:
\begin{eqnarray}
 &&|\widetilde\Gamma_q|
 =\frac{|\sin(\pi q)|}{4\pi^2}
 \quad\hbox{(generic)}\ ,
\nonumber\\
 &&|\widetilde\Gamma_q|
 =\frac{|\sin(\pi q)|}{2\pi^2}
 \quad\hbox{(resonant)}\ .
 \label{eq:N2-Gamma-general}   
\end{eqnarray}
For non-zero integral $q$ the residue vanishes, while $q=0$ is instead the
hard-edge $E^{-1/2}$ case.  In the ${\mathcal Q}\to\infty$ limit this  BPS
zero-mode sector remains and becomes the familiar fixed-charge ${\cal N}=2$
JT contribution.

Note that this procedure, while fixing nicely the normalization of the discrete part of the density,  does not determine the finite allowed charge range~(\ref{eq:visible-range}) for the BPS states. That all comes from careful analysis of the extended characters. Indeed the
characteristic $\sin(\pi q)$ dependence that emerged is the functional form of the modular transform coefficient ${\cal D}_0$ of ref.~\cite{Eguchi:2003ik}, which for the seed family is in fact $r^\prime$ independent:
\begin{eqnarray}
&&{\cal D}_{0}(r',s')=
\frac{2}{N}
\sin\left[\frac{\pi(s'-K)}{N}\right],
\nonumber\\
\label{eq:N=2-density-form-two-discrete}
\end{eqnarray}
and  writing this in terms of the  centered charges (\ref{eq:N2-short-centered-charge}) yields the form $\sin(\pi q_{s^\prime})$ seen above in the residue computation.

\subsection{Defining ${\cal N}{=}2$ Strings}
The proposal  is simply that there is a new kind of critical string here (or a family of them), fully defined by taking density (\ref{eq:N=2-density-form-two}), together with the BPS  contribution (\ref{eq:N2-BPS-density}), as the spectral data of the positive matrix models of the  kinds we've extensively  discussed for the ${\cal N}{=}1$ cases. There is presumably a worldsheet string theory approach involving ${\cal N}{=}2$ Liouville,  ghosts, and presumably a timelike ${\cal N}{=}2$ Liouville, combined such that $c_{\rm tot}{=}0$ (generalizing the prototype VMS case) but we will not pursue that here. Using this matrix model approach as a definition has proven to be  robust in the ${\cal N}{=}1$ case, with later approaches confirming the structures found~\cite{Rangamani:2025wfa,Muhlmann:2025ngz,Eberhardt:2026hfh} as we've discussed, and so we proceed with confidence.

 It is very natural to define analogues of our previous type 0A models, and we shall do that here.\footnote{Whether we can define type~0B model is unclear here, since the simplest kinds are made by combining $\Gamma=\pm$ solutions, but here~$\Gamma$ is not  free, but tied to the BPS content.}  Two edge types (the local threshold behaviour of the density) will appear: the cases with $\rho(E){\sim}(E-E_0)^{1/2}$ have a soft edge, while those with $\rho(E){\sim}(E-E_0)^{-1/2}$ have a hard edge  threshold, which can occur even at non-zero $E_0$. It bears repeating that having a soft edge has {\it \'a priori} nothing to do with bosonic models--both edge possibilities are available in these classes of matrix model.

The path for the methods used in this paper is very clear. We simply need to determine the  function $u_0(x)$   that corresponds to our ${\cal N}{=}2$ density through the integral transform~(\ref{eq:leading-representation-2}), and all amplitudes can be built in terms of it using the formulae of ref.~\cite{Johnson:2026twg}. Sometimes the knowledge of $u_0(x)$ is simply given as a closed form formula for the $t_k$. Such a  formula for the $t_k$ (in terms of Bessel functions) for ${\cal N}{=}2$ JT supergravity was originally derived in ~\cite{Johnson:2023ofr}, and a refined method for determining them, as well as for computing the form of the BPS formula, was developed in ref.~\cite{Johnson:2025oty}.

It is not hard to write generalizations of the  ${\cal N}=2$  formulae here, although a closed form is difficult to extract in general.
The reconstruction can be organized image by spectral image.  Since the density is
even under $q\to-q$, we can choose the positive
square root of the threshold $E_0$ and write:
\begin{equation}
 R_n\equiv\sqrt{E_0}+n\frac{{\cal Q}^2}{2}
 =\frac{|q|}{2}+n\frac{{\cal Q}^2}{2}\ .
 \label{eq:N2-image-definitions}
\end{equation}
Recall that the relation between the leading spectral density and the function $u_0(x)$ is through an Abel transform~(\ref{eq:spectral-density-leading}), which can be written as a $u_0$ integral:
\begin{equation}
\label{eq:nice-form-u0-2}
    \rho_0(E) = \frac{1}{2\pi \hbar} \int_{E_0}^E \frac{-x^\prime(u_0) du_0}{\sqrt{E-u_0}} \ , 
\end{equation}
where $x^\prime(u_0)$  is the Jacobian. We want to invert this relation and find the function $u_0(x)$ associated to a given density. The inverse Abel integral is what we need:
\begin{eqnarray}
    I(u_0)&=&\int_{E_0}^{u_0}\frac{\rho_0(E)}{\sqrt{u_0-E}}\,dE\nonumber\\
    &=&\frac{1}{2\pi\hbar}\int_{E_0}^{u_0}\frac{dE}{\sqrt{u_0-E}}\left(\int_{E_0}^E \frac{-x^\prime(v) dv}{\sqrt{E-v}}\right)\nonumber\\
    &=&\frac{1}{2\pi\hbar}\int_{E_0}^{u_0}[-x^\prime(v)]{dv}\left(\int_{v}^{u_0} \frac{ dE}{\sqrt{u_0-E}\sqrt{E-v}}\right)\nonumber\\
    &=&\frac{1}{2\hbar}[\mu-x(u_0)]\ ,
    \label{eq:en-abel}
\end{eqnarray}
where after the change of order of integrals, the $E$ integral simply yielded
a factor of $\pi$, and we also used that $x(E_0)=\mu$.  Recall also that the classical string equation~(\ref{eq:compact-string-equation}) can be  used to write:
\begin{equation}
    x(u_0)=\frac{\widetilde\Gamma}{\sqrt{u_0}}-\sum_{k=1}^\infty t_ku_0^k\ ,
    \label{eq:classical-implication}
\end{equation}
and hence, specializing to our particular density:
\begin{equation}
2\hbar \int_{E_0}^{u_0}\frac{\rho_{0,q}^{({\mathcal Q})}(E)}{\sqrt{u_0-E}}\,dE
    =\mu-\frac{\widetilde\Gamma_q}{\sqrt{u_0}}+\sum_{k=1}^\infty t_{k,q}^{(\mathcal Q)}u_0^k\ ,
\end{equation}
so we learn~\cite{Johnson:2023ofr,Johnson:2025oty} that any $u_0^{-\frac12}$ parts produced by the integral are to be separated out and associated with the BPS sector while the regular contribution is the multicritical decomposition of $u_0$.
From this point on, one can work with the Abel transform of  the $n$th image in the sum, getting an expression for the $n$th contribution to the $t_{k,q}^{(\mathcal Q)}$ and then sum the result. It is straightforward to see that the zeroth image (which survives in the classical ${\cal Q}\to\infty$ limit) recovers the $t_k$ expression for ${\cal N}{=}2$ JT supergravity, but the general result does not seem to have a closed form so we will not display it here.

Note that it will be useful later  to take a derivative and write:
 \begin{equation}
 -x'(u_0)=2\hbar\frac{d}{du_0}
\int_{E_0}^{u_0}\frac{\rho_{0,q}^{({\mathcal Q})}(E)\,dE}{\sqrt{u_0-E}}\ .
 \label{eq:N2-Abel-inverse}
\end{equation}

A next step  is  to extract expressions for the endpoint data, {\it i.e.,} $u_0(x)$ and
its derivatives at $x=\mu$, since these can be  used in the universal correlator
formulae of~\cite{Johnson:2026twg}.  Since $u_0(\mu)=E_0$, equation
(\ref{eq:classical-implication}) implies:
\begin{equation}
 u_0(\mu)=E_0\ ,
 \qquad
 \mu=\frac{\widetilde\Gamma_q}{\sqrt{E_0}}
 -\sum_{k\geq1}t_{k,q}^{({\mathcal Q})}E_0^k\ .
 \label{eq:N2-endpoint-equation}
\end{equation}
For a soft channel define the first few derivatives of $x(u_0)$ at the
endpoint by:
\begin{eqnarray}
 &&A_{{\mathcal Q}}(E_0)\equiv-x'(E_0)\ ,\quad
 B_{{\mathcal Q}}(E_0)\equiv x''(E_0)\ ,\nonumber\\
 &&C_{{\mathcal Q}}(E_0)\equiv-x'''(E_0)\ .
 \label{eq:N2-endpoint-defintions}
\end{eqnarray}

Near the threshold energy $E_0$, expanding
(\ref{eq:N=2-density-form-one}) in small
$p=\sqrt{E-E_0}$ is a useful avenue, since:
\begin{eqnarray}
\frac{1}{
p^2+
\left(
\frac q2+n\frac{{\cal Q}^2}{2}
\right)^2
}
=
\frac{1}{R_n^2}
-\frac{p^2}{R_n^4}
+\frac{p^4}{R_n^6}
+\cdots\ ,
\end{eqnarray}
recalling the definition~(\ref{eq:N2-image-definitions}) of $R_n$ above.
Expanding also the factor $\sinh(2\pi p)$ and summing over the spectral
images naturally introduces the soft-edge lattice moments
\begin{equation}
 {\cal S}_{2m}\equiv\sum_{n\in\mathbb Z}\frac{1}{R_n^{2m}},
 \qquad m=1,2,3,\ldots .
 \label{eq:N2-lattice-moments}
\end{equation}
For a soft-edge sector none of the $R_n$ vanish.  Moreover, these lattice
sums can be performed exactly!  Some example results are:\footnote{\label{fn:summing-homework}These re-summations follow from the standard partial-fraction expansion of the form:
$ \sum_{n\in{\mathbb Z}}(n+a)^{-2}
 =\pi^2\csc^2(\pi a)\ ,
$ along  with repeated double-differentiation with respect to $a$. See {\it e.g.}, refs.~\cite{NIST:DLMF,Abramowitz:1964}.}
\begin{widetext}
\begin{align}
 {\cal S}_2
 &=
 \frac{4\pi^2}{{\cal Q}^4}\csc^2\theta\ ,\qquad
 {\cal S}_4
 =
 \frac{16\pi^4}{{\cal Q}^8}
 \left(
 \csc^4\theta-\frac23\csc^2\theta
 \right)\ ,\qquad
 {\cal S}_6
 =
 \frac{64\pi^6}{{\cal Q}^{12}}
 \left(
 \csc^6\theta-\csc^4\theta
 +\frac{2}{15}\csc^2\theta
 \right)\ ,
 \label{eq:N2-lattice-moments-closed}
\end{align}
where  we have defined $\theta\equiv\frac{\pi q}{{\cal Q}^2}$.
Using the differentiated inverse Abel relation~(\ref{eq:N2-Abel-inverse}) then gives,
after some algebra,
\begin{align}
 A_{{\mathcal Q}}(E_0)
 &=
 \frac{{\cal S}_2}{4\pi}
 =
 \frac{\pi}{{\cal Q}^4}\csc^2\theta\ ,
 \nonumber
 \\
 B_{{\mathcal Q}}(E_0)
 &=
 \frac{3{\cal S}_4}{8\pi}
 -\frac{\pi{\cal S}_2}{4}
 =
 \frac{\pi^3}{{\cal Q}^8}\csc^2\theta
 \left[
 6\csc^2\theta-\left({\cal Q}^4+4\right)
 \right]\ ,
  \nonumber
 \\
 C_{{\mathcal Q}}(E_0)
 &=
 \frac{15{\cal S}_6}{16\pi}
 -\frac{5\pi{\cal S}_4}{8}
 +\frac{\pi^3{\cal S}_2}{8}
 =
 \frac{\pi^5}{6{\cal Q}^{12}}\csc^2\theta
 \left[
 360\csc^4\theta
 -60\left({\cal Q}^4+6\right)\csc^2\theta
 +3{\cal Q}^8+40{\cal Q}^4+48
 \right]\ .
 \label{eq:N2-ABC}
\end{align}
This form is also well behaved for the resonant soft sectors discussed
above: the isolated BPS contribution is separated at $E=0$, while the
continuum threshold expansion at $E=E_0$ remains regular.  Inverting
$x(u_0)$ at the endpoint gives:
\begin{equation}
 \begin{aligned}
 u_0'(\mu)
 &=-\frac{1}{A_{{\mathcal Q}}(E_0)}
 =-\frac{{\cal Q}^4}{\pi}
 \sin^2\!\left(\frac{\pi q}{{\cal Q}^2}\right),\\[1mm]
 u_0''(\mu)
 &=
 \frac{B_{{\mathcal Q}}(E_0)}
 {A_{{\mathcal Q}}(E_0)^3}\ ,\quad
 u_0'''(\mu)
=
 \frac{
 A_{{\mathcal Q}}(E_0)C_{{\mathcal Q}}(E_0)
 -3B_{{\mathcal Q}}(E_0)^2}
 {A_{{\mathcal Q}}(E_0)^5}\ .
 \end{aligned}
 \label{eq:N2-u-endpoint-derivatives}
\end{equation}
\end{widetext}
Higher derivatives follow in the same way since expanding the  density near the threshold to higher orders introduces the higher ${\cal S}_{2m}$ which are
generated recursively by:
\begin{equation}
 {\cal S}_{2m+2}
 =
 \frac{1}{(2m)(2m+1)}
 \frac{\partial^2}{\partial(\sqrt{E_0})^2}
 {\cal S}_{2m}\ .
 \label{eq:N2-lattice-moment-recursion}
\end{equation}
So in this way the complete endpoint data of $u_0(x)$ can be obtained
in closed form and substituted directly into the universal formulae
of ref.~\cite{Johnson:2026twg} to get the amplitudes.

Notice the first line of~(\ref{eq:N2-u-endpoint-derivatives})  is particularly  attractive since, following the universal formulae of ref.~\cite{Johnson:2026twg}, we immediately get our fundamental three-point correlator:
\begin{equation}
 W_{0,3}(z_1,z_2,z_3)
 =
 -\frac12\frac{u_0'(\mu)}{z_1^2z_2^2z_3^2}
 =
 \frac{{\cal Q}^4}{2\pi}
 \sin^2\!\left(\frac{\pi q}{{\cal Q}^2}\right)
 \frac{1}{z_1^2z_2^2z_3^2}\ .
 \label{eq:N2-W03-charge}
\end{equation}
It is set by just the central charge and the $U(1)_R$ charge of the sector under consideration. We shall see it in action in examples presently.

It is useful to organize the remaining endpoint information into the
dimensionless quantities:
\begin{equation}
 v\equiv \frac{\pi}{4A_{{\mathcal Q}}(E_0)}
 =\frac{{\cal Q}^4}{4}\sin^2\theta\ ,
 \quad
 \kappa\equiv
 \frac{B_{{\mathcal Q}}}{\pi^2 A_{{\mathcal Q}}}
 =
 \frac{3}{2v}-1-\frac{4}{{\cal Q}^4}\ ,
 \label{eq:N2-v-kappa}
\end{equation}
and
\begin{equation}
 \eta\equiv
 \frac{C_{{\mathcal Q}}}{\pi^4 A_{{\mathcal Q}}}
 =
 \frac{15}{4v^2}
 -\left(\frac52+\frac{15}{{\cal Q}^4}\right)\frac1v
 +\frac12+\frac{20}{3{\cal Q}^4}
 +\frac{8}{{\cal Q}^8}\ .
 \label{eq:N2-eta}
\end{equation}
Equivalently,  rewriting  as endpoint curvature data:
\begin{equation}
 \kappa=
 \frac{u_0''(\mu)}
 {\pi^2u_0'(\mu)^2}\ ,
 \qquad
 \eta=
 3\kappa^2-
 \frac{u_0'''(\mu)}
 {\pi^4u_0'(\mu)^3}\ .
\end{equation}

To convert  correlators into amplitudes we use the momentum convention:
\begin{equation}
 P_i\equiv\frac12\sqrt{E_i-E_0}\ ,
 \label{eq:N2-momentum-convention}
\end{equation}
together with the same Laplace convention used for the ${\cal N}{=}1$ SVMS
$0A^+$  in~(\ref{eq:quantum-volume-conversion-plus}).  With these
conventions some useful low-genus examples of the universal formulae become:
\begin{widetext}
\begin{align}
 &V_{0,3}=v,
 \nonumber\\
 &V_{1,1}(P)
 =\frac{v}{24}\left(P^2-\kappa\right),
 \nonumber\\
 &V_{0,4}(\{P_i\})
 =v^2\left(\sum_{i=1}^{4}P_i^2-\kappa\right),
 \nonumber\\
 &V_{1,2}(P_1,P_2)
 =v^2
 \left[
 \frac{2\kappa^2-\eta}{24}
 -\frac{\kappa}{12}\left(P_1^2+P_2^2\right)
 +\frac{1}{48}\left(P_1^2+P_2^2\right)^2
 \right].
 \label{eq:N2-Vgn-general}
\end{align}
\vfill\eject
\end{widetext}
The factor $v$ therefore carries the charge-dependent normalization of the
three-point amplitude, while $\kappa$ and $\eta$ encode the next local
shape data of $u_0(x)$ at the soft endpoint.  Higher-genus and higher-point amplitudes
follow in exactly the same fashion from the higher endpoint derivatives
and the general expressions of ref.~\cite{Johnson:2026twg}.

An interesting feature that jumps out (including in more complicated examples) is that the overall power of~$v$ is minus the Euler number $\chi{=}2{-}2g{-}n$ associated to the amplitude. Aside from that universal factor (which can be removed by a redefinition of $\hbar$) the  $V_{g,n}$ have a universal form for the top degree part (in powers of momentum). It is the lower degree parts that contain the details $(q,{\cal Q})$ of a particular model. This is reminiscent of a familiar structure from the volumes of  ${\cal N}{=}2$ JT supergravity in  ref.~\cite{Turiaci:2023jfa}, where the top degree of a given $V_{g,n}$ (in the lengths $b_i$) was universally the Weil-Petersson volumes. In fact, it is easy to see that if we translate the Weil-Petersson $b_i\to\sqrt{2}P_i$, what we have is precisely:
\begin{equation}
V_{g,n}^{\rm top}(\{P_i\})
=
v^{\,2g-2+n}\,
V_{g,n}^{{\rm WP},{\rm top}}
\!\left(\{\sqrt{2}\,P_i\}\right)\ .
\end{equation}
This intriguing connection  certainly deserves further exploration.

Our  formulae in~(\ref{eq:N2-Vgn-general}), and the more general ones that readily follow from~\cite{Johnson:2026twg},  apply to the soft-edge branches, for which no $R_n$ vanishes.
If
$ \frac{q}{{\cal Q}^2}\in{\mathbb Z}$,
one of the spectral images instead has $R_n{=}0$ and the corresponding
continuum has a hard edge.  Such sectors must use the hard-edge
correlator formulae of~\cite{Johnson:2026twg} instead.

Rather than write more general expressions, we will go in  a different direction. To learn new features about our new strings, it seems  valuable to examine some key special cases. 
We will examine the cases $\mathcal{Q}^2{=}1$ and $2$, corresponding to $\hat c{=}2$ and $\hat c{=}3$. These  of course have an interesting history as the  types of superconformal  field theories that arise for (non-compact) hyper-K\"ahler manifolds, and Calabi--Yau 3-folds respectively. Here they will play a quite different role in how they define string theories.

\subsection{Some $\hat c=3$ strings}
\label{eq:chat-equals-3-strings}
For $\hat c=3$ we have rational ${\cal Q}^2{=}2$ and the pair $(K,N)$ can be  $(m,m)$ where $m\in\mathbb{Z}_{>0}$.
As $m$ increases the available lattice of charges becomes finer (the spacing is set by~$1/N$) and the spectrum of possibilities richer. For $\mathcal Q^2{=}2$, we have the general form for the leading spectral density:
\begin{equation}
 \rho_{0,q}^{(\widehat c=3)}(E)
 =\frac{1}{8\pi^2\hbar}
 \frac{\sinh^2(2\pi p)}
 {p\,[\cosh(2\pi p)-\cos(\pi q)]}
 \Theta\!\left(E-\frac{q^2}{4}\right)\ ,
 \label{eq:hc3-general-density-1}
\end{equation}
where $p\equiv\sqrt{E-E_0}$, and $E_0=\frac{q^2}{4}$. So
 the various values of $q$ from the various sectors will result in different shapes for $\rho_{0,q}^{(\widehat c=3)}(E)$.

In fact, something interesting happens for this special case of $\hat c=3$ that directly pertains to ${\cal N}{=}1$.
The spectral density~(\ref{eq:hc3-general-density-1}) can be re-written  for integer $q$ as:
\begin{equation}
 \rho_{0,q}^{(\widehat c=3)}(E)
 =\frac{1}{2\pi^2\hbar}
 \frac{\sinh^2(\pi p)\cosh^2(\pi p)}
 {p\,[\cosh(2\pi p)-(-1)^q)]}
 \Theta\!\left(E-\frac{q^2}{4}\right)\ ,
 \label{eq:hc3-general-density-2}
\end{equation}
but the denominator factor $\cosh(2\pi p){-}(-1)^q)$ is either $2\sinh^2(\pi p)$ or $2\cosh^2(\pi p)$, depending upon whether $q$ is even or odd. Then one or other of these cases cancels the corresponding combination in the numerator, leaving (up to normalization factors) precisely the form of one or other of the Cardy densities of the ${\cal N}{=}1$ theory, when written in terms of $2p$ (see (\ref{eq:super-NSminus-density}) or (\ref{eq:super-NSplus-density})), for the case $b{=}1$! So the distinct $\pm$ choices of vacua in ${\cal N}{=}1$ are directly related to even or odd $U(1)_R$ charge in the ${\cal N}{=}2$ theory. We shall see such even and odd charge  sectors appear naturally in the lattice of charges of particular rational theories. There will also be quite new behaviour when $q$ takes fractional values.

We will next explore the cases of $m{=}1$ and $m{=}2$ for illustration.

\subsubsection{The primitive case of $m=1$.}
\label{N2-chat3-primitive}
Since ${\cal Q}^2{=}2$, the charge increment under spectral flow is two units (recall equation~(\ref{eq:q-family})).  According to equation~(\ref{eq:n-and-r}), $r\in\mathbb Z_1$, which is trivial, so each family has only a single  representative. Meanwhile $j_0$ runs over just two seed families, according to~(\ref{eq:jzero-range}). Table~\ref{tab:primitive-c3-massive} summarizes this short list of complete massive sector charges.
\begin{table}[h]
\begin{center}
{
\setlength{\arrayrulewidth}{0.4pt}
\setlength{\tabcolsep}{5pt}
\renewcommand{\arraystretch}{1.35}
\setlength{\extrarowheight}{1.5pt}

\begin{tabular}{
!{\vrule width 1.1pt}
c | c | c | c
!{\vrule width 1.1pt}
}
\noalign{\hrule height 1.1pt}
$j_0$ & $Q_R=j_0$ & $q=j_0-1$ & $\{q+2r\}$, $r\in\mathbb{Z}_1$ \\
\hline
0 & $0$ & $-1$ & $-1$ \\
1 & $1$ & $0$ & $0$ \\
\noalign{\hrule height 1.1pt}
\end{tabular}
}
\end{center}
\caption{The massive sector charge table for the $\hat c=3$, $m=1$ (primitive)  model.}\label{tab:primitive-c3-massive}
\end{table}
The charge-conjugate $Q_R{=}{-}1$ state lies in the same fully integral-flow
extended family as the $Q_R{=}{+}1$ representative, so the two massive series are  written as $Q_R{=}0$ and~$|Q_R|{=}1$.

In the centered charge convention of
(\ref{eq:N=2-density-form-one}) the $Q_R{=}0$ channel is represented by
$q{=}-1$, and so  the threshold is $E_0=\frac14$.  
The resulting continuum part of the density has a soft edge, and  is:
\begin{eqnarray}
 \rho_{0,\mathrm{soft}}^{(\hat c=3)}(E)
 &=&
 \frac{1}{4\pi^2\hbar}
 \frac{\sinh^2\left(\pi p\right)}{p}\ ,
 \label{eq:N2-hc3-soft-density}
\end{eqnarray}
with $p{\equiv} \sqrt{E-\frac14}$, and 
where the $\Theta\!\left(E-\frac14\right)$ factor is understood. As already noticed above, except for the shift by $\frac14$, this is the form of the ${\cal N}=1$ Cardy density.

The $|Q_R|=1$ channel is equivalent to  $q=0$, and the continuum  hard-edge density is:
\begin{equation}
 \rho_{0,\mathrm{hard}}^{(\hat c=3)}(E)
 =\frac{1}{4\pi^2\hbar}
 \frac{\cosh^2(\pi p)}{p} \ ,
 \label{eq:N2-hc3-hard-density}
\end{equation}
with $p= \sqrt E$. Indeed, this is the form of the ${\cal N}=1$ Cardy density defined with $(-1)^F$ in the vacuum.

In this case, the ${\cal N}{=}2$ vacuum transform has no discrete massless term in either channel, as can be seen from the fact that while 
the full vacuum family is indexed by~$s{=}1,2$ (according to (\ref{eq:full-massless-range})),  the massless vacuum index~$s^\prime$, in the range given by (\ref{eq:visible-range}), $K+1...K+N-1$ is empty (since here $K+1>K+N-1$, {\it i.e.} $2>1$).\footnote{The corresponding short representations for $s{=}1,2$ do exist in the full CFT. In fact, after half-unit spectral flow their NS representatives map to Ramond ground states with $Q_R=\pm\frac12$.  They simply do not appear in the vacuum modular transform, so they do not appear in our construction.}
This is confirmed directly by the formulae~(\ref{eq:N2-Gamma-general}) where the choices of $q=-1$ or $0$ give $\widetilde\Gamma=0$: The soft-edge density~(\ref{eq:N2-hc3-soft-density}) is finite at $E=0$, while the hard channel~(\ref{eq:N2-hc3-hard-density}) has  the usual $E^{-1/2}$ hard wall branch point and hence no $1/E$ BPS pole.

We next turn to the determination of $u_0(x)$, which once known determines all scattering amplitudes. Of course, the relation to the ${\cal N}=1$ densities means that we already morally know the answer, but it is useful (especially because of the normalization change and energy shift) to directly work through the details.
For the soft-edge channel the classical string equation follows directly by
applying the Abel inversion to (\ref{eq:N2-hc3-soft-density}):
\begin{equation}
\int_{1/4}^{u_0}\frac{\rho_{0,\mathrm{soft}}(E)\,dE}{\sqrt{u_0-E}}
=
\frac{1}{8\pi\hbar}
\left[
I_0\left(2\pi\sqrt{u_0-\frac14}\right)-1
\right]\ ,
\end{equation}
giving, according to~(\ref{eq:en-abel}):
\begin{equation}
\mu-x(u_0)
=
\frac{1}{4\pi}
\left[
I_0\left(2\pi\sqrt{u_0-\frac14}\right)-1
\right].
\end{equation} Since we have $x(0)=0$, using $I_0(i\pi)=J_0(\pi)$ gives:
\begin{equation}
\mu=\frac{J_0(\pi)-1}{4\pi}\ ,
\end{equation}
and hence the classical string equation (in the $x{<}0$ region) can be written
\begin{equation}
{\cal R}_{0}
\equiv x+
\frac{
I_0\!\left(2\pi\sqrt{u_0-\frac14}\right)-J_0(\pi)}
{4\pi}=0.
\end{equation}
Expanding the regular term in powers of $u$ therefore gives
\begin{equation}
 t_k^{(3,\mathrm{soft})}
 =
 \frac{2^{k-2}\pi^{k-1}}{k!}J_k(\pi)\ ,\quad (k\geq1)
 \label{eq:N2-hc3-soft-tk}
\end{equation}
For the hard channel with spectral density~(\ref{eq:N2-hc3-hard-density}), we can separate the universal hard-wall term by writing $\cosh^2(\pi \sqrt{E})=1+\sinh^2(\pi\sqrt{E})$, giving $\rho_{\rm hw}{=}\frac{1}{4\pi^2\hbar\sqrt{E}}$. This  comes from the part of the integral~(\ref{eq:spectral-density-leading}) from~$0$ to~$\mu$  with $u_0(x){=}0$, and hence $\mu{=}\frac{1}{2\pi}$. The remaining part of the  computation  gives:
\begin{equation}
 {\cal R}_{0}
 \equiv x+\frac{I_0(2\pi\sqrt u_0)-1}{4\pi}=0\ ,
 \quad
 t_k^{(3,\mathrm{hard})}
 =\frac{\pi^{2k-1}}{4(k!)^2}\ .
 \label{eq:N2-hc3-hard-tk}
\end{equation}
Indeed, both these formulae for $t_k$ could have been deduced directly from the formulae~(\ref{eq:tk-formulae}) with $\widehat{Q}$ set to zero (since $b{=}1$), along with implementing a rescaling and a shift.

Now that we know $u_0(x)$, (recall that for the soft edge case it is the solution of ${\cal R}_0=0$ for $x<0$), writing down the amplitudes for these cases is simple. The $Q_R{=}0$ soft-edge lattice moments~(\ref{eq:N2-lattice-moments-closed}) work out to be:
\begin{equation}
 {\cal S}_2=\pi^2,\qquad
 {\cal S}_4=\frac{\pi^4}{3},\qquad
 {\cal S}_6=\frac{2\pi^6}{15}\ ,
\end{equation} and the endpoint data~(\ref{eq:N2-ABC}) simplify to:
\begin{equation}
 A=\frac{\pi}{4},\qquad
 B=-\frac{\pi^3}{8},\qquad
 C=\frac{\pi^5}{24}.
\end{equation}
Equivalently, the dimensionless endpoint quantities of~(\ref{eq:N2-v-kappa}) and~(\ref{eq:N2-eta}) are
\begin{equation}
 v=1,\qquad \kappa=-\frac12,\qquad \eta=\frac16.
\end{equation}
Using these in~(\ref{eq:N2-Vgn-general}), we get (in terms of the physical threshold momentum $P=\frac12\sqrt{E-\frac14}$):
\begin{equation}
\begin{aligned}
 &\biggl\{\hat c=3,\;m=1,\;q=-1\biggr\}
 \\[1mm]
 &V_{0,3}
 =1\ ,
 \\
 &V_{1,1}(P)
 =\frac1{48}+\frac{P^2}{24}\ ,
 \\
 &V_{0,4}(\{P_i\})
 =\frac{1}{2}
 +\sum_{i=1}^{4}P_i^2\ ,
 \\
 &V_{1,2}(P_1,P_2)
 =\frac{1}{72}
 +\frac{1}{24}\left(P_1^2+P_2^2\right)
 +\frac{1}{48}\left(P_1^2+P_2^2\right)^2\ .
\end{aligned}
\label{eq:N2-hc3-Vgn}
\end{equation}
Notice that these amplitudes are precisely those that can be obtained from our $b{=}1$, ${\cal N}{=}1$ type~0A$^+$ SVMS string, by putting $c{=}\frac{27}{2}$ into equations~(\ref{eq:N1-quantum-volumes-0Aplus}). This explains why the convention choices we made in equations~(\ref{eq:N2-momentum-convention}) and~(\ref{eq:quantum-volume-conversion-plus}) were the natural ones to use.

Turning to the hard-edge
$|Q_R|{=}1$ channel the $t_k$ of (\ref{eq:N2-hc3-hard-tk}) can just be  inserted into the general hard-edge amplitudes of ref.~\cite{Johnson:2026twg}, adjusted to our conventions. 
Some are given in equations~(\ref{eq:amplitudes-hard-edge}), but now use the $t_k^{(3,\mathrm{hard})}$ instead of $t_k^-$ of~(\ref{eq:tk-minus-in-terms-of-c}). In fact there is a simple translation:
\begin{equation}
t_k^{(3,\mathrm{hard})}
=
\frac{1}{8\sqrt2\,\pi^2}\,
\frac{1}{4^k}\,
\left.t_k^-\right|_{c=27/2}\ ,
\end{equation} converting the previous results for the ${\cal N}{=}1$ 0A$^-$ case.

\subsubsection{The  case of $m=2$.}
\label{N2-chat3-next}

This case will begin to introduce very new behaviour. Now that   $K{=}2$, equations~(\ref{eq:jzero-range}) and~(\ref{eq:q-family}) show that the four seed families have
charges:
\begin{equation}
 q_{j_0,0}=-1,-\frac12,0,+\frac12,
 \qquad j_0=0,1,2,3 \ .
\end{equation}
$N{=}2$  gives two spectral images per family, and since $\mathcal Q^2{=}2$, the second member of each family is shifted by two units.

\begin{table}[h]
\begin{center}
{
\setlength{\arrayrulewidth}{0.4pt}
\setlength{\tabcolsep}{5pt}
\renewcommand{\arraystretch}{1.35}
\setlength{\extrarowheight}{1.5pt}

\begin{tabular}{
!{\vrule width 1.1pt}
c | c | c | c
!{\vrule width 1.1pt}
}
\noalign{\hrule height 1.1pt}
$j_0$ & $Q_R=\frac{j_0}{2}$ & $q=\frac{j_0-2}{2}$ & $\{q+2r\}$, $r\in\mathbb{Z}_2$ \\
\hline
0 & $0$ & $-1$ & $-1,1$ \\
1 & $\tfrac12$ & $-\tfrac12$ & $-\tfrac12,\tfrac32$ \\
2 & $1$ & $0$ & $0,2$ \\
3 & $\tfrac32$ & $\tfrac12$ & $\tfrac12,\tfrac52$ \\
\noalign{\hrule height 1.1pt}
\end{tabular}
}
\end{center}
\caption{The massive sector charge table for the $\hat c=3$, $m=2$  model.}\label{tab:next-c3-massive}
\end{table}

Table~\ref{tab:next-c3-massive} shows the available charges for the massive 
case, including flow copies (last column), making 8 in total since $j\in\mathbb{Z}_8$. While the simpler $m=1$ theory of the previous subsection  had only the two $j$ labels giving $q=-1,0$, this $m=2$ case   adds half-integer $q=\pm\frac12$ as well as their flow images.

These four distinct choices will result in three different functions for the continuum (the shape for the charge conjugates $q=\pm\frac12$ are the same):
\begin{align}
 q\equiv 1\pmod 2:\qquad
 \rho(E)
 &=\frac{1}{4\pi^2\hbar}
 \frac{\sinh^2(\pi p)}{p},
 \label{eq:hc3-odd-density}\\
 q\equiv 0\pmod 2:\qquad
 \rho(E)
 &=\frac{1}{4\pi^2\hbar}
 \frac{\cosh^2(\pi p)}{p},
 \label{eq:hc3-even-density}\\
 q\equiv \pm\frac12\pmod 2:\qquad
 \rho(E)
 &=\frac{1}{8\pi^2\hbar}
 \frac{\sinh^2(2\pi p)}{p\cosh(2\pi p)}.
 \label{eq:hc3-half-density}
\end{align}
Here we've again used the shorthand $p=\sqrt{E-q^2/4}$.

The $q=0$ case is the hard edge already present in the primitive $\widehat c=3$ example of last subsection, while the $q=1$ case is the soft edge also seen there. One might have  expected something new from the flowed cousin of the $q=0$ at $q=2$ since there is now a threshold of $E_0=1$, however it is again a hard edge, just shifted by unity compared to the standard case, {\it i.e.,} $\rho\sim 1/\sqrt{E-1}$, because of the cosh numerator.

It is the $q{=}{\pm}\frac12$ cases (and their flowed cousins $q{=}\frac32,\frac52$) that present the new kind of spectral density seen in~(\ref{eq:hc3-half-density}). We will discuss it below.

Turning to the massless representatives, the index $s^\prime$ takes the value 3, meaning there is a single BPS state. Recall that the primitive theory did not support a BPS state. For this case $s'=3$, $q_{s'}=\frac12$.  Here $q_{s'}$ labels the short representation in the NS description; after half-unit spectral flow it becomes a Ramond ground state with $Q_R{=}0$.
This state is non-resonant since $2q/{\cal Q}^2{=}q{\notin}\mathbb{Z}$, and we get: 
\begin{equation}
 |\widetilde\Gamma_{\frac12}|=\frac{1}{4\pi^2}\ .
\end{equation}

For the new half-integer continuum sector it is interesting to see how far we can take the Abel inversion  procedure.  Taking $q=\frac12$, so that $E_0=\frac{1}{16}$, define:
\begin{equation}
 U\equiv u_0-\frac1{16}.
\end{equation}
Using:
\begin{equation}
 \frac{\sinh^2(2\pi p)}{\cosh(2\pi p)}
 =\cosh(2\pi p)-\operatorname{sech}(2\pi p),
\end{equation}
the Abel inversion integral appearing in~(\ref{eq:N2-Abel-inverse}) becomes:
\begin{widetext}
\begin{align}
\int_{1/16}^{u_0}\frac{\rho_{0,1/2}^{(\hat c=3)}(E)\,dE}{\sqrt{u_0-E}}
 &=\frac{1}{4\pi^2\hbar}\int_0^{\pi/2}d\theta\,
 \left[
 \cosh\!\left(2\pi\sqrt U\sin\theta\right)
 -\operatorname{sech}\!\left(2\pi\sqrt U\sin\theta\right)
 \right]\nonumber\\
 &=\frac{1}{8\pi\hbar}I_0(2\pi\sqrt U)
 -\frac{1}{4\pi^2\hbar}\int_0^{\pi/2}d\theta\,
 \operatorname{sech}\!\left(2\pi\sqrt U\sin\theta\right).
 \label{eq:N2-hc3-half-Abel}
\end{align}
Consequently, our equation for~$u_0$ is:
\begin{equation}
 -x'(u_0)=
 \frac{1}{4\pi}\frac{d}{du_0}I_0(2\pi\sqrt U)
 -\frac{1}{2\pi^2}\frac{d}{du_0}
 \int_0^{\pi/2}d\theta\,
 \operatorname{sech}\!\left(2\pi\sqrt U\sin\theta\right)\ .
 \label{eq:N2-hc3-half-xprime}
\end{equation}
\end{widetext}
We see that the $\operatorname{sech}$ term  prevents us from writing it in the simple Bessel form found for the $q=\pm1$ soft channel.  Nevertheless, this relation can be used to compute the values of  the $t_k$ numerically by expanding around $u_0=0$, as we shall see later.
Moreover, for these $q=\pm\frac12$ cases the lattice moments~(\ref{eq:N2-lattice-moments-closed}) are:
\begin{equation}
 {\cal S}_2=2\pi^2,\qquad
 {\cal S}_4=\frac{8\pi^4}{3},\qquad
 {\cal S}_6=\frac{64\pi^6}{15}\ ,
 \label{eq:qhalf-moments}
\end{equation}
giving the
endpoint quantities:
\begin{equation}
 A=\frac{\pi}{2},\qquad
 B=\frac{\pi^3}{2},\qquad
 C=\frac{31\pi^5}{12}\ .
 \label{eq:qhalf-endpoints}
\end{equation}
Equivalently,
\begin{equation}
 v=\frac12,\qquad \kappa=1,\qquad \eta=\frac{31}{6}.
\end{equation}
Using~(\ref{eq:N2-Vgn-general}), the corresponding soft-edge amplitudes are:
\begin{equation}
\begin{aligned}
 &\left\{\hat c=3,\;m=2,\;q=\pm\frac12\right\}
 \\[1mm]
 &V_{0,3}
 =\frac12\ ,
 \\
 &V_{1,1}(P)
 =-\frac1{48}+\frac{P^2}{48}\ ,
 \\
 &V_{0,4}(\{P_i\})
 =-\frac14+\frac14\sum_{i=1}^{4}P_i^2\ ,
 \\
 &V_{1,2}(P_1,P_2)
 =-\frac{19}{576}
 -\frac1{48}\left(P_1^2+P_2^2\right)
 +\frac1{192}\left(P_1^2+P_2^2\right)^2\ .
\end{aligned}
\label{eq:N2-hc3-m2-half-Vgn}
\end{equation}
The flowed members of the same $\frac12$-integer families have the same polynomials in the physical  momentum $P{=}\frac12\sqrt{E-E_0}$, with the appropriate shifted value of~$E_0$.

\subsection{Some $\hat c=2$ strings}
For $\hat c=2$ we now have rational ${\cal Q}^2{=}1$ and the pair $(K,N)$ can be  $(m,2m)$ where $m\in\mathbb{Z}_{>0}$.
Again, increasing $m$ enriches the theory considerably. We will explore the cases $m=1$ and $2$ in turn, finding some new features.

\subsubsection{The primitive case of $m=1$}
\label{sec:N2-chat2-primitive}

The two basic  families, corresponding to $j_0=0,1$, are in table~\ref{tab:primitive-c2-massive}, along with their flowed cousins that differ in charge by one unit.

\begin{table}[h]
\begin{center}
{
\setlength{\arrayrulewidth}{0.4pt}
\setlength{\tabcolsep}{5pt}
\renewcommand{\arraystretch}{1.35}
\setlength{\extrarowheight}{1.5pt}

\begin{tabular}{
!{\vrule width 1.1pt}
c | c | c | c
!{\vrule width 1.1pt}
}
\noalign{\hrule height 1.1pt}
$j_0$ & $Q_R=\frac{j_0}{2}$ & $q=\frac{j_0-1}{2}$ & $\{q+r\}$, $r\in\mathbb{Z}_2$ \\
\hline
0 & $0$ & $-\tfrac12$ & $-\tfrac12,\,+\tfrac12$ \\
1 & $\tfrac12$ & $0$ & $0,\,1$ \\
\noalign{\hrule height 1.1pt}
\end{tabular}
}
\end{center}
\caption{The massive sector charge table for the $\hat c{=}2$, $m{=}1$  (primitive) model.}\label{tab:primitive-c2-massive}
\end{table}

The density for the $q=0$ case is:
\begin{equation}
 \rho_{0,\mathrm{hard}}^{({\mathcal Q}^2=1)}(E)
 =
 \frac{1}{4\pi^2\hbar}
 \frac{\cosh(2\pi\sqrt E)}{\sqrt E}\ ,
 \qquad
 t_k^{(2,\mathrm{hard})}
 =\frac{\pi^{2k-1}}{2(k!)^2}\ .
 \label{eq:N2-hc2-hard-unrecombined}
\end{equation}
and we have written the $t_k$ that follow from it since this is (up to a normalization) the classic density associated to ${\cal N}=1$ JT gravity and first analyzed in terms of $t_k$ in ref.~\cite{Johnson:2020heh}.

It is useful to compare this $q=0$ hard-edge channel with the primitive $\hat c=3$ hard edge in (\ref{eq:N2-hc3-hard-density}).  The common universal hard-wall term is:
\begin{equation}
 \rho_{\rm hw}(E)\equiv\frac{1}{4\pi^2\hbar\sqrt E}\ ,
\end{equation}
and the two densities obey:
\begin{equation}
 \rho_{0,\mathrm{hard}}^{(\hat c=2)}(E)-\rho_{\rm hw}(E)
 =2\left[\rho_{0,\mathrm{hard}}^{(\hat c=3)}(E)-\rho_{\rm hw}(E)\right],
\end{equation}
which follows immediately from $\cosh(2x)-1=2\sinh^2x$.  We see that the full densities are not simply related by an overall factor of two: the hard-wall piece is common to both.  This makes sense from the perspective of  the spectral-flow image lattice.  At $E_0=0$, the image positions are $R_n=n{\cal Q}^2/2$, so they are $R_n=n$ for $\hat c=3$ but $R_n=n/2$ for $\hat c=2$.  The latter  has the two $q=0,1$  representatives, while the primitive $\hat c=3$  family has only the single $q=0$  representative.  Both lattices contain the same single zero image, which gives the universal $\rho_{\rm hw}$ term, but  the remaining regular contribution is doubled in the finer $\hat c=2$ lattice.  

The density for $q=-\frac12$ is, writing $p=\sqrt{E-\frac{1}{16}}$:
\begin{equation}
\rho_{0,\mathrm{soft}}^{(\hat c=2)}(E)
=
\frac{1}{4\pi^2\hbar}
\frac{\sinh^2(2\pi p)}
{p\,\cosh(2\pi p)}
\Theta\!\left(E-\frac1{16}\right).
\label{eq:N2-hc2-soft-density}
\end{equation}
a form we encountered in the previous section for the $m=2$ $\hat c=3$ theory (see (\ref{eq:hc3-half-density})). It  is a factor of two larger here though, and it is also reflected in the analysis of the  BPS sector. Now the massless vacuum label~$s^\prime$ takes one allowed value, which is 2, and $q_{s^\prime}=+\frac12$.  This is again the NS short-state label; after half-unit spectral flow the corresponding Ramond ground state has $Q_R^{\rm BPS}=0$. This time this state is resonant, since 
$\frac{2q}{{\mathcal Q}^2}=1\in{\mathbb Z}$
and as a consequence it  has twice the residue it had in the $\hat c=3$ case, giving:\begin{equation}
 |{\widetilde\Gamma}_{\frac12}|
 =\frac{1}{2\pi^2}\ .
 \label{eq:N2-hc2-Gamma}
\end{equation}
This factor of two will take on extra significance in Section~\ref{sec:surprise-N=4}.

As we saw previously, there is no  simple  expression for the regular soft
multicritical tower because of the $\operatorname{sech}(2\pi p)$ structure, but the endpoint data are straightforward to work out. 
The soft-edge lattice moments are double what is in~(\ref{eq:qhalf-moments}), and  the endpoint data are double those in~(\ref{eq:qhalf-endpoints}).
This follows from comparing (\ref{eq:N2-hc2-soft-density}) with~(\ref{eq:hc3-half-density}):
\begin{equation}
 \rho_{0,q=\pm\frac12}^{(\hat c=2,m=1)}(E)
 =2\,\rho_{0,q=\pm\frac12}^{(\hat c=3,m=2)}(E)\ .
\end{equation}
 The corresponding values of $\kappa$ and $\eta$ are unchanged while~$v$ is halved, and~(\ref{eq:N2-Vgn-general}) therefore gives the translation dictionary:
\begin{equation}
 V_{g,n}^{(\hat c=2,m=1)}
 =2^{\chi}V_{g,n}^{(\hat c=3,m=2;q=\pm\frac12)},
 \qquad \chi=2-2g-n.
 \label{eq:N2-hc2-from-hc3-half}
\end{equation}
So $V_{0,3}$ and $V_{1,1}$ are one half of those given in~(\ref{eq:N2-hc3-m2-half-Vgn})  while $V_{0,4}$ and $V_{1,2}$ are one quarter of  the ones displayed there. Indeed, using~(\ref{eq:N2-Vgn-general}), we have:
\begin{equation}
\begin{aligned}
 &\left\{\hat c=2,\;m=1,\;q=\pm\frac12\right\}
 \\[1mm]
 &V_{0,3}
 =\frac14\ ,
 \\
 &V_{1,1}(P)
 =-\frac1{96}+\frac{P^2}{96}\ ,
 \\
 &V_{0,4}(\{P_i\})
 =-\frac1{16}+\frac1{16}\sum_{i=1}^{4}P_i^2\ ,
 \\
 &V_{1,2}(P_1,P_2)
 =-\frac{19}{2304}
 -\frac1{192}\left(P_1^2+P_2^2\right)
\\&\hskip5cm+\frac1{768}\left(P_1^2+P_2^2\right)^2\ ,
\end{aligned}
\label{eq:N2-hc2-m1-half-Vgn}
\end{equation}
written  out explicitly for later ease of comparison.

\subsubsection{The case of $m=2$}
\label{sec:N2-chat2-next}
Here $(K,N)=(2,4)$.  The charge spacing is therefore refined to $\Delta Q_R=\Delta q=1/4$.  There are now four basic massive seed families:
\begin{equation}
 q_{j_0,0}=-\frac12,-\frac14,0,+\frac14,
 \qquad j_0=0,1,2,3,
\end{equation}
and since ${\cal Q}^2=1$ the remaining (spectral image) members of each family differ in charge by one unit.  They are summarized in table~\ref{tab:next-c2-massive}.
\begin{table}[h]
\begin{center}
{
\setlength{\arrayrulewidth}{0.4pt}
\setlength{\tabcolsep}{5pt}
\renewcommand{\arraystretch}{1.35}
\setlength{\extrarowheight}{1.5pt}

\begin{tabular}{
!{\vrule width 1.1pt}
c | c | c | c
!{\vrule width 1.1pt}
}
\noalign{\hrule height 1.1pt}
$j_0$ & $Q_R=\frac{j_0}{4}$ & $q=\frac{(j_0-2)}{4}$ &  $\{q+r\}$, $r\in\mathbb{Z}_4$ \\
\hline
0 & $0$ & $-\tfrac12$ & $-\tfrac12,\tfrac12,\tfrac32,\tfrac52$ \\
1 & $\tfrac14$ & $-\tfrac14$ & $-\tfrac14,\tfrac34,\tfrac74,\tfrac{11}{4}$ \\
2 & $\tfrac12$ & $0$ & $0,1,2,3$ \\
3 & $\tfrac34$ & $\tfrac14$ & $\tfrac14,\tfrac54,\tfrac94,\tfrac{13}{4}$ \\
\noalign{\hrule height 1.1pt}
\end{tabular}
}
\end{center}
\caption{The massive sector charge table for the $\hat c{=}2$, $m{=}2$ model.}
\label{tab:next-c2-massive}
\end{table}

\begin{widetext}
So the rational vacuum transform now contains sixteen massive labels, $j\in\mathbb Z_{16}$.  As in the $\hat c=3$, $m=2$ example, we get new features: the quarter-integer charge sectors produce a  new continuum shape.

For ${\cal Q}^2=1$ the general density becomes:
\begin{equation}
 \rho_{0,q}^{(\widehat c=2)}(E)
 =\frac{1}{4\pi^2\hbar}
 \frac{\sinh(2\pi p)\sinh(4\pi p)}
 {p\,[\cosh(4\pi p)-\cos(2\pi q)]}
 \Theta\!\left(E-\frac{q^2}{4}\right),
 \qquad p=\sqrt{E-\frac{q^2}{4}}.
 \label{eq:N2-hc2-m2-general-density}
\end{equation}
The four seed families reduce to three distinct  forms:
\begin{align}
 q\equiv\frac12\pmod1:\qquad
 \rho(E)
 &=\frac{1}{4\pi^2\hbar}
 \frac{\sinh^2(2\pi p)}{p\cosh(2\pi p)},
 \label{eq:N2-hc2-m2-half-density}\\
 q\equiv0\pmod1:\qquad
 \rho(E)
 &=\frac{1}{4\pi^2\hbar}
 \frac{\cosh(2\pi p)}{p},
 \label{eq:N2-hc2-m2-int-density}\\
 q\equiv\pm\frac14\pmod1:\qquad
 \rho(E)
 &=\frac{1}{4\pi^2\hbar}
 \frac{\sinh(2\pi p)\tanh(4\pi p)}{p}.
 \label{eq:N2-hc2-m2-quarter-density}
\end{align}
\end{widetext}
The first of these is the same form of density already present in the primitive $\hat c{=}2$ soft sector, while the $q{=}0$ member of the second is the familiar hard edge.  Its non-zero integer flow copies have shifted hard-edge-type thresholds, since the density still behaves as $1/p$ at the endpoint.  The $\frac14$-integer family is new to the examples considered so far; it is a soft edge, since its density vanishes linearly in $p$ at threshold.

The full massless family has $1\leq s\leq7$, but the vacuum modular transform retains only
\begin{equation}
 s'=3,4,5,
 \qquad r'\in\mathbb Z_4.
\end{equation}
The $q_{s'}$ are the short-state labels in the NS description.  After half-unit spectral flow there are three distinct vacuum-visible Ramond BPS charges, with four partial-flow representatives for each $s'$.  Their charge and residue data are shown in table~\ref{tab:next-c2-BPS}.
\begin{table}[h]
\begin{center}
{
\setlength{\arrayrulewidth}{0.4pt}
\setlength{\tabcolsep}{5pt}
\renewcommand{\arraystretch}{1.35}
\setlength{\extrarowheight}{1.5pt}
\begin{tabular}{
!{\vrule width 1.1pt}
c | c | c | c
!{\vrule width 1.1pt}
}
\noalign{\hrule height 1.1pt}
$s'$ & $q_{s'}=\frac{s'-2}{4}$ & $Q_R^{\rm BPS}=q_{s'}-\frac12$ & $|\widetilde\Gamma_{q_{s'}}|$ \\
\hline
3 & $\tfrac14$ & $-\tfrac14$ & $\tfrac{\sqrt2}{8\pi^2}$ \\
4 & $\tfrac12$ & $0$ & $\tfrac{1}{2\pi^2}$ \\
5 & $\tfrac34$ & $\tfrac14$ & $\tfrac{\sqrt2}{8\pi^2}$ \\
\noalign{\hrule height 1.1pt}
\end{tabular}
}
\end{center}
\caption{The vacuum-connected BPS sectors for the $\hat c=2$, $m=2$  model.}
\label{tab:next-c2-BPS}
\end{table}
The middle state is the same resonant mechanism already present in the primitive $m=1$ theory, while the neighboring quarter-charge states are non-resonant and are new.  In particular the $m=2$ theory has visible Ramond BPS charges $Q_R^{\rm BPS}=-\frac14,0,+\frac14$, rather than only the single charge 0 state of the primitive case.

For the  new $\frac14$-integer continuum family,  the  Abel transform can be reduced to a Bessel term plus an angular integral, but again the latter does not simplify further, so we will not display it here.  
For this $q=\frac14$ model, and indeed the $q=\frac12$ model, the $t_k$ can be computed numerically from the Abel methods, and the first eight values will be useful in the next section for numerical analysis, and so they are listed in table~\ref{tab:hc2-m2-numerical-tk}.
\begin{table}[h]
\begin{center}
{
\setlength{\arrayrulewidth}{0.4pt}
\setlength{\tabcolsep}{8pt}
\renewcommand{\arraystretch}{1.35}
\setlength{\extrarowheight}{1.0pt}

\begin{tabular}{
!{\vrule width 1.1pt}
c | c | c
!{\vrule width 1.1pt}
}
\noalign{\hrule height 1.1pt}
$t_k$ & $q=\frac12$ & $q=\frac14$ \\
\hline
$t_0$ & $0.12207844642148$  & $0.117466240615828$ \\
$t_1$ & $1.07240439284848$  & $1.61008752094708$ \\
$t_2$ & $3.23938258107813$  & $2.55175666082976$ \\
$t_3$ & $3.45832881158930$  & $12.0089222144296$ \\
$t_4$ & $2.62578282990482$  & $-53.7776178106353$ \\
$t_5$ & $0.370306584519403$ & $408.545066442474$ \\
$t_6$ & $1.29326118775459$  & $-2993.18932818940$ \\
$t_7$ & $-1.86757807038127$ & $22249.1445479435$ \\
\noalign{\hrule height 1.1pt}
\end{tabular}
}
\end{center}
\caption{The numerical multicritical couplings $t_k$ for the
$m{=}2$, $\widehat c{=}2$ models with $q=\frac12$ and $q=\frac14$.}
\label{tab:hc2-m2-numerical-tk}
\end{table}

For the representatives $q=\pm\frac14$, the endpoint expansion instead readily  gives for the lattice moments~(\ref{eq:N2-lattice-moments-closed}):
\begin{equation}
 {\cal S}_2=8\pi^2,\qquad
 {\cal S}_4=\frac{128\pi^4}{3},\qquad
 {\cal S}_6=\frac{4096\pi^6}{15}\ ,
\end{equation}
and endpoint data:
\begin{equation}
 A=2\pi,\qquad
 B=14\pi^3,\qquad
 C=\frac{691\pi^5}{3}.
\end{equation}
Equivalently,
\begin{equation}
 v=\frac18,\qquad \kappa=7,\qquad \eta=\frac{691}{6}.
\end{equation}
Using~(\ref{eq:N2-Vgn-general}), the resulting amplitudes are
\begin{equation}
\begin{aligned}
 &\left\{\hat c=2,\;m=2,\;q=\pm\frac14\right\}
 \\[1mm]
 &V_{0,3}
 =\frac18\ ,
 \\
 &V_{1,1}(P)
 =-\frac7{192}+\frac{P^2}{192}\ ,
 \\
 &V_{0,4}(\{P_i\})
 =-\frac7{64}
 +\frac1{64}\sum_{i=1}^{4}P_i^2\ ,
 \\
 &V_{1,2}(P_1,P_2)
 =-\frac{103}{9216}
 -\frac7{768}\left(P_1^2+P_2^2\right)
 \\&\hskip5cm+\frac1{3072}\left(P_1^2+P_2^2\right)^2\ .
\end{aligned}
\label{eq:N2-hc2-m2-quarter-Vgn}
\end{equation}
The other members of the same quarter-integer families have the same polynomials in $P=\frac12\sqrt{E-E_0}$, with their shifted thresholds.

\subsection{An enhancement to an ${\cal N}=4$ string}
\label{sec:surprise-N=4}
Revisiting the primitive $(K,N)=(1,2)$ case of $\hat c=2$, {\it i.e.,} $c=6$ there is actually a potential enhancement of the 
${\cal N}=2$ algebra, as can be seen by noting that spectral flow~(\ref{eq:N2-flow-weights}) by
$\eta$ acts on the NS vacuum ($h{=}0$, $Q_R{=}0$)  to give a state with:
\begin{equation}
h_\eta=\frac{c}{6}\eta^2,
\qquad
Q_{R,\eta}=\frac{c}{3}\eta.
\end{equation}
For unit spectral flow, $\eta=\pm1$, this therefore gives new NS states with:
\begin{equation}
h_{\pm1}=1\ ,
\qquad
Q_{R,\pm1}=\pm2\ ,
\end{equation}
{\it i.e.}, precisely at $\hat c{=}2$, these states have conformal dimension one. If the corresponding unit spectral-flow fields (denoted $U_{\pm1}$) are present as mutually local holomorphic operators, they can therefore join the original
$U(1)_R$ current as additional currents of the chiral algebra.  Together they generate the
$SU(2)_1$ current algebra, giving the characteristic R-symmetry
of the  ``small'' ${\cal N}=4$ algebra at level one.  Notice that this does {\it not} happen automatically for the non-primitive theories.  The construction of ref.~\cite{Eguchi:2003ik} only requires the $N$-unit spectral-flow fields $U_{\pm N}$ to be mutually local.  So, for the non-primitive $\hat c=2$ theories $(K,N)=(m,2m)$, it guarantees locality of $U_{\pm 2m}$, but not of $U_{\pm1}$.  The finer fractional-charge sectors can have non-trivial monodromy with $U_{\pm1}$, so the unit-flow fields need not belong to the local chiral algebra. 

The enhancement is also visible in the organization of the vacuum
modular transform.  The  small-${\cal N}=4_1$ vacuum combines
the two  ${\cal N}{=}2$  spectral-flow orbits in Table~\ref{tab:primitive-c2-massive}.  Their
massive parts recombine into a single continuum channel with $Q_R=0$,
while their massless parts give a single fully extended massless
character appearing with coefficient two.  This factor of two dovetails nicely with the doubling of the BPS sector, already discussed in Section~\ref{sec:N2-chat2-primitive}, due to the resonant pole (see equation~(\ref{eq:N2-hc2-Gamma})).

Before the small ${\cal N}{=}4$ recombination, the 
${\cal N}{=}2$  bookkeeping also had a  $q{=}0$ hard-edge
representative. While it is still a sector of the underlying theory,  it is not mutually local with the additional unit spectral-flow currents and therefore does not define an ordinary local sector of the extended small ${\cal N}{=}4$ theory, and hence won't appear in the vacuum modular transformations~\cite{Eguchi:2003ik}.

Put differently, in the primitive $(K,N)=(1,2)$ bookkeeping this sector has $Q_R{=}1/2$ and hence its OPE with $U_{\pm1}$ carries a half-integral power (see refs.~\cite{Schwimmer:1986mf,Lerche:1989uy}), giving non-trivial monodromy with the fields that extend the ${\cal N}{=}2$ algebra to small ${\cal N}{=}4_1$, and so cannot be part  of the extended local chiral algebra.  The $q=-\frac12$ continuum  instead has $Q_R=0$ and is local with respect to~$U_{\pm1}$, and so will appear in the extended chiral algebra and hence the vacuum modular transformations. This will become much more explicit in Section~\ref{sec:an-N=4-generalization} when we directly construct the small ${\cal N}{=}4$ strings that follow naturally from  our approach.

\subsection{Non-Perturbative Physics}
\label{sec:non-perturbative-studies}
Of course, another powerful feature of the string equation methods used here is that they readily allow non-perturbative features (beyond the genus expansion)  to be studied. 
We have not checked that all possible ${\cal N}{=}2$ models  defined by these methods are guaranteed to be non-perturbatively well defined. This would require a demonstration, for all models, that the leading solutions $u_0(x)$ their leading spectral densities  provide (through the Abel transform~(\ref{eq:spectral-density-leading})) can be lifted to a full solution of non-perturbative string equation~(\ref{eq:big-string-equation}). That  is an interesting general problem  for a later project. 

Nevertheless, it is instructive to study properties of the full solutions for $u(x)$ of the string equations for the particular models we've studied in detail so far. When they can be found, the associated Schr\"odinger problem:
\begin{equation}
\left[-\hbar^2\frac{d^2}{dx^2}+u(x)\right]\psi_E(x)=E\psi_E(x)\ ,
\label{eq:schrodinger}
\end{equation}
can be studied, from which the wavefunctions can be extracted and from them the complete non-perturbative~$\rho(E)$ can be built, according to:
\begin{equation}
    \rho(E)=\int_{-\infty}^\mu \psi(E,x)^2dx
\label{eq:exact-density}\ .
\end{equation}
This approach (based on structures found long ago~\cite{Gross:1990aw,Banks:1990df}) was developed in ref.~\cite{Johnson:2020exp}, and refined over the years in subsequent papers such as ref.~\cite{Johnson:2022wsr}.  

The high order and high nonlinearity of the string equation necessitates the use of numerical methods to extract  $u(x)$. Using {\tt MATLAB}'s {\tt bvp4c} solver, the string equation~(\ref{eq:big-string-equation}) was solved in a truncation with $t_1,\ldots,t_7$ non-zero,\footnote{\label{fn:truncation}Ref.~\cite{Johnson:2020exp} had a partial explanation for why this truncation did not lose any essential physics. For one thing, in the cases studied there  the $t_k$, which bring in increasingly higher $x$-derivatives to the ODE for higher $k$, themselves rapidly shrink with increasing~$k$. This is enhanced further by the decreasing coefficient of the higher derivatives in the Gel'fand-Dikii polynomials. Also important here is the fact that each derivative comes with a factor of $\hbar$ so that for small $\hbar$, which we are working with here, their effective coefficient gets smaller with increasing order.}   Further refinements performed for this project now allow the numerical analysis to be controlled, with the aid of  continuation, at much smaller values of $\hbar$ than was possible before, permitting more direct comparisons alongside the classical limit than previously done. The key improvements  are described in Appendix~\ref{app:numerical-recipes-stringeq}.  The results presented below are for  $\hbar{=}10^{-3}$.

Next, the Schr\"odinger problem~(\ref{eq:schrodinger})
is  solved numerically directly using the full numerical~$u(x)$ as potential.  The  normalization of the numerical wavefunction is computed by matching onto  the exact Bessel form of the wavefunctions of the asymptotic form of the potential to the right: $u(x)\to \hbar^2(\Gamma^2-\frac14)/x^2$. The integration~(\ref{eq:exact-density}) then 
gives the  non-perturbative spectral density~$\rho(E)$. Some details are given in Appendix \ref{app:numerical-recipes-schrodinger}.

\begin{widetext}
We start by studying  the $q{=}{-}1$ 
case of the $\hat c{=}3$ $m{=}1$ model from Section~\ref{N2-chat3-primitive}. The leading spectral densities of that section have already been noted to have the form of the Cardy densities for ${\cal N}{=}1$ case, with the soft threshold shifted to $E{=}\frac14$ here. Full matrix model corrections for densities of this type  have been studied in our previous work~\cite{Johnson:2024bue}, but only for larger $\hbar$.  The new  $\hbar{=}10^{-3}$ non-perturbative solution for $u(x)$ is given on the left in figure~\ref{fig:sample-potentials}.

\begin{figure}[h]
    \centering
    \includegraphics[width=0.5\linewidth]{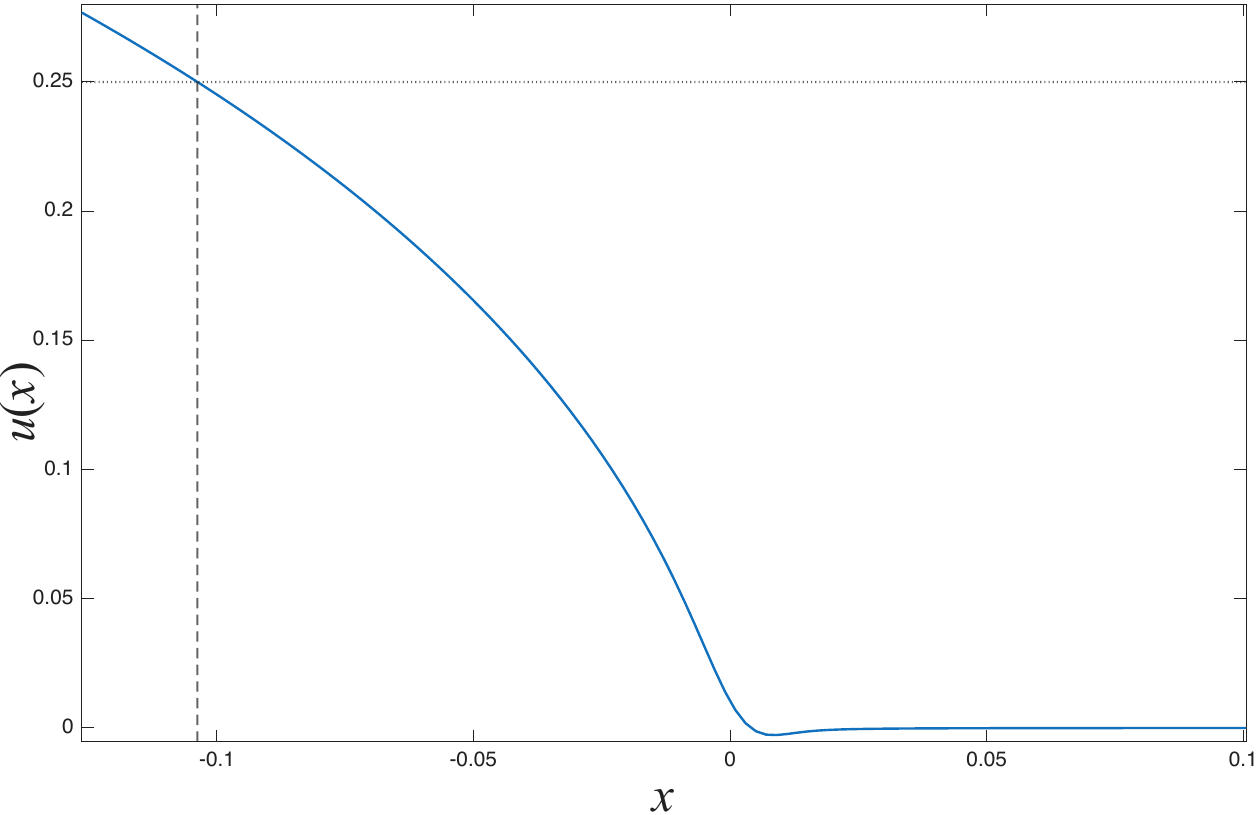}\includegraphics[width=0.50\linewidth]{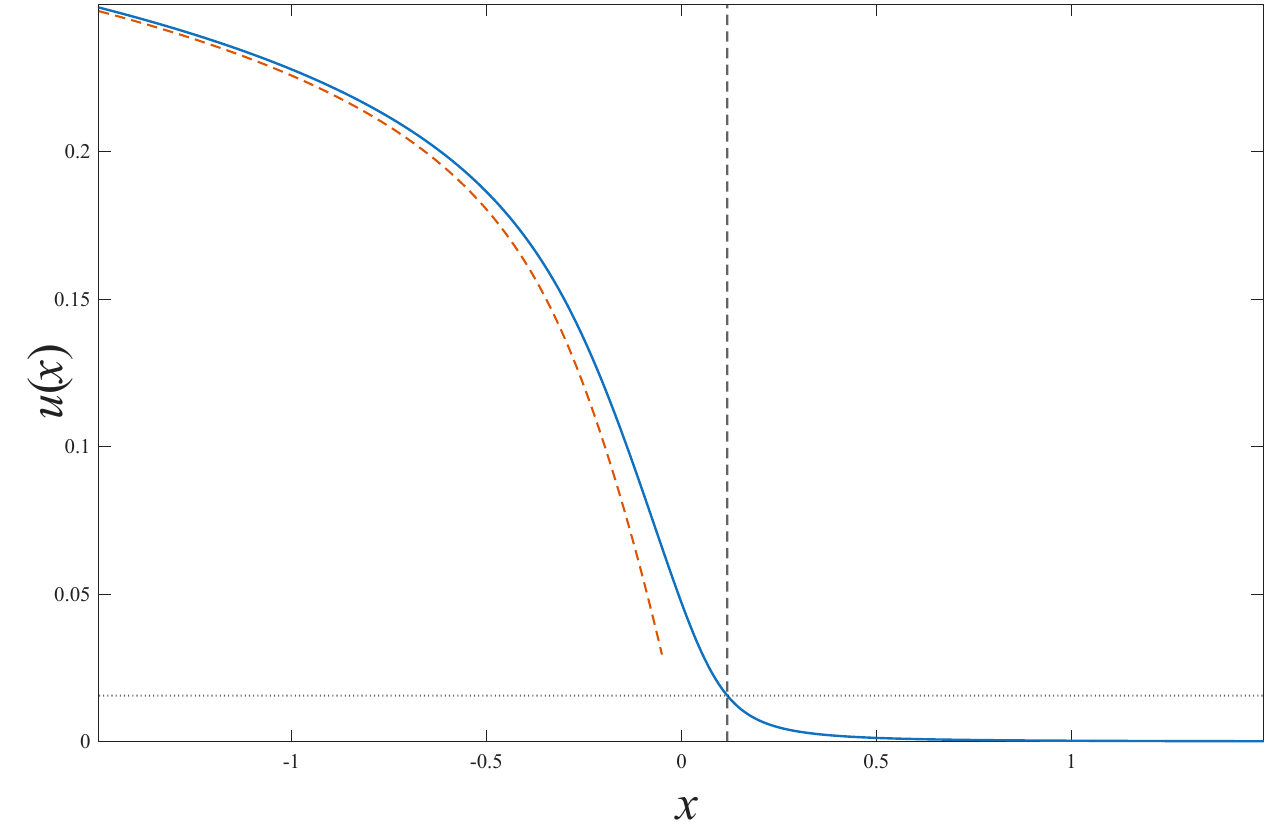}
    \caption{On the left is the solution of the string equation~(\ref{eq:big-string-equation}) for potential $u(x)$ for the   $\hat c=3$ model with $q=-1$. Here,~$\Gamma{=}0$.  The vertical dashed line is at $\mu{=}\frac{J_0(\pi)-1}{4\pi}$, at which $u_0(x){=}E_0{=}\frac14$. On the right is $u(x)$ for the   $\hat c=2$ model with $q=\frac14$. Here,~$\widetilde{\Gamma}{=}\frac{\sqrt{2}}{8\pi^2}$, and $E_0{=}\frac{1}{64}$ with $\mu\simeq 0.11747$. The dashed curve shows the classical asymptote to the left.}
    \label{fig:sample-potentials}
\end{figure}

From this $u(x)$ the Schr\"odinger-type spectral problem can be implemented, with the results for the fully non-perturbative spectral density  given  in figure~\ref{fig:non-perturbative-plots-0}.\footnote{Indeed, this can be compared to the plot for the $b{=}1$ model of ref.~\cite{Johnson:2024bue}, where it was noticed that (in the context of the matrix model of the ordinary VMS theory) it had rather nice non-perturbative behaviour. The models are  essentially the same, up to a shift of the threshold energy.} The wiggles have a precise understanding as underlying probability peaks of the discrete energy levels.~\cite{Johnson:2021zuo}.

\begin{figure}[h]
    \centering \includegraphics[width=0.5\linewidth]{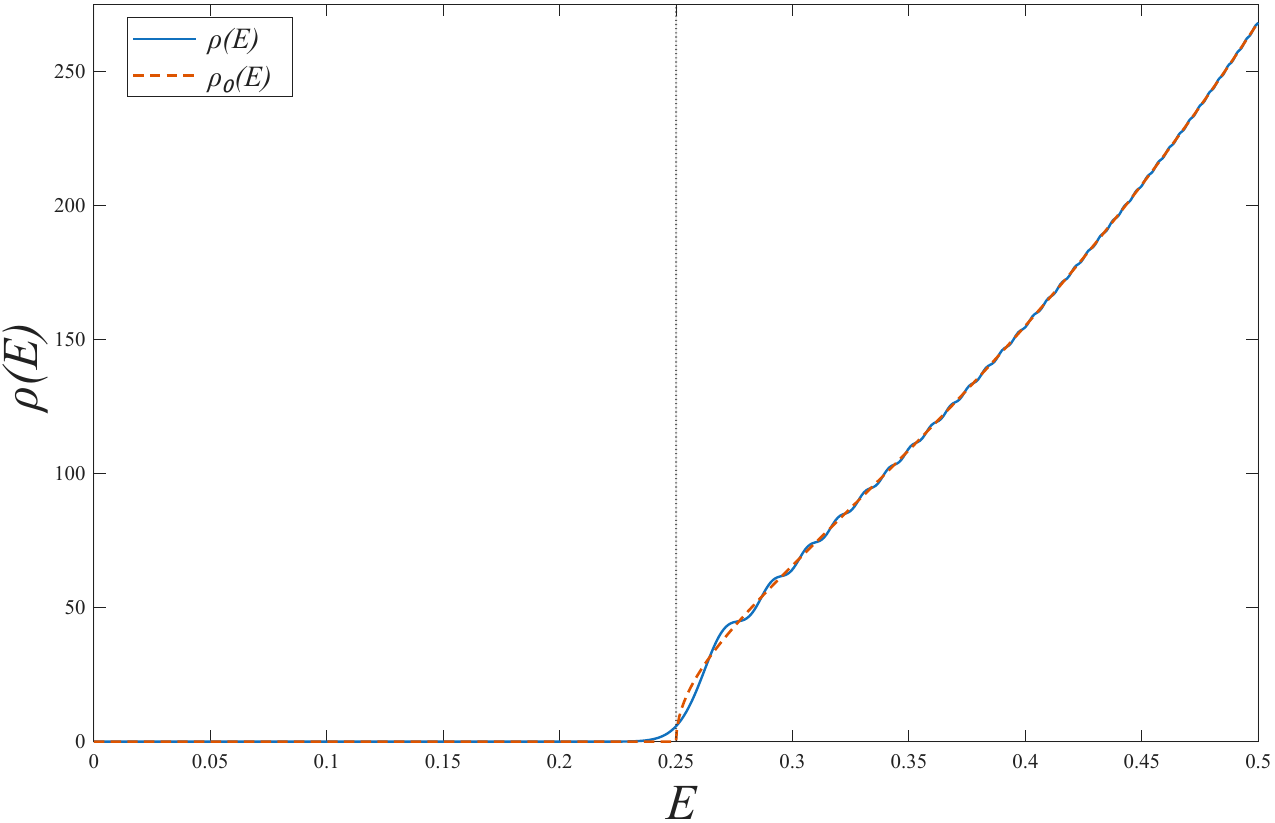}\includegraphics[width=0.5\linewidth]{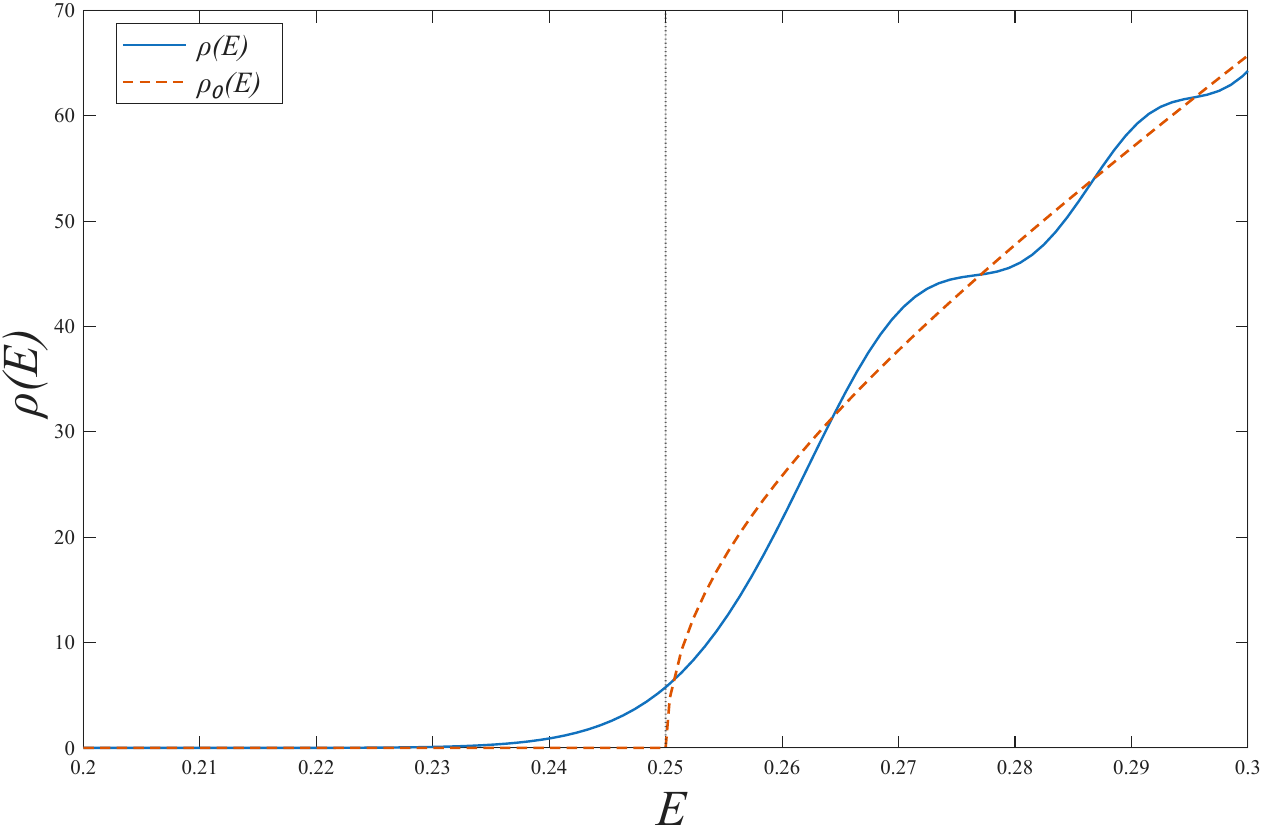}
    \caption{Left: The non-perturbative spectral density computed for the soft edge  $\hat c\,{=}3$ model with $q{=}{-}1$ (left). Right: A zoom into the (resolved) threshold region. Here, $\hbar{=}10^{-3}$. The classical spectral density is plotted as a dashed line for comparison.}
    \label{fig:non-perturbative-plots-0}
\end{figure}

The next example is the $q{=}0$ model from the $\hat c{=}2$ primitive model of Section~\ref{sec:N2-chat2-primitive} (or of Section~\ref{sec:N2-chat2-next}), with classical density given in~(\ref{eq:N2-hc2-hard-unrecombined}). This kind of model was treated fully non-perturbatively in ref.~\cite{Johnson:2020exp}, since it is analogous to a ${\cal N}{=}1$ JT supergravity model. Here,  $\hbar{=}10^{-3}$ is low enough to give a very useful read-across to the classical model, and they are plotted together in figure~\ref{fig:non-perturbative-plots-1}.

\begin{figure}[h]
    \centering \includegraphics[width=0.5\linewidth]{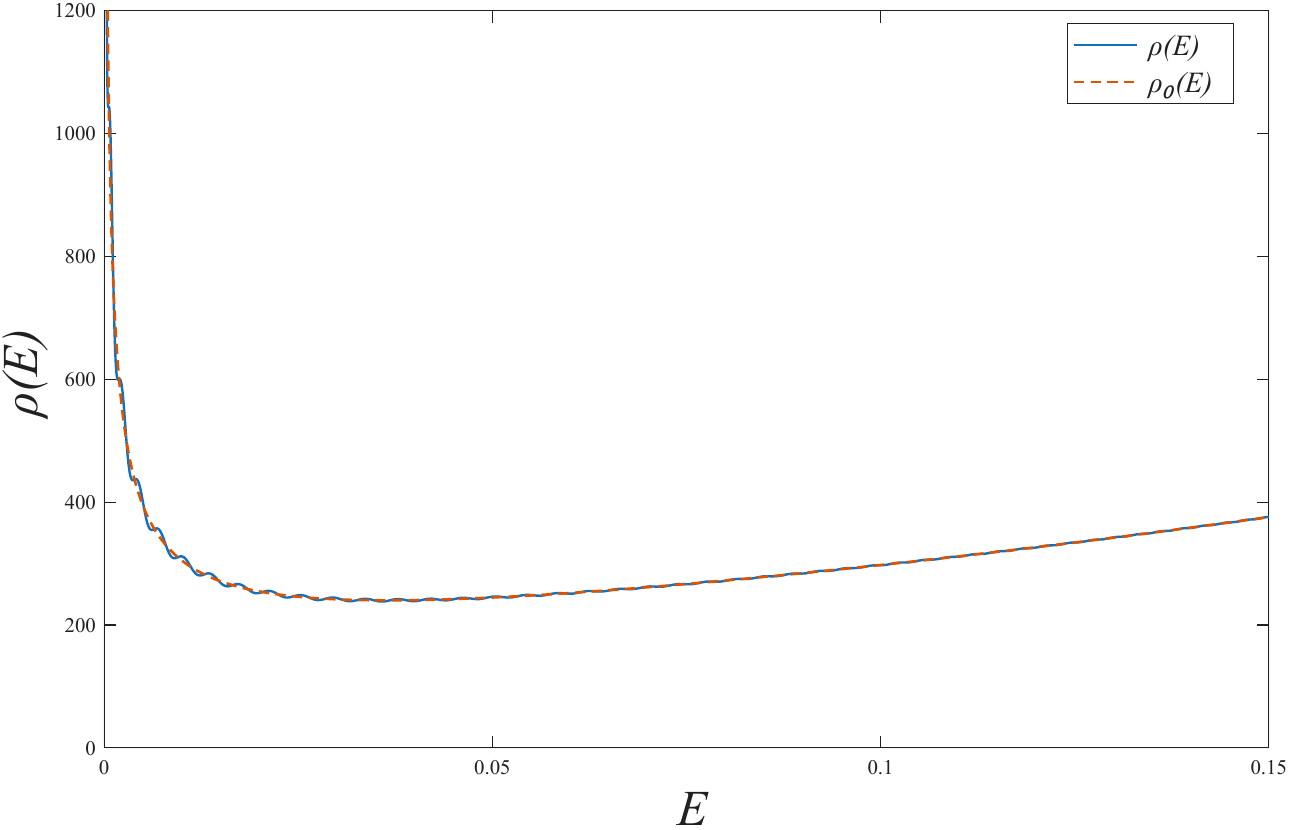}\includegraphics[width=0.5\linewidth]{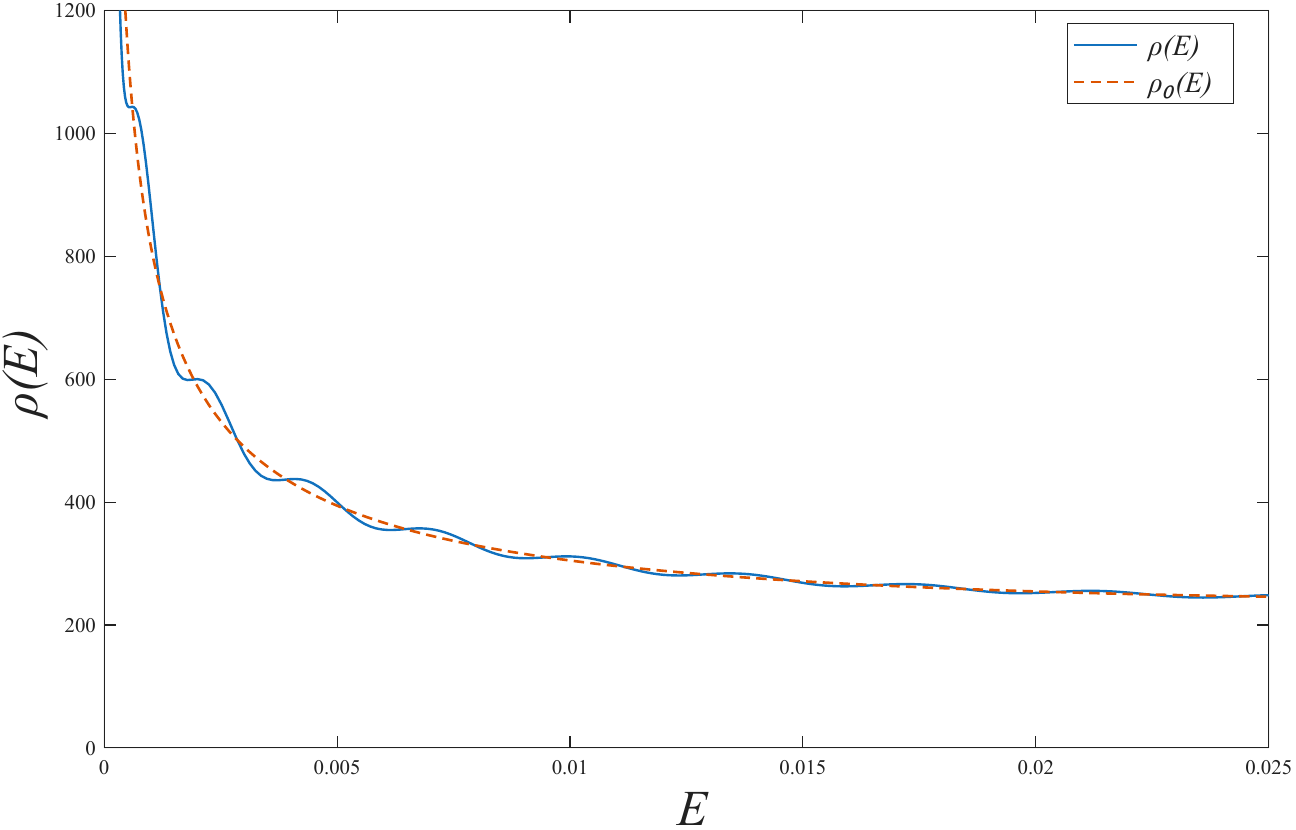}
    \caption{The non-perturbative spectral density computed for the hard edge  $\hat c{=}2$ model with $q{=}0$ (left). On the right is a closeup of the hard edge tail. Here, $\hbar{=}10^{-3}$. The classical spectral density is plotted as a dashed line for comparison.}
    \label{fig:non-perturbative-plots-1}
\end{figure}

Next to be treated non-perturbatively are the two new kinds of spectral density seen for $q{=}\frac12$ and $q{=}\frac14$, with classical densities given in~(\ref{eq:N2-hc2-soft-density}) (or \ref{eq:N2-hc2-m2-half-density}) and (\ref{eq:N2-hc2-m2-quarter-density}) respectively. These   have not previously been explored with these methods, and the non-perturbative spectral densities are displayed in figure~\ref{fig:non-perturbative-plots-2} and~\ref{fig:non-perturbative-plots-3}. Note that, for the $q{=}\frac14$ model, the  non-perturbative solution for $u(x)$ is presented on the right of figure~\ref{fig:sample-potentials} for comparison with the $\hat c{=}3$ $q{=}{-}1$ case that has no BPS states. The main difference is the presence of  the slight potential well of the latter. Turning on non-zero~$\widetilde\Gamma$ ``fills'' it up.

\begin{figure}[h]
    \centering \includegraphics[width=0.5\linewidth]{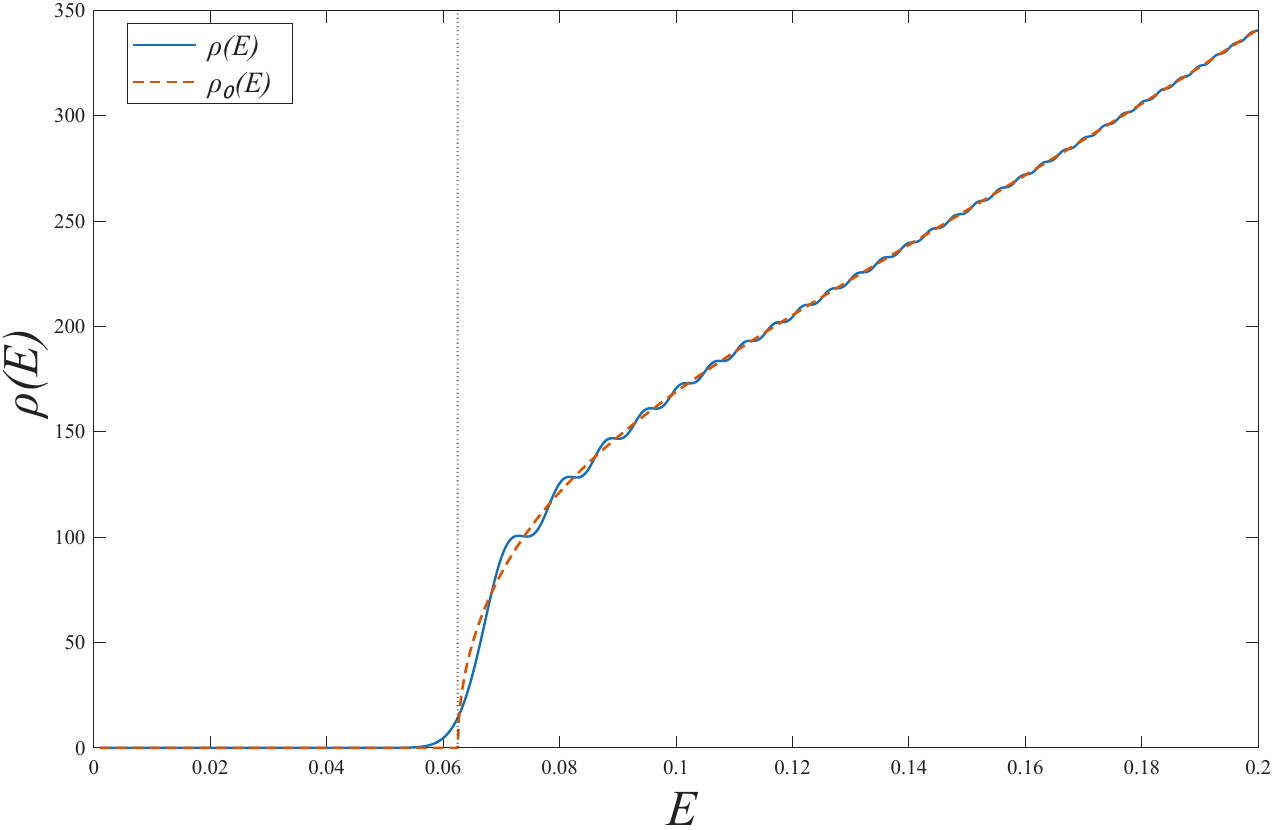}\includegraphics[width=0.5\linewidth]{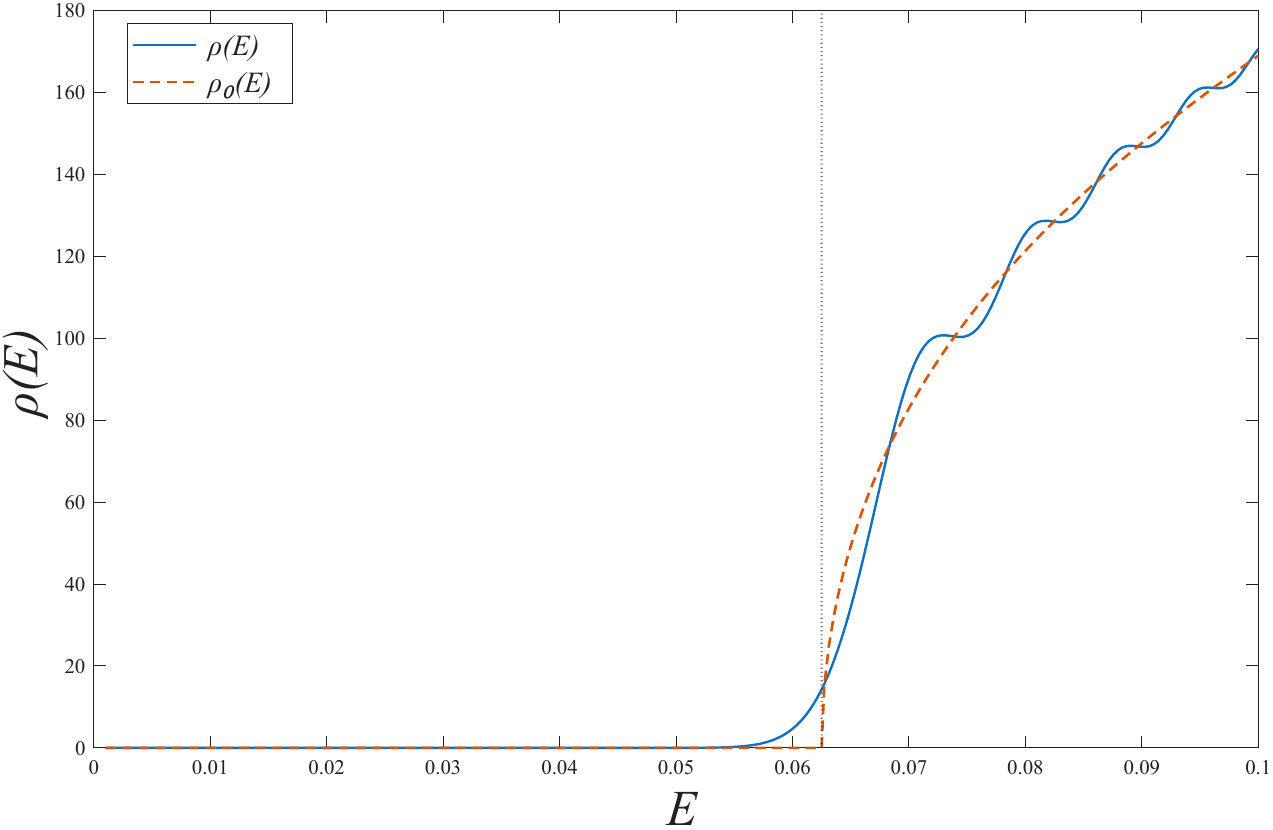}
    \caption{Left: The non-perturbative spectral density computed for the soft edge   $\hat c{=}2$ model with $q{=}\frac12$, using $\hbar{=}10^{-3}$. The classical spectral density~(\ref{eq:N2-hc2-soft-density}) or~(\ref{eq:N2-hc2-m2-half-density}) is plotted as a dashed line for comparison. Right: A zoom-in on the same model data.}
    \label{fig:non-perturbative-plots-2}
\end{figure}

\begin{figure}[h]
    \centering \includegraphics[width=0.5\linewidth]{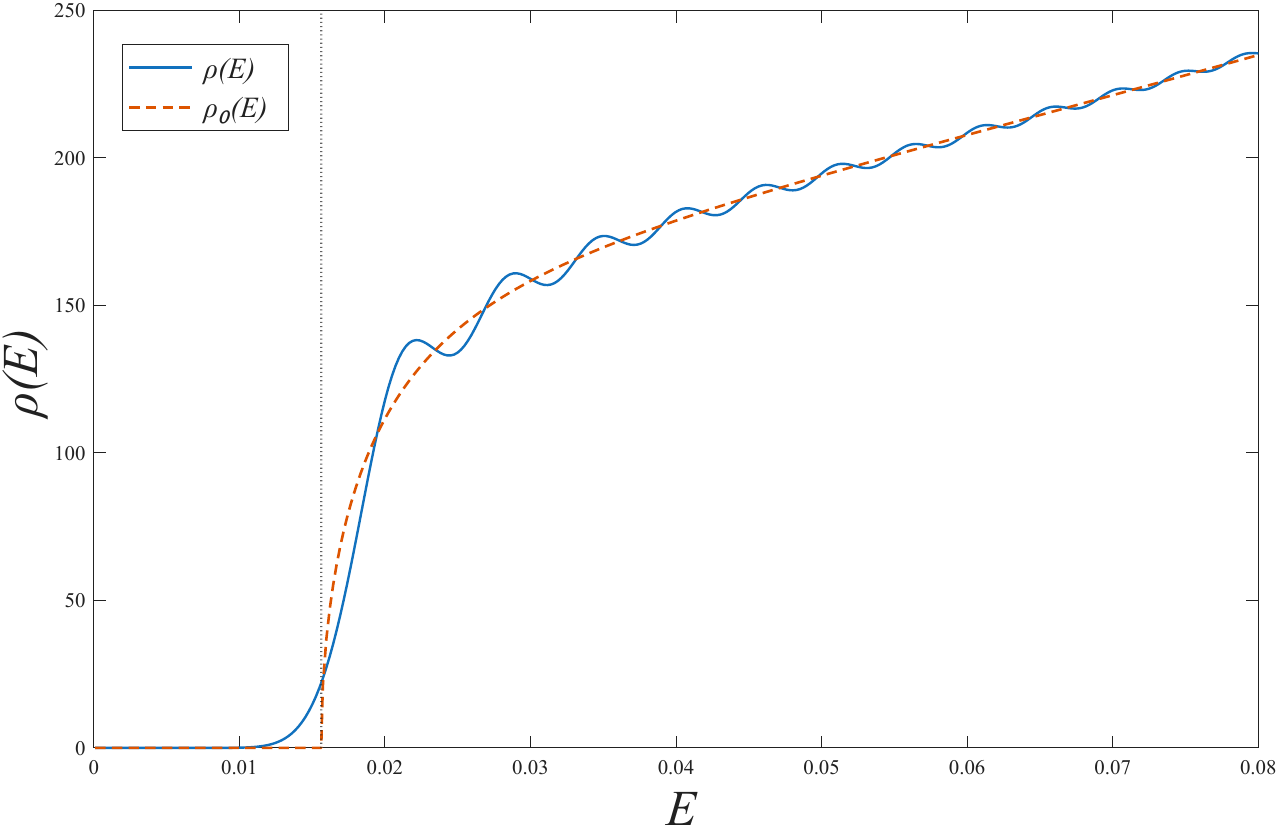}\includegraphics[width=0.5\linewidth]{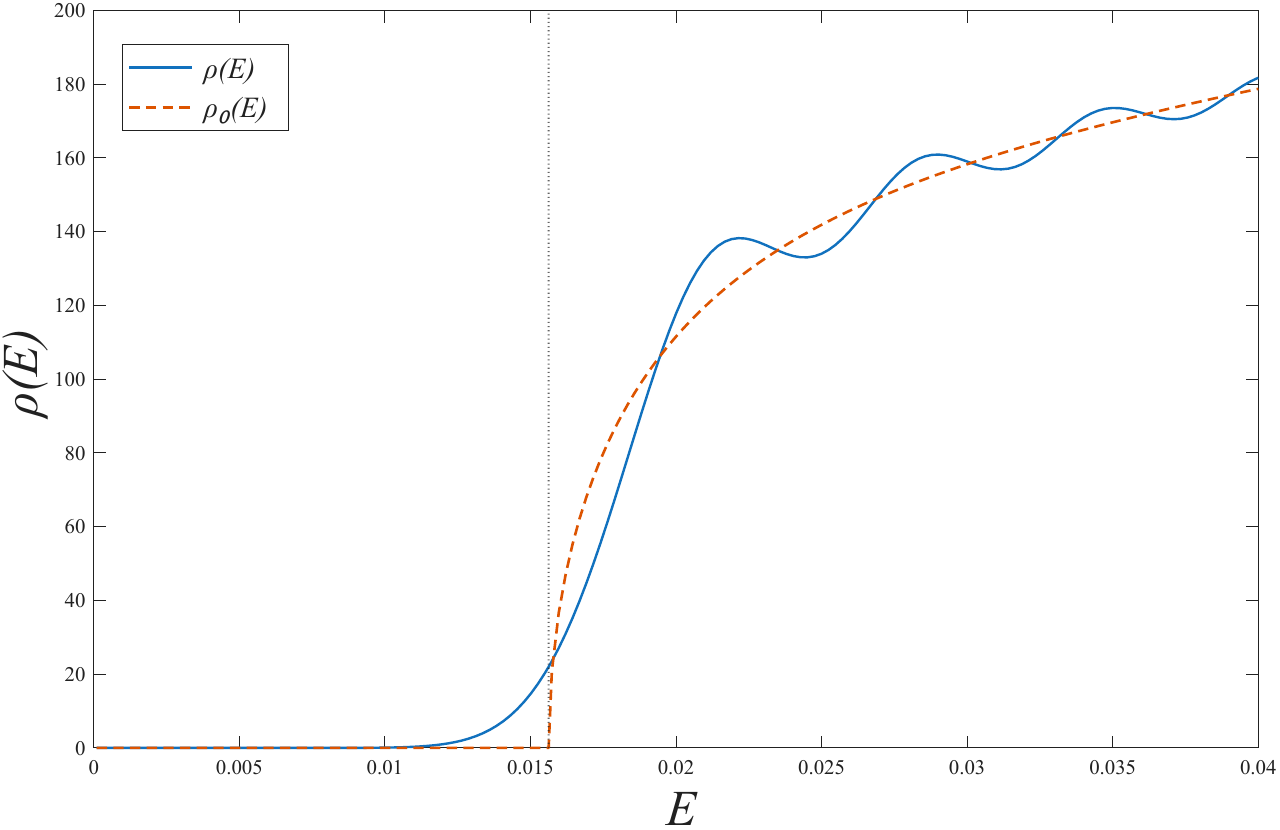}
    \caption{Left: The non-perturbative spectral density computed for the  soft edge $\hat c{=}2$ model with $q{=}\frac14$, using $\hbar{=}10^{-3}$. The classical spectral density~(\ref{eq:N2-hc2-m2-quarter-density}) is plotted as a dashed line for comparison. Right: A zoom-in on the same model data.}
    \label{fig:non-perturbative-plots-3}
\end{figure}

\end{widetext}

\section{New  ${\cal N}=4$ Supersymmetric Strings}
\label{sec:an-N=4-generalization}

The enhanced supersymmetry encountered in Section~\ref{sec:surprise-N=4} invites a
natural extension of our successful  ${\cal N}{=}2$ construction. Here we will formulate new ``small'' ${\cal N}{=}4$ string theories at
finite level. (Small here means that the R-symmetry is a single $SU(2)$, instead of  the full $SU(2){\times} SU(2){\sim }SO(4)$ allowed for the whole ${\cal N}{=}4$ superconformal algebra.)

The idea is again (following what happens for the prototype VMS) that underlying is a critical string computation that results in the disc/D-brane amplitude becoming the small ${\cal N}{=}4$ vacuum ``Cardy'' density. The construction here will be purely following, as we did before, the random matrix model approach: Having the leading (disc) spectral density in the correct type of matrix model allows the formalism to self-consistently build the scattering amplitudes of the resulting string theory to all orders in perturbation theory, and potentially even define things non-perturbatively (provided solutions to the string equation exist). 

Since we have done all the hard work in the last sections to  set up normalizations,  extract the BPS sector, as well as the endpoint data from which to build the amplitudes, this Section can get to the point very swiftly.

Denote the level of the affine $SU(2)_R$ R-symmetry algebra by non-zero positive integer $\hat k$ (a hat because we've been using   $k$ for the index of the couplings $t_k$). There will be~${\hat k}$ massive channels involved in the modular transformation. They will be labelled by $j=0,\ldots,\hat k-1$, and they have  $SU(2)$ spin $j/2$. The central charge of the theory and  conformal weight of the NS $j$th state  are:
\begin{eqnarray}
 &&c=6\hat k\ ,\quad h
 =2p^2
 +\frac{j(j+2)}{4(\hat k+1)}
 +\frac{\hat k^2}{4(\hat k+1)}\ ,\nonumber\\
 \label{eq:N4-level-def}
\end{eqnarray}
where $2p$ is the continuum momentum in the CFT.

There are NS massless states in the CFT as well, labelled by integer $\ell=0,\cdots,\hat k$, with conformal weight and spin given by $\ell/2$. The vacuum modular transform to be written below only involves the  single massless character with $\ell{=}\hat k$.  It appears in the vacuum modular transform with coefficient
$\hat k+1$.

The exact modular transformation of 
vacuum character (and more background on the facts quoted above) was presented by Eguchi, Sugawara and Taormina
in refs.~\cite{Eguchi:2008ct,Eguchi:2006tu}, and we will write it shortly.
First, it is useful to define an energy and a shifted spin (motivated by spectral flow to the R-sector):
\begin{equation}
 E\equiv\frac12\left(h-\frac j2\right)\ ,
 \qquad
 J\equiv\frac{\hat k-j}{2}= \frac12,1,\frac32,\ldots,\frac{\hat k}{2}\ ,
 \label{eq:N4-energy-J-def}
\end{equation}
giving:
\begin{equation}
 E=p^2+\frac{J^2}{2(\hat k+1)},
 \qquad
 E_0=\frac{J^2}{2(\hat k+1)},
 \qquad
 p=\sqrt{E-E_0}.
 \label{eq:N4-energy-threshold}
\end{equation}
This means that the massless/BPS state mentioned above has~$J{=}0$. It is also  worth remarking here that later, when writing the string amplitudes, the  physical momentum $P$ convention
already introduced in~(\ref{eq:N2-momentum-convention}) will be used. That $P=\frac p2$. It won't be needed yet.

As before, an overall normalization will  be
chosen when going from the CFT modular transform conventions to those of the random matrix model.
Our  choices are such that the density is:
\begin{widetext}
\begin{equation}
 \rho^{(\hat k)}_{0,J}(E)
 =
 \frac{1}{2\pi^2(\hat k+1)\hbar}\,
 \frac{
 \sin\!\left(\frac{2\pi J}{\hat k+1}\right)
 \sinh(2\pi s)
 \sinh\!\left(\frac{2\pi s}{\hat k+1}\right)
 }{
 s\left[
 \cosh\!\left(\frac{2\pi s}{\hat k+1}\right)
 -\cos\!\left(\frac{2\pi J}{\hat k+1}\right)
 \right]^2
 }
 \Theta(E-E_0),
 \qquad
 s=\sqrt{2(\hat k+1)(E-E_0)}.
 \label{eq:N4-density}
\end{equation}

A first interesting observation is that near $s=0$, expanding shows that the scaling is like $s$, {\it i.e.,}
$\rho_{0,J}(E)\sim\sqrt{E-E_0}$,  so there is  a soft edge for {\it all allowed states} because there's no $J=0$ continuum component. (The first sine factor ensures that the continuum vanishes for  $J{=}0$.)
There is also a useful spectral-image representation analogous to the one
used for the ${\cal N}=2$ case.  Define the image quantity analogous to~(\ref{eq:N2-image-definitions}):
\begin{equation}
 R_n\equiv J+n(\hat k+1),\qquad n\in{\mathbb Z}\ ,
 \label{eq:N4-images}
\end{equation}
and then in fact:
\begin{equation}
 \rho^{(\hat k)}_{0,J}(E)
 =
 \frac{\hat k+1}{2\pi^4\hbar}\,
 \sinh(2\pi s)
 \sum_{n\in{\mathbb Z}}
 \frac{R_n}{(R_n^2+s^2)^2}
 \Theta(E-E_0)\ .
 \label{eq:N4-density-images}
\end{equation}
Interestingly, while the ${\cal N}=2$ construction involved the image sum
$\sum_n(s^2+R_n^2)^{-1}$,  (see equation~(\ref{eq:N=2-density-form-one})), here the sum
involves a double-pole structure.

The large $\hat k$ classical limit should return us to small ${\cal N}{=}4$ JT supergravity.  This is most easily done with the image form.  Holding  fixed the JT energy $ E_{\rm JT}\equiv 2(\hat k+1)E$  as well as the spin $J$, we see again that all images except
$n=0$  decouple. Remembering the Jacobian factor 
$(2(\hat k+1))$, equation~(\ref{eq:N4-density-images})
yields:
\begin{equation}
 \rho^{({\cal N}=4\,{\rm JT})}_{0,J}(E_{\rm JT})
 =
 \frac{J}{\pi^2\hbar_{\rm JT}E_{\rm JT}^2}
 \sinh\!\left(2\pi\sqrt{E_{\rm JT}-J^2}\right)
 \Theta(E_{\rm JT}-J^2)\ ,
 \label{eq:N4-JT-limit}
\end{equation}
where we have defined  $\hbar_{\rm JT}=4\pi^2\hbar$. Indeed this is the fixed-$J$ small-${\cal N}=4$ JT density in the conventions used in 
refs.~\cite{Turiaci:2023jfa,Johnson:2024tgg,Johnson:2025oty,Heydeman:2025vcc}.

\end{widetext}
At this point the matrix-model reconstruction proceeds exactly as it did
for ${\cal N}=2$.  Applying the Abel inversion
(\ref{eq:N2-Abel-inverse}) determines (implicitly or explicitly) the multicritical couplings $t_{k,J}$, and the
classical string equation is:
\begin{equation}
x+\sum_{k=1}^{\infty}t_{k,J}^{(\hat k)}u_0^k
 =
 \frac{\widetilde\Gamma}{\sqrt{u_0}}.
 \label{eq:N4-classical-string}
\end{equation}
The resulting $t_k$ and $\widetilde\Gamma$ can then simply be inserted
into the full positive-matrix string equation
(\ref{eq:big-string-equation}) if a study of the full non-perturbative physics is to be done.

The BPS study follows in the same way as in
(\ref{eq:N2-residue-Gamma}), by analytically continuing the continuum
density to the origin following ref.~\cite{Turiaci:2023jfa,Johnson:2025oty}.  The result is particularly simple:
\begin{equation}
 \widetilde\Gamma
 =-\frac{\cos(2\pi J)}{2\pi^2},
 \qquad
 |\widetilde\Gamma|=\frac{1}{2\pi^2}\ .
 \label{eq:N4-Gamma}
\end{equation}
So half-integral~$J$ selects the plus branch while integral~$J$ selects
the minus branch.  In the convention of ref.~\cite{Johnson:2025oty}, there is an extra overall factor of $4\pi^2$ relating the $\hbar$s, and so we match the overall $|\widetilde\Gamma|$ result there, which is 2. It is worth emphasizing here that while there is always a non-zero $|\widetilde\Gamma|$ turned on, for any $J$, it is {\it only} for $J{=}0$ that there is a BPS state~\cite{Johnson:2025oty}.

The endpoint data can likewise be obtained directly from the image
representation, just as the even lattice moments
(\ref{eq:N2-lattice-moments}) organized the ${\cal N}{=}2$ calculation.
Here the natural quantities are the odd moments we'll define as:
\begin{equation}
 {\cal T}_{2m+1}(J;\hat k)
 \equiv
 \sum_{n\in{\mathbb Z}}\frac{1}{(J+n(\hat k+1))^{2m+1}},
 \qquad m=1,2,\ldots .
 \label{eq:N4-odd-moments}
\end{equation}
Writing everything in terms of the parameter:
\begin{equation}
 \theta\equiv\frac{\pi J}{\hat k+1},
\end{equation}
the first three are:\footnote{Differentiate the  identity in footnote~\ref{fn:summing-homework} once with respect to $a$ to get: 
$ \sum_{n\in{\mathbb Z}}(n+a)^{-3}
 =\pi^3\cot(\pi a)\csc^2(\pi a).
$ Then higher odd moments come from successive double $a$-differentiations.}
\begin{align}
 {\cal T}_3
 &=
 \frac{\pi^3}{(\hat k+1)^3}\cot\theta\,\csc^2\theta\ ,
 \nonumber\\
 {\cal T}_5
 &=
 \frac{\pi^5}{3(\hat k+1)^5}\cot\theta\,\csc^2\theta
 \left(3\csc^2\theta-1\right)\ ,
 \nonumber\\
 {\cal T}_7
 &=
 \frac{\pi^7}{45(\hat k+1)^7}\cot\theta\,\csc^2\theta
 \left(45\csc^4\theta-30\csc^2\theta+2\right)\ .
 \label{eq:N4-first-moments}
\end{align}
The resulting  first three endpoint
quantities analogous to those defined in~(\ref{eq:N2-endpoint-defintions}) then work out to be:
\begin{widetext}
\begin{align}
 A&=[2(\hat k+1)]^{3/2}\frac{{\cal T}_3}{2\pi^2}\ ,
\quad
 B=[2(\hat k+1)]^{5/2}
 \left(\frac{3{\cal T}_5}{2\pi^2}-\frac12{\cal T}_3\right)\ ,
 \quad
 C=[2(\hat k+1)]^{7/2}
 \left(
 \frac{45{\cal T}_7}{8\pi^2}
 -\frac52{\cal T}_5
 +\frac{\pi^2}{4}{\cal T}_3
 \right)\ .
 \label{eq:N4-ABC}
\end{align}
As a result,  the same useful dimensionless combinations~(\ref{eq:N2-v-kappa}) and~(\ref{eq:N2-eta}) introduced for the 
${\cal N}=2$ analysis yield:
\begin{align}
 v
 &=
 \frac{(\hat k+1)^{3/2}}{4\sqrt2}
 \frac{\sin^3\theta}{\cos\theta}\ ,
 \nonumber\\
 \kappa
 &
 =
 \frac{2}{\hat k+1}
 \left(3\csc^2\theta-1-(\hat k+1)^2\right)\ ,
 \nonumber\\
 \eta
 &=
 \frac{45\csc^4\theta-30\csc^2\theta+2}{(\hat k+1)^2}
 -20\csc^2\theta+\frac{20}{3}+2(\hat k+1)^2\ .
 \label{eq:N4-v-kappa-eta}
\end{align}
\end{widetext}
Since every continuum channel is soft edged, the sample formulae for the scattering amplitudes are
therefore precisely those already displayed in
(\ref{eq:N2-Vgn-general}), with the values
(\ref{eq:N4-v-kappa-eta}) substituted into them.  (Indeed we use the same physical momentum~(\ref{eq:N2-momentum-convention}) and the same Laplace
transform convention~(\ref{eq:quantum-volume-conversion-plus})).  Since there are no hard-edge sectors, our amplitude work is  done. Just as with the ${\cal N}{=}2$ case, higher derivatives of $u_0(\mu)$ can be readily computed at the Reader's leisure in order to work out higher point and higher genus amplitudes, using the simple formulae of ref.~\cite{Johnson:2026twg}. The observations about the structure of the amplitudes/volumes made in the paragraphs below~(\ref{eq:N2-Vgn-general}) apply here too. 

As a final task, let us examine the case of level one, putting into all our formulae the values:
\begin{equation}
 \hat k=1\ ,\qquad j=0\ ,\qquad J=\frac12\ .
\end{equation}
There is only one massive continuum channel here, and after some quick algebra,  equation
(\ref{eq:N4-density}) reduces to:
\begin{equation}
 \rho^{(\hat k=1)}_{0,1/2}(E)
 =
 \frac{1}{4\pi^2\hbar}
 \frac{\sinh^2(2\pi p)}
 {p\cosh(2\pi p)}
 \Theta\!\left(E-\frac1{16}\right)\ ,
 \label{eq:N4-level-one-density}
\end{equation}
where $p\equiv\sqrt{E-\frac1{16}}$.
This is exactly the  $\hat c=2$, $m=1$, $q=\pm\frac12$,
${\cal N}=2$ density~(\ref{eq:N2-hc2-soft-density}).  Moreover, the $J$ independent value  we have for $|\widetilde\Gamma|=\frac{1}{2\pi^2}$ means that for the BPS sector we have 
in exact agreement with~(\ref{eq:N2-hc2-Gamma}) (where recall the $\frac12$ subscript there is the $U(1)_R$ $q$-charge of the BPS state).  

The endpoint formulae indeed yield the same values as that model:
\begin{equation}
 A=\pi,\qquad
 B=\pi^3,\qquad
 C=\frac{31\pi^5}{6},
\end{equation}
and hence:
\begin{equation}
 v=\frac14\ ,\qquad
 \kappa=1\ ,\qquad
 \eta=\frac{31}{6}\ .
\end{equation}
Using these in~(\ref{eq:N2-Vgn-general}) therefore gives
%
{\it exactly} the amplitudes already obtained in
(\ref{eq:N2-hc2-m1-half-Vgn})!  This confirms the supersymmetry enhancement discussed in Section~\ref{sec:surprise-N=4}, and cleanly shows the absence of the  $q=0$ hard-edge sector once the  
${\cal N}=2$ model is  extended to a 
small-${\cal N}=4_1$ theory.

Note that the non-perturbative analysis of the primitive $N{=}2$ model done in Section~\ref{sec:non-perturbative-studies} (see {\it e.g.,} figure~\ref{fig:non-perturbative-plots-2}) constitutes a first non-perturbative analysis of one of our ${\cal N}{=}4$ models. It would be interesting to explore more such aspects of these models for other $J$, especially since the results of ref.~\cite{Johnson:2025oty} strongly suggest that in the ${\cal N}{=}4$ JT supergravity limit, the string equation does not admit the right kind of non-perturbative solutions for $J{>}\frac12$.\footnote{Indeed, applying this paper's improved numerical methods to the classical limit, while a good $J{=}\frac12$ solution for $u(x)$ is readily found at $\hbar{=}10^{-3}$, a study of $J{=}1$ does not find a (well-behaved) solution.} We will leave this exploration as a future project.

 \section{Conclusion}
 \label{sec:closing}
 The original VMS construction~\cite{Collier:2023cyw} pointed out something quite remarkable:  A critical string with the right conformal field theory ingredients  results in a disc amplitude that (after a Laplace transform) yields a density of states that is simply  the vacuum Cardy density of states of a  non-compact Liouville-type conformal field theory. On the other hand, the appropriate double-scaled random matrix model is a marvelous machine for taking that density of states and self-consistently constructing the full family of amplitudes $V_{g,n}(\{P_i\})$ of the string theory, with $n$ external states ($i{=}1,\cdots,n$) at loop order given by genus~$g$. The random matrix model is even capable of yielding data well beyond genus perturbation theory.

This paper (continuing what we began in ref.~\cite{Johnson:2024fkm}) simply takes the two key ingredients of this striking  picture and pushes it in a natural direction: Vacuum densities for CFTs with supersymmetry naturally have a rich and intricate structure and should therefore define, through random matrix model techniques,  similarly rich families of new kinds of string. (This relies on identifying the correct kind of random matrix model--ensembles of positive matrices emerge naturally as the right approach.)

For ${\cal N}{=}1$ this immediately paid off,  with the powerful ``string equation'' framework allowing for a variety of choices of string construction (explored recently in~\cite{Eberhardt:2026hfh}) to find a  home in a simple, unified framework, while predicting new kinds of string as well. 

The completely new ${\cal N}{=}2$ and ${\cal N}{=}4$ families of string theory, constructed and explored here for the first time, yielded a diverse array of new models, and also revealed that some of the various ${\cal N}{=}1$ strings can  appear together as subsectors. Discrete choices such as inserting $(-1)^F$ in the ${\cal N}{=}1$ setting become natural consequences of having spectral flow and $U(1)_R$ charge sectors in ${\cal N}{=}2$, which was satisfying to see. An extra feature is that in addition to the continuous momenta, there are the BPS states, a subset of which appear in the theories too. These would appear to be special ``discrete states'' in the string theory at imaginary momentum, and are worth further study.

It is worth remarking here that even if the results of this paper are not of interest as string theories, the amplitudes $V_{g,n}$ they define represent a particularly natural family of ``quantum'' deformations of the Weil-Petersson volumes of JT supergravity theories with extended supersymmetry~\cite{Turiaci:2023jfa,Ahmed:2025lxe}.  Such objects are interesting to study in their own right, for both  mathematical and physical reasons.

Non-perturbative analysis of some of the simplest of the ${\cal N}{=}2$ string models showed that they are well defined, and it would be interesting to extend this analysis  further to (perhaps all) of the models in some comprehensive way.  One of the ${\cal N}{=}4$ models was also shown to be well defined non-perturbatively, but this followed because it shared features with a special ${\cal N}{=}2$ model at an enhanced symmetry point. Overall, there is much left to explore in that regard for both the  ${\cal N}{=}{2}$ family  and the small ${\cal N}{=}4$ family (see also   the  comments at the very end of Section~\ref{sec:an-N=4-generalization}).

The powerful techniques recently introduced in ref.~\cite{Johnson:2026twg} for computing correlators  $W_{g,n}$ and amplitudes $V_{g,n}$ were particularly useful here. They make use of the fact that the leading density $\rho_0(E)$ is equivalent to a function $u_0(x)$, whose derivatives at the endpoint $x{=}\mu$ can be combined in a simple systematic way to build arbitrary correlators. In some cases, this approach allows for powerful closed-form formulae for amplitudes to be derived. Moreover, it allows (through the parent string equation) for a natural way to derive formulae for the cases of Ramond insertions, and even derivation of the underlying spectral curve describing such sectors~\cite{Johnson:2026jls}.  

Incidentally, in connection with spectral curves and topological recursion~\cite{Chekhov:2006vd,Eynard:2007kz},  it would be interesting to see how far that approach can go here. There, the  density~$\rho_0(E)$ itself   defines a spectral curve by analytic continuation, from which the machinery computes the $W_{g,n}$ recursively. An interesting question is whether all of the various new densities uncovered by our new constructions  can be fully analyzed as spectral curves and successfully yield correlators this way. This would be interesting to explore. This would provide interesting connections to the intersection theory and perhaps 3D gravity connections that the VMS and its generalizations are known to provide~\cite{Collier:2023fwi,Eberhardt:2026hfh}.

In closing, this project still feels closer to the beginning of an exciting program for critical string theory than it does to being near the end of an amusing recreational foray into extended supersymmetry. Many remarkable and un-looked-for simplifications occurred and computational avenues opened up during the course of the project, and it is to be  hoped that this promises more to come.

 \section*{acknowledgments}
 
 CVJ  thanks  the  US Department of Energy for funds (under award \#\protect{DE-SC} 0011702). This manuscript was finished at the Aspen Center for Physics, which is funded under NSF grant \#\protect{NSF PHY-2309135}. CVJ also thanks the University of California Santa Barbara for support, and Amelia for her support and patience.    

\appendix
\section{Numerical Methods}

\subsection{Numerical solution of the string equation.}
\label{app:numerical-recipes-stringeq}
Task one of the non-perturbative study is obtaining a full
non-perturbative solution $u(x)$ of the string equation
(\ref{eq:big-string-equation}). Such studies by the Author go back many years to {\tt FORTRAN} work presented in ref.~\cite{Dalley:1991qg} (using NAG  solver {\tt D02RAF}), a flurry of studies in {\it e.g.,} refs~\cite{Carlisle:2005mk,Carlisle:2005wa}, and then improved powerful techniques in {\tt MATLAB} for studying 15th order versions of the equation in various JT gravity studies beginning with ref.~\cite{Johnson:2020exp}, and refined through to~ref.~\cite{Johnson:2022wsr}. While those latter physics applications in principle have all the $t_k$ turned on, in practice the system is
truncated at some finite $k{=}K$. The calculations presented in this paper use $K{=}7$. See footnote~\ref{fn:truncation}.

All of those numerical approaches used the physical form of the equation in terms of coordinate $x$ and parameter $\hbar$. Writing everything as a first order system introduces a weakness, however: The division by high derivatives to write that form also divides by high powers of $\hbar$. Hence, the study of small $\hbar$ becomes numerically challenging.
It is therefore numerically advantageous to introduce the scaled coordinate
\begin{equation}
 s=\frac{x}{\hbar}\ .
 \label{eq:app-scaled-coordinate}
\end{equation}
With the normalization of the Gel'fand--Dikii polynomials used here (see (\ref{eq:GD-polynomials})), there is an
$\hbar$ accompanying every derivative, and so their $\hbar$ content is absorbed
into the change of coordinate.  Differentiating the string equation gives
the convenient form (factor of $\cal R$ is common to all terms):
\begin{equation}
 u_s{\cal R}+2u{\cal R}_s-\frac12{\cal R}_{sss}=0,
 \qquad
 {\cal R}=\hbar s+\sum_{k=1}^{K}t_k R_k[u] ,
 \label{eq:app-differentiated-string}
\end{equation}
where subscripts denote derivatives with respect to $s$.  This is an equation of order $2K+1$, and it is treated as a boundary
value problem for $u,u_s,\ldots,\partial_s^{2K}u$.

The boundary data are supplied by the known asymptotics of the
string equation described in the body of the paper.  At large negative $x$ the appropriate classical branch
for the model under consideration is used, together with its derivatives,
while at large positive $x$ we have:
\begin{equation}
 u(x)\simeq
 \frac{\hbar^2\left(\Gamma^2-\frac14\right)}{x^2}.
 \label{eq:app-string-right-tail}
\end{equation}
The finite numerical boundaries, denoted $[x_L,x_R]$ are chosen sufficiently far into these
asymptotic regions, and  stability under their displacement provides one of the 
checks done on the calculation.

The resulting boundary value problem is solved using {\tt MATLAB}'s
{\tt bvp4c}.  For the small values of $\hbar$ required here it is essential  to
proceed by continuation, starting at $\hbar{=}1$ and gradually reducing it: a converged solution at one value of $\hbar$ is
used as the initial approximation for the next, smaller value, and so on,  with the step size refined as needed at smaller $\hbar$. Continuation could also be done on the parameters $\{t_k,{\widetilde\Gamma}\}$, but here they (as well as other  physical model parameters) are held fixed throughout.  

Since equation~(\ref{eq:app-differentiated-string}) was obtained by taking a derivative, the integration constant is independently monitored by using the original first integral:
\begin{equation}
 {\cal I}(s)
 =
 u{\cal R}^2-\frac12{\cal R}{\cal R}_{ss}
 +\frac14{\cal R}_s^2
 =
 \widetilde\Gamma^2 .
 \label{eq:app-first-integral}
\end{equation}
The constancy of ${\cal I}(s)$, the reported residuals of the boundary-value solver,
and the value of $u(\mu)$ for small~$\hbar$ all provide useful numerical diagnostics.  

At the end of the computation, the result is converted back to the physical coordinate $x{=}\hbar s$, giving the~$u(x)$  needed for the next stage of the computation.

\subsection{Numerical construction of $\rho(E)$}
\label{app:numerical-recipes-schrodinger}
Once a full non-perturbative solution $u(x)$ of the string equation has
been obtained numerically on some interval $[x_L,x_R]$, the corresponding spectral density can be constructed by
solving the Schr\"odinger problem~(\ref{eq:schrodinger}). This appendix describes some of the methods used to do so.

First, the numerical string-equation solution is converted into a continuous potential using a shape-preserving cubic interpolating routine.

Note that for the energies of interest, $x_L$ lies deep in the classically forbidden region, where $u(x_L)>E$. Direct evolution of $\psi_E$ from  far into this region is numerically challenging, since its
absolute magnitude is exponentially small.  Instead, following the
notation used in the numerical implementation, define its logarithmic
derivative:
\begin{equation}
 z(x)\equiv \frac{\psi_E'(x)}{\psi_E(x)}\ ,
 \label{eq:app-zdef}
\end{equation} and the Schr\"odinger equation then becomes the Riccati equation:
\begin{equation}
 z'(x)=\frac{u(x)-E}{\hbar^2}-z(x)^2.
 \label{eq:app-riccati}
\end{equation}
At $x_L$, the WKB solution for $\psi_E$ sets the
initial condition:
\begin{equation}
 z_L\equiv z(x_L)
 =
 \frac{\sqrt{u(x_L)-E}}{\hbar}
 -\frac{u'(x_L)}{4\left[u(x_L)-E\right]}
 +\cdots \ ,
 \label{eq:app-zL}
\end{equation}
thereby avoiding numerical representation of the exponentially small normalization of the wavefunction itself. 

We must also keep track of the integrated wavefunction norm that will eventually enter the spectral density.  It is convenient to introduce the normalization-independent quantity:
\begin{equation}
 R(x)\equiv
 \frac{\displaystyle\int_{-\infty}^{x}dx'\,
             \psi_E(x')^2}
      {\psi_E(x)^2}\ ,
 \label{eq:app-Rdef}
\end{equation}
which itself evolves according to:
\begin{equation}
 R'(x)=1-2z(x)R(x)\ ,
 \label{eq:app-Requation}
\end{equation}
with WKB initial condition:
\begin{equation}
 R_L\equiv R(x_L)\simeq \frac{1}{2z_L}.
 \label{eq:app-RL}
\end{equation}
Equations~(\ref{eq:app-riccati}) and~(\ref{eq:app-Requation}) are
hence evolved together from~$x_L$.  In our {\tt MATLAB} implementation this is carried out using the solver {\tt ode15s}, with tight relative and absolute tolerances.  

This all works well in the
forbidden region, but when the solution enters
the oscillatory regime it will become problematic since $z=\psi_E'/\psi_E$ develops poles at zeros
of~$\psi_E$. So the numerical evolution is switched at a point~$x_s$ chosen just inside the forbidden region, normally using:
\begin{equation}
 u(x_s)=E+\delta ,
\end{equation}
where $\delta$ is a small positive number. The root defining~$x_s$ is located numerically; when no such crossing occurs before $x=\mu$, $x_s$ is instead chosen slightly to the left of $\mu$, while remaining in the forbidden region.  With the definitions:
\begin{equation}
 z_S=z(x_s),\qquad R_S=R(x_s),
\end{equation}
we are  free to choose a convenient normalization:
\begin{equation}
 \psi_E(x_s)=1\ ,\qquad
 \psi_E'(x_s)=z_S\ .
 \label{eq:app-switchdata}
\end{equation}
With this choice, $R_S$ is  the accumulated norm to the left of
$x_s$ in the same arbitrary normalization.

From $x_s$ onward the ordinary Schr\"odinger system is evolved,
along with a third equation that accumulates the remaining norm. As a first order system:
\begin{align}
 \psi_E'&=\phi_E\ , \qquad 
 \phi_E'=\frac{u(x)-E}{\hbar^2}\psi_E\ , \qquad
 N'&=\psi_E^2\ ,
 \label{eq:app-directsystem}
\end{align}
with:
\begin{equation}
\psi_E(x_s)=1\ ,\qquad
 \phi_E(x_s)=z_S\ ,\qquad
 N(x_s)=0\ .
\end{equation}
This  stage is evolved with {\tt MATLAB}'s {\tt ode113} solver.
The numerical integration is carried through to $x_R$, while the 
output of {\tt ode113} for $N(x)$ is used to evaluate the accumulated norm at the
Fermi endpoint $x{=}\mu$.  So we have:
\begin{equation}
 N_\mu
 =
 R_S+N(\mu)
 =
 \int_{-\infty}^{\mu}dx\,\psi_E(x)^2\ ,
 \label{eq:app-Nmu}
\end{equation}
but still in the arbitrary normalization chosen at~$x_s$.

So the remaining task is to fix the continuum normalization of the wavefunction. For this we use that  the
string-equation solution has the universal asymptotic positive $x$ behavior
given in~(\ref{eq:app-string-right-tail}),
so beyond $x_R$ the potential is continued using the asymptotic form (\ref{eq:app-string-right-tail}), for which  the corresponding exact  Schr\"odinger basis of wavefunctions is~\cite{Carlisle:2005wa,Johnson:2020heh}:
\begin{equation}
 \phi_J(x)=\sqrt{x}\,
 J_{|\Gamma|}\left(\frac{\sqrt{E}\,x}{\hbar}\right)\ ,
 \quad
 \phi_Y(x)=\sqrt{x}\,
 Y_{|\Gamma|}\left(\frac{\sqrt{E}\,x}{\hbar}\right)\ .
 \label{eq:app-besselbasis}
\end{equation}
At the  boundary $x_R$, the computed wavefunction and its
derivative are therefore matched to:
\begin{equation}
 \psi_E(x)=A_E\phi_J(x)+B_E\phi_Y(x)\ .
 \label{eq:app-besselmatch}
\end{equation}
The two coefficients $A_E$ and $B_E$ are determined directly from
$\psi_E(x_R)$ and $\psi_E'(x_R)$. 

The standard normalization of the Bessel functions is used, while the
physical absolute normalization is fixed below by imposing the continuum energy condition $\langle E|E'\rangle=\delta(E-E')$. As a result, the full non-perturbative
spectral density is finally:
\begin{equation}
 \rho(E)=
 \frac{N_\mu}
 {2\hbar^2\left(A_E^2+B_E^2\right)}\ .
 \label{eq:app-rhofinal}
\end{equation}
A key point is that the arbitrary normalization introduced at $x_s$ cancels, since
under $\psi_E\to C\psi_E$, both $N_\mu$ and
$A_E^2+B_E^2$ acquire the same $C^2$ factor. (In the numerical code the
final expression is evaluated in logarithmic form in order to provide
additional protection against overflow and underflow.)

 The
procedure is carried out independently for each desired value of $E$,
producing the complete non-perturbative $\rho(E)$.

Example {\tt MATLAB} code packages implementing both the solution of the string equation for $u(x)$
 and the construction of $\rho(E)$ have been made publicly
available~\cite{Johnson_N2VMSNumerics_2026}. We acknowledge here the utility
of OpenAI's {\tt GPT-5.6} in assisting with the implementation and debugging
of the numerical improvements and procedures, building on the original artisanally
hand-written codes developed in
refs.~\cite{Johnson:2020exp,Johnson:2022wsr}, as well as documentation.

\bibliographystyle{apsrev4-1}
\bibliography{references,extra_references,extrarefs}

\end{document}